\documentclass{aa} 

\pdfoutput=1
\usepackage{longtable}
\usepackage{multirow}
\usepackage{natbib,afterpage}
\usepackage{color}
\usepackage{amssymb,times}
\usepackage[english]{babel}
\usepackage{subfigure}
\usepackage{ifpdf}
\usepackage{morefloats}
\usepackage[rightcaption]{sidecap}
\usepackage{lscape}
\usepackage{multicol}
\usepackage{booktabs}
\usepackage{stackengine}
\newcommand\xrowht[2][0]{\addstackgap[.5\dimexpr#2\relax]{\vphantom{#1}}}
\usepackage{arydshln}
\usepackage{placeins}

\newif\ifpdf \ifx\pdfoutput\undefined\pdffalse\else\pdftrue\fi
\ifpdf\pdfoutput=1
        \usepackage{graphicx}
        \usepackage{epstopdf}
        \DeclareGraphicsExtensions{.pdf,.png,.gif,.jpg}
        \else \usepackage[dvips]{color,graphicx} \fi

\def\Msun{\hbox{M$_{\odot}$}}               %% solar mass
\def\Rsun{\hbox{R$_{\odot}$}}               %% solar mass
\def\Lsun{\hbox{L$_{\odot}$}}               %% solar luminosity
\def\Rstar{\hbox{R$_{\star}$}}              %% stellar radius
\def\Mdot{\hbox{$\dot{M}$}}               %% Mdot      
\def\arcsec{\hbox{$^{\prime\prime}$}}

\def\Al2O3{\hbox{Al$_2$O$_3$}}

\def\farcs{\hbox{$.\!\!^{\prime\prime}$}}

\begin{document} 

\selectlanguage{english}
\newcommand{\red}{\textcolor[rgb]{1,0,0}}
\newcommand{\blue}{\textcolor[rgb]{0,0,1}}

\newlength{\fntxvi} \newlength{\fntxvii}
\newcommand{\chemical}[1]
{{\fontencoding{OMS}\fontfamily{cmsy}\selectfont
\fntxvi\the\fontdimen16\font
\fntxvii\the\fontdimen17\font
\fontdimen16\font=3pt\fontdimen17\font=3pt
$\mathrm{#1}$
\fontencoding{OMS}\fontfamily{cmys}\selectfont
\fontdimen16\font=\fntxvi \fontdimen17\font=\fntxvii}}

\title{ALMA detection of CO rotational line emission in red supergiant stars of the massive young star cluster RSGC1}
\subtitle{Determination of a new mass-loss rate prescription for red supergiants}
 
 \author{L. Decin\inst{1}
 	%\and B. Davies\inst{2}
	\and A.M.S.~Richards\inst{2}
	\and P. Marchant \inst{1}
	\and H. Sana \inst{1}
   }

  \offprints{Leen.Decin@kuleuven.be}

  \institute{
  %1
  Institute of Astronomy, KU Leuven, Celestijnenlaan 200D, 3001 Leuven, Belgium  
  \email{Leen.Decin@kuleuven.be}
  \and
  %2
  %Astrophysics Research Institute, Liverpool John Moores University, 146 Brownlow Hill, Liverpool L3 5RF, UK
  % \and
  %3
  JBCA, Department Physics and Astronomy, University of Manchester, Manchester M13 9PL, UK  
  %\and
  %4
  }

   \date{Received date; accepted date}

% \abstract{}{}{}{}{} 
% 5 {} token are mandatory
 
  \abstract
   {The fate of stars largely depends on the amount of mass lost during the end stages of evolution. For single stars with an initial mass between $\sim$8--30\,\Msun,  most mass is lost during the red supergiant (RSG) phase, when stellar winds deplete the H-rich envelope. However, the RSG mass-loss rate (\Mdot) is poorly understood theoretically, and so stellar evolution models rely on empirically derived mass-loss rate prescriptions. However, it has been shown that these empirical relations differ largely, with differences up to 2 orders of magnitude.}  
   {We aim to derive a new mass-loss rate prescription for RSGs that is not afflicted with some uncertainties inherent in preceding studies. }
   {We have observed CO rotational line emission towards a sample of RSGs in the open cluster RSGC1 that all are of a similar initial mass. The ALMA CO(2-1) line detections allowed us  to retrieve the gas mass-loss rates (\Mdot$_{\rm{CO}}$). In contrast to 
   	mass-loss rates derived from the analysis of dust spectral features (\Mdot$_{\rm{SED}}$), the data allowed us a direct determination of the wind velocity and no uncertain dust-to-gas correction factor was needed.}
  {Five RSGs in RSGC1 have been detected in CO(2-1). The retrieved \Mdot$_{\rm{CO}}$ values are systematically lower than \Mdot$_{\rm{SED}}$. Although only five RSGs in RSGC1 have been detected, the data allow us to propose a new mass-loss rate relation for M-type red supergiants with effective temperatures between $\sim$3200\,--\,3800\,K that is dependent on the luminosity and initial mass, and that is valid during the phase where nuclear burning determines the evolution along the RSG branch. The new mass-loss rate relation is based on the new \Mdot$_{\rm{CO}}$ values for the RSGs in RSGC1 and on prior \Mdot$_{\rm{SED}}$ values for RSGs in four clusters, including RSGC1. The new \Mdot-prescription yields a good prediction for the mass-loss rate of some well-known Galactic RSGs that are observed in multiple CO rotational lines, including $\alpha$ Ori, $\mu$~Cep and VX~Sgr. Moreover, there are indications that a stronger, potentially eruptive,  mass-loss process  is occurring during some fraction of the RSG lifetime, suggesting that RSGs might experience a phase change in mass loss leading to the wind mass-loss rate dominating the RSG evolution at that stage.}
   {Implementing a lower mass-loss rate in evolution codes for massive stars has important consequences as to the nature of their end-state. A reduction of the RSG mass-loss rate implies that quiescent RSG mass loss is not enough to strip  a single star's hydrogen-rich envelope. Upon core collapse such single stars would explode as RSGs. Mass-loss rates of order $\sim$6 times higher would be needed to strip the H-rich envelope and produce a Wolf-Rayet star while evolving back to the blue side of the Hertzsprung-Russell diagram. Future observations of a larger sample of RSGs in open clusters should allow a more stringent determination of the \Mdot$_{\rm{CO}}$-luminosity relation and a sharper diagnostic as to when the phase change in mass loss is occurring.}

   \keywords{Stars: evolution, Stars: massive, Stars: mass-loss, Stars: supergiants, Galaxies: clusters:
individual: RSGC1, Instrumentation: interferometers}
   
\titlerunning{Mass loss from CO rotational line emission of red supergiants in RSGC1}
  \maketitle

\section{Introduction} \label{Sec:Introduction}

The evolution of massive stars up to the point of supernovae (SNe) remains poorly
understood. The steepness of the initial mass function and their short lifetimes ($\sim$15\,Myr)
make such stars rare, whilst the brevity of their post main-sequence (MS) evolution makes
the direct progenitors of SNe rarer still. The pre-SN mass-loss behaviour is the key
property that determines the appearance of the SN, since it dictates the extent to which
the envelope is stripped prior to explosion. It also determines the nature of the end-state,
that is complete disruption, neutron star, black hole, or total implosion with no supernova \citep[e.g.][]{Heger2003ApJ...591..288H}.

The most common of the core-collapse SNe are of type IIP, which are observed to have
red supergiants (RSGs) as their direct progenitors. 
\citet{Smartt2009ARA&A..47...63S} noted that the range of initial masses of these SN progenitors inferred from pre-explosion photometry, 8$\le$$M$/\Msun$\le$17, is at odds with conventional theory, which predicts that
the upper mass limit should be closer to $\sim$30\,\Msun; referred to as the `red supergiant problem' \citep[e.g.][]{Ekstrom2012A&A...537A.146E}. 
A potential explanation for this discrepancy is that the missing RSGs (i.e.\ those with an initial mass between 17\,--\,30\,\Msun) collapse to form black holes with no observable SNe. Later on, \citet{Davies2020MNRAS.493..468D} cautioned that this observational cutoff (of 17\,\Msun) is more likely to be higher and is fraught with large uncertainties (19$^{+4}_{-2}$\,\Msun) and that  also  the upper mass limit from theoretical  models should be shifted downwards to $\sim$25\,--\,27\,\Msun.  One of the main uncertainties in both the data analysis and stellar evolution predictions is our relatively poor knowledge of RSG mass-loss rates.

Mass loss during the RSG phase can affect the progenitors of SNe in two ways. Firstly,
increased mass loss can strip the star of a substantial fraction of the envelope, causing the
star to evolve back to the blue before SN \citep{Georgy2012A&A...538L...8G}, and possibly depleting the
stellar envelope of hydrogen (hence changing what would have been a Type-II SN into a
Type-I SN). Secondly, the mass ejected can enshroud the star in dust, increasing the visual
extinction by several magnitudes \citep[e.g.][]{deWit2008ApJ...685L..75D, Beasor2018MNRAS.475...55B}, and causing
the observer to underestimate the pre-SN luminosity of the star, or perhaps preventing the
progenitor from being detectable at all \citep{Walmswell2012MNRAS.419.2054W}. Hence, accurate
knowledge of RSG mass-loss rates is crucial to our understanding of stellar evolution and
SN progenitors.

Despite this, the mass-loss rates (\Mdot) of RSGs are relatively poorly known, in comparison
to the winds of hot massive stars which have been studied extensively \citep[e.g.][]{Sundqvist2011A&A...528A..64S, Hawcroft2021A&A...655A..67H, Rubio2022A&A...658A..61R}. The mass-loss rate
recipes most often used in evolutionary models are those from \citet{deJager1988A&AS...72..259D} and \citet{Nieuwenhuijzen1990A&A...231..134N}, which are somewhat antiquated as they are basically scaled up from
red giants, and which can only predict \Mdot\ of field RSGs to within $\pm$1 dex (see Fig.~\ref{Fig:Mdot} below).
These recipes assume that \Mdot\ scales with mass, luminosity and temperature,  but do not
take into account how \Mdot\ may change as the opacity of the circumstellar material builds
up over time. What is required is a study of the mass-loss rates of samples of RSGs with
uniform initial abundances and masses, where the evolutionary phase is the only variable.

There are several established methods for measuring \Mdot\ in cool stars. Arguably the best is
to monitor the spectrum of a companion star as it passes behind the primary’s wind \citep[e.g.][]{Kudritzki1978A&A....70..227K}, but unfortunately there are very few such systems. Until
recently, the only way to study large numbers of RSGs was to observe and model the
infrared excess arising from the circumstellar dust \citep[e.g.][]{vanLoon1999A&A...351..559V, Bonanos2010AJ....140..416B}. However, most of the material is in molecular gas, and so a large ($\sim$$\times$200--500),
uncertain correction factor must be applied to convert the measured dust mass-loss rate into a gas mass-loss rate (hereafter referred to as \Mdot$_{\rm{SED}}$), while there is no information on the outflow
speed or radial density profile (required to get \Mdot). 
	Alternatively,  OH masers can be used to measure the gas wind speed. The modelling of the spectral energy distribution (SED) for an assumed gas-to-dust ratio yields a predicted wind speed that can be compared to the observed one. The scaling of the SED models to account for the difference between expected and measured expansion velocity, then yields the gas-to-dust ratios and mass-loss rates of the sample under study \citep[see, e.g.,][]{Goldman2017MNRAS.465..403G}.
A much better way is to  observe the gas using CO molecular line transitions to derive the expansion velocity and gas mass-loss rate directly from the CO line profiles --- hereafter referred to as \Mdot$_{\rm{CO}}$ \citep[e.g.,][]{Knapp1985ApJ...292..640K, Loup1993A&AS...99..291L, Josselin1998A&AS..129...45J, Decin2006A&A...456..549D, Ramstedt2008A&A...487..645R, Danilovich2015A&A...581A..60D}.
The faintness of these lines has meant that, until now, such observations were only possible for
nearby bright RSGs. One exception for the detection of CO from an extragalactic RSG is IRAS~05280$-$6910 situated in the Large Magellanic Cloud, but the spectral resolution of the data acquired with the \textit{Herschel Space Observatory} did not allow the expansion velocity to be measured \citep{Matsuura2016MNRAS.462.2995M}.
Now, with the immense gain in sensitivity provided by the Atacama Large  Millimeter/submillimeter Array (ALMA), we can directly measure the expansion velocity and gas mass-loss rates of homogeneous samples of RSGs with well-constrained distance and stellar parameters.

With this study, we aim to provide the first measurements of the gas mass-loss rates (\Mdot$_{\rm{CO}}$) of RSGs as a function of the specific RSG age. To do this, we have identified  a sample of RSGs which have roughly the
same masses and identical initial chemical compositions. Such samples are uniquely
found in star clusters. The cluster RSGC1 at a distance, $d$, of $\sim$6\,600\,pc contains 14 RSGs and one post-RSG, all with initial masses $\sim$25\,\Msun\ \citep{Davies2008ApJ...676.1016D, Beasor2020MNRAS.492.5994B}. The cluster is effectively coeval, since even a large cluster age spread (0.5\,Myr) would be short compared to the cluster age of 12\,Myr \citep{Davies2008ApJ...676.1016D}. This means that the range of initial masses of the RSGs in RSGC1 must be narrow, and since all stars with this initial mass must follow virtually the same evolutionary track on the Hertzsprung-Russell (HR)
diagram, the differences in the stars’ luminosities are entirely due to how evolved they are.
By measuring the mass-loss rates of these stars, we will be able to determine not only
accurate mass-loss rates for a sample of RSGs, but also how mass-loss
behaviour changes as the star evolves. In  a next step, using these measurements in conjunction with model estimates of the RSG lifetimes, one can then integrate over all stars in the
cluster to find the total mass lost during the RSG phase, a key property when estimating
the fate of a star and the type of SN it will produce. 

\section{Observations and data reduction}\label{Sec:Observations}

%Figure produced using /Users/leen/leda/ALMA/RSGs/programs/plot_lineprofiles_new_data_final.pro
\begin{figure*}[htp]
	\centering	\includegraphics[angle=270,width=.8\textwidth]{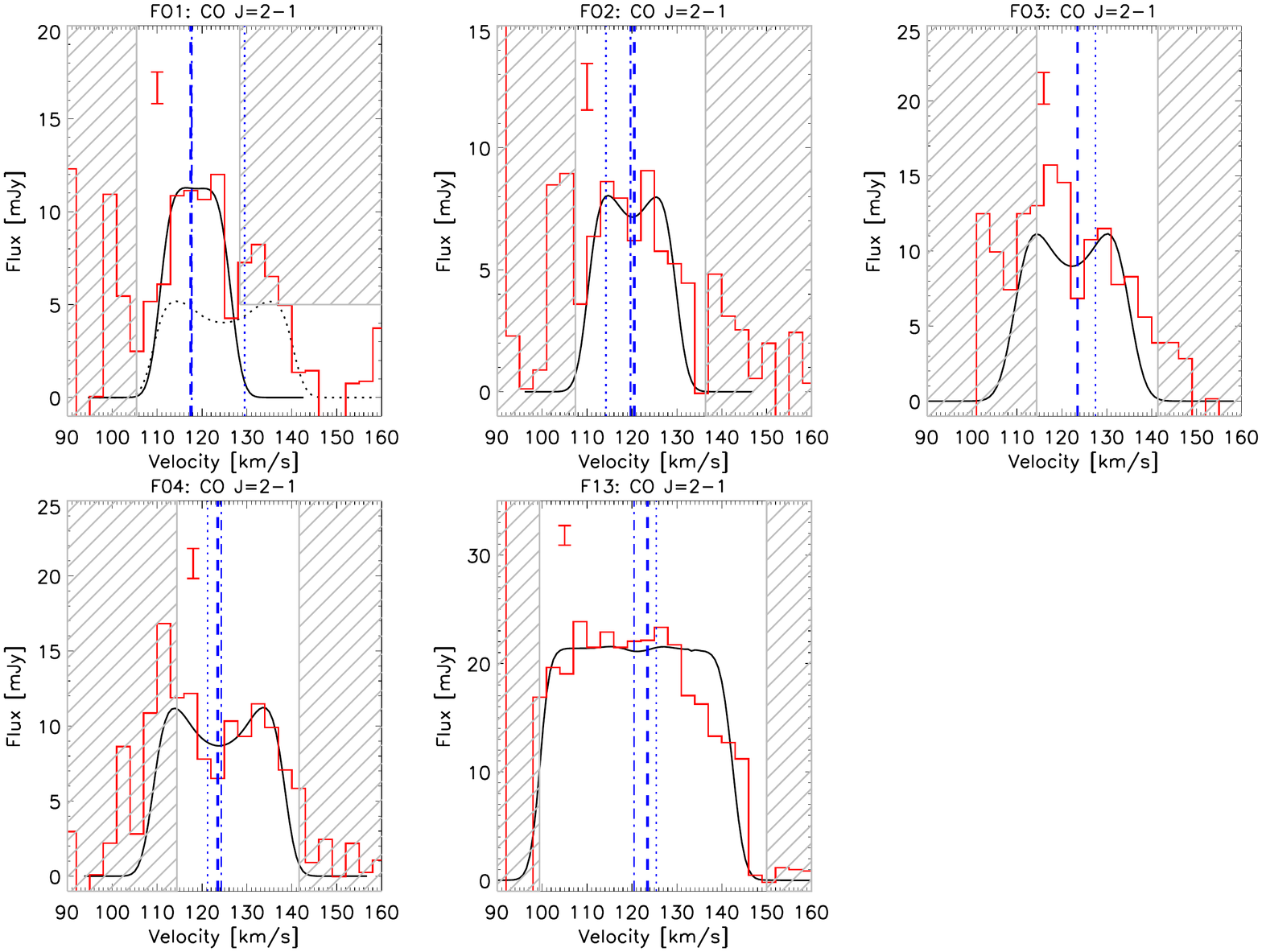}
	\caption{CO(2-1) line profiles of 5 red supergiants in RSGC1. The ALMA data are plotted as red histograms. Synthetic line profiles (see Sect.~\ref{Sec:Analysis}) are overplotted as black solid lines. The grey dashed regions indicate frequency regions that are contaminated by the ISM, a strong noise background, or other genuine emission (see App.~\ref{App:ALMA}).  The vertical dashed blue lines indicate the local standard of rest velocity, $v_{\rm{LSR}}$, as deduced from the ALMA CO data (see App.~\ref{App:ALMA}). The dotted and dashed-dotted blue lines indicate the $v_{\rm{LSR}}$ value as determined by \citet{Davies2008ApJ...676.1016D} and \citet{Nakashima2006ApJ...647L.139N}, respectively (see Table~\ref{Tab:targets}).
		The noise in the spectrum is 1.2\,--\,1.4\,mJy (see text for derivation) and is shown as (red) error bar in each panel. The dotted black line in the upper left panel is an alternative fit to the CO(2-1) line profile of F01, as discussed in App.~\ref{App:CO}.}
	\label{Fig:CO}
\end{figure*}

The 14 RSGs (F01--F14) and 1 post-RSG (F15) in RSGC1 \citep[see Table 2 in ][]{Davies2008ApJ...676.1016D} were observed with ALMA on 2015 June 9 and 11 for proposal code 2013.1.01200.S. We requested observations at both band 9, centred on the CO v=0 J=6-5 rotational line transition and band 6, centred on the CO v=0 J=2-1 rotational line transition, but only the
latter were obtained.  Positions and other details are given in Table~\ref{Tab:targets}. The observations have three spectral windows (spw); one `line' spw with a width of 1.875\,GHz and 3840 channels to cover the CO(2-1) transition, and two `continuum' 2\,GHz spw with 128 channels each centred at 228.5\,GHz and 213\,GHz. The recorded line channels are not independent and the minimum effective spectral resolution of 0.977\,MHz  is approximately double the channel width. The spectral resolution of the continuum data is 15.625\,MHz. 36 antennas were used with minimum and
maximum baselines of 63--783\,m, providing a maximum recoverable scale of $\sim$4\farcs4 in a field-of-view of 26\arcsec. The total integration per target source was 4.8\,min.  Standard ALMA Cycle 2 observing and quality control procedures were
used\footnote{https://almascience.eso.org/documents-and-tools/cycle-2/cycle-2}. The flux scale was set relative to Titan (excluding its atmospheric lines).  Compact quasars J1733-1304 and J1832-1035 were used for bandpass calibration and phase-referencing, respectively.

The data were calibrated using the ALMA Quality Assurance scripts implemented in CASA (the Common Software Applications package) \citep{McMullin2007ASPC..376..127M}. The estimated accuracy of the flux scale as applied to the targets is $\sim$7\%. The target-phase-reference separation is $\sim$3.7$^{\circ}$ (depending on target). Inspection of the (small) slopes in the phase-reference phase solutions, along with the probable antenna position uncertainties in 2015 \citep{Hunter2016SPIE.9914E..2LH}, suggests an absolute astrometric accuracy $\gtrsim1/16$ of the synthesised beam, depending on the target signal to noise ratio (S/N).

We inspected the continuum spw for each target and excluded several channels covering the SiO v=3 J=5-4 line seen around 212.582\,GHz for some sources. We imaged each target, achieving a noise from the full 1.7\,GHz range of $\sigma_{\mathrm{rms}}$$\sim$0.05\,mJy.  The synthesised beam in all images is about (0\farcs49$\times$0\farcs37) at position angle $\sim$--65$^{\circ}$, depending on the frequency.  Table~\ref{Tab:targets} shows that the continuum peaks are less than half the rms in a 3\,km s$^{-1}$ spectral channel, so we did not perform continuum subtraction. Image cubes were made for the spw covering the CO(2-1) line for each source at 3\,km~s$^{-1}$ velocity resolution (approximately 4 input channels), adjusted to constant velocity in the Local Standard of Rest frame ($v_{\mathrm{LSR}}$) with respect to the CO(2-1) rest frequency of 230.538\,GHz. We obtained $\sigma_{\mathrm{rms}}$$\sim$1.9\,mJy. We also made images at 1.3 and 10\,km~s$^{-1}$ resolution but these do not reveal any more detections or significant details.  We imaged the SiO v=3 J=5-4 line which, where detected, covered  2--4  continuum channels (width $\sim$22km~s$^{-1}$).  In these low spectral resolution continuum spw the per-channel   $\sigma_{\mathrm{rms}}$ is $\sim$0.6\,mJy and no other lines were detected.   The parameters for all detections are given in  App.~\ref{App:ALMA}--App.~\ref{App:CO}.

 For five RSGs (F01, F02, F03, F04, and F13), the CO(2-1) line emission was detected, with a spatial extent $\la$1\arcsec\ (see App.~\ref{App:CO}). These are the first detections of spectrally and spatially resolved CO rotational line emission of sources in an open cluster, in this particular case CO emission arising from the stellar wind of red supergiants located in RSGC1. For each of those five RSGs, the CO(2-1) line profile was extracted for a circular aperture of  0\farcs75 centred on the peak of the continuum emission. Fig.~\ref{Fig:CO} shows the observed line profiles.  
The noise in the spectrum (clear of ISM contamination) is given by the $\sigma_{\mathrm{rms}}$ values from Table~\ref{Tab:targets} divided by the square root of the number of beams, resulting in a spectral noise of $\sim$1.2\,--\,1.4\,mJy.
 As visible both in the CO(2-1) channel maps (Fig.~\ref{Fig:CO_F01}--Fig.~\ref{Fig:CO_F13}) and line profiles, there is contamination by interstellar medium (ISM) emission at specific frequencies; see the dashed regions in Fig.~\ref{Fig:CO}. In general, there is much less (or no) ISM contamination visible on the high velocity side, although there might be a noise background. Therefore, the red part of the line profile can also be used to assess the ISM contamination at lower velocities, since  the CO line profiles are expected to be symmetric\footnote{\label{footnote_vel}The blue side of the line profile can be of slightly lower intensity when the source function around optical depth $\tau_{\nu}$$\sim$1 probes more outer regions of the CSE.}. For each source, the ISM contamination and alternative fits are discussed in App.~\ref{App:CO}.

We made total intensity (zeroth moment) maps for each CO line over the uncontaminated channels. The rms-noise in the total intensity maps ranges between 21\,--\,29\,mJy/beam km/s, with corresponding peak signal-to-noise ratio between $\sim$12\,--\,34 (see Sect.~\ref{App:CO}).
We measured the azimuthally averaged flux in annuli 200-mas thick, taking the minimum of the rms or the median average deviation as the error \citep[see][see Fig.~\ref{Fig:CO_annuli}]{Decin2018A&A...615A..28D}. These gave flux distributions with full width half maximum sizes of 460, 400, 480, 600 and 460\,mas for F01, F02, F03, F04, F13, respectively, uncertainty  being $\sim$50\,mas. This gives an indication of the relative sizes of the brightest $\sim$60\% of the emission, rather than the true size, since the flux distribution is not necessarily Gaussian and may be irregular. It is more challenging to estimate the total size of the CO emission, since our observations are sensitivity-limited and provide a lower limit. We estimated where 3$\times$ the rms noise (3$\times$25 mJy beam$^{-1}$ km s$^{-1}$) intersected the azimuthal average profiles. This gave diameters of 700, 530, 770, 650 and  900\,mas for F01, F02, F03, F04, F13, respectively but the  position uncertainty is proportional to the phase noise, $\sim$140 mas at S/N\,=\,3, and the weaker sources, in particular, may be more extended.

\section{Analysis and results}\label{Sec:Analysis}

\begin{table*}[htp]
	\caption{Stellar and CSE parameters for the five RSGs in RSGC1 for which CO(2-1) emission was detected. }
	\label{Tab:outcome}
	\vspace*{-1.5ex}
	\setlength{\tabcolsep}{.7mm}
	\begin{tabular}{l|ccccc|cccc|cc}
		\hline
		\multicolumn{1}{c}{}	& \multicolumn{5}{|c|}{} & \multicolumn{4}{c}{} \\[-2ex]	
		\multicolumn{1}{c}{}	& \multicolumn{5}{|c|}{Stellar parameters} & \multicolumn{6}{c}{CSE parameters} \\
		\hline
		\xrowht[()]{8pt}Star & $\log(L_{\rm{bol}}/L_\odot)$$^{(a,e)}$ & $T_{\rm{eff}}$ $^{(a)}$& Spectral$^{(a)}$  & \Rstar$^{(a)}$ & $v_{\rm{LSR}}$$^{(b)}$ & $R_{\rm{dust}}$ & $v_\infty$$^{(b)}$ & \Mdot$_{\rm{CO}}$$^{(b)}$ & \Mdot$_{\rm{SED}}$$^{(c)}$ & \Mdot$_{\rm{CO}}/v_\infty$ & \Mdot$_{\rm{SED}}$/25  \\
		& & [K] & type & [R$_\odot$] & [km~s$^{-1}$] & [\Rstar]   & [km~s$^{-1}$] & [10$^{-6}$\,M$_\odot$/yr] & [10$^{-6}$\,M$_\odot$/yr] &  &  \\
		\hline
		& & & & & & & & & & & \\[-2ex]
		F01 & 5.42$^{+0.12}_{-0.13}$ & 3450$\pm$127 & M5 & 1450 &  117.5 & 3 & 8 & 2.0$^{(e)}$ & 5.57$^{+2.37}_{-2.17}$ & 0.25 & 0.22  \\
		F02 & 5.56$^{+0.12}_{-0.13}$ & 3660$\pm$127 & M2 & 1500 & 120.5 & 3 & 10 & 1.8 & 5.18$^{+2.72}_{-1.75}$  & 0.18 & 0.20 \\
		F03 & 5.24$^{+0.12}_{-0.13}$ & 3450$\pm$127 & M5 & 1200 & 123.5 & 3 & 13 & 3.2$^{(e)}$ &  4.18$^{+3.08}_{-0.84}$ & 0.25 & 0.17\\
		F04 & 5.32$^{+0.11}_{-0.13}$ & 3752$\pm$117 & M1 & 1100 & 123.5 & 3 & 15 & 4.0 &  $-$ & 0.27 & $-$\\
		F13 & 5.45$^{+0.12}_{-0.13}$ & 3590$\pm$45 & \ \ \ M3$^{(d)}$ & 1430  & 123.5 & 4 & 22 & 42 & $-$ & 1.9 & $-$ \\
		\hline
	\end{tabular}
	\tablefoot{Listed are the stellar luminosity $L_{\rm{bol}}$, the effective temperature $T_{\rm{eff}}$, the spectral type, the stellar radius \Rstar, the local standard of rest velocity $v_{\rm{LSR}}$ (see discussion in App.~\ref{App:ALMA}), the dust condensation radius $R_{\rm{dust}}$, the terminal wind velocity $v_\infty$, and the mass-loss rate as deduced from the CO(2-1) line, \Mdot$_{\rm{CO}}$, and as deduced from an SED analysis by \citet{Beasor2020MNRAS.492.5994B}, \Mdot$_{\rm{SED}}$. The last two columns compare a density measure, \Mdot/$v_\infty$, based on the values deduced in this study (column~11) and used in \citet{Beasor2020MNRAS.492.5994B} who assumed a terminal wind velocity of 25\,km~s$^{-1}$ (column~12).\\
		\tablefoottext{a}{From \citet{Davies2008ApJ...676.1016D}.}
		\tablefoottext{b}{As deduced from the ALMA CO(2-1) lines. The uncertainties on \Mdot$_{\rm{CO}}$ are discussed in App.~\ref{App:CO}\,--\,\ref{App:sigmas}. %are of the order of 30%, see app -- for details
		}
		\tablefoottext{c}{From \citet{Beasor2020MNRAS.492.5994B}.}
		\tablefoottext{d}{F13 was classified as K2 by \citet{Davies2008ApJ...676.1016D}. However, later on it became clear that  the CO-spectral type correlation is flawed if a star has a strong stellar wind, and that under these circumstances also the spectral-type -- $T_{\rm{eff}}$ relation from \citet{Levesque2006ApJ...645.1102L} does not hold. We therefore use the effective temperature and spectral type classification from \citet{Messineo2021AJ....162..187M}.}
		\tablefoottext{e}{\citet{Beasor2020MNRAS.492.5994B} have updated the luminosities of the RSGC1 sources published by \citet{Davies2008ApJ...676.1016D}. In particular, \citet{Beasor2020MNRAS.492.5994B} derived $\log L_{\rm{bol}}$(F01) = 5.58 and $\log L_{\rm{bol}}$(F03) = 5.33. Changing the luminosities to the values from \citet{Beasor2020MNRAS.492.5994B} induces an increase in \Mdot$_{\rm{CO}}$ of 10\% for F01 and of 3\% for F03.}
	}
\end{table*}

The five  sources detected in CO(2-1) include (i) the three M-type RSGs with the highest \Mdot$_{\rm{SED}}$ amongst the sample of red supergiants in RSGC1 analysed by \citet{Beasor2020MNRAS.492.5994B} (F01, F02, and F03; with \Mdot$_{\rm{SED}}$ $\ga$4$\times$10$^{-6}$\,\Msun/yr); and (ii) the peculiar RSG F13, which is anomalously red compared to the other RSGs in the cluster \citep{Davies2008ApJ...676.1016D}, and for which the true luminosity is difficult to determine \citep{Beasor2020MNRAS.492.5994B}. Ten sources (F05, F06, F07, F08, F09, F10, F11, F12, F14, F15) remain undetected at a CO(2-1) rms noise value of $\sim$1.7\,mJy/beam. For five of those sources, \citet{Beasor2020MNRAS.492.5994B} derived a mass-loss rate between 1.8$\times$10$^{-7} \le$\Mdot$_{\rm{SED}} \le$8.7$\times$10$^{-7}$\,\Msun/yr, implying that a line sensitivity of at least a factor 5--20 better is required to detect those sources with a similar ALMA setup. 
The Beasor et al.\ \Mdot\ estimates were not yet available at the moment these observations were proposed and a general value of 1$\times$10$^{-6}$\,\Msun/yr was used for calculating the line sensitivities in the proposal. Moreover,  the CO outer envelope radius was then estimated based on \citet{Mamon1988ApJ...328..797M}. However, both the \Mdot-estimate (for the undetected sources) and the CO outer envelope radius (for the detected sources) turned out to be smaller  (see Sect.~\ref{Sec:CO_analysis}), implying an overestimate of the actual CO(2-1) line strength.

For the five sources for which CO(2-1) rotational emission was detected, we derive the properties of the red supergiant's wind from a radiative transfer analysis.  The outcomes, in particular the wind mass-loss rates, \Mdot$_{\rm{CO}}$, are then compared to (literature) parameters retrieved from an analysis of the dust spectral features visible in the SED, \Mdot$_{\rm{SED}}$.

\subsection{Radiative transfer analysis of the CO(2-1) emission}\label{Sec:CO_analysis}

To retrieve the stellar wind parameters from the CO(2-1) line profiles, we used the non-local thermodynamic equilibrium (non-LTE) radiative transfer code {\sc{GASTRoNOoM}} \citep{Decin2006A&A...456..549D} based on a multi-level approximate Newton-Raphson (ANR) operator. The molecular line data are as specified in Appendix A of \citet{Decin2010A&A...516A..69D}. When ray-tracing the modelled circumstellar envelope (CSE), we used a circular model beam with the same extraction aperture (of 0\farcs75 diameter) to allow direct comparisons. The modelled CSEs were divided into 150 shells, evenly spaced on a logarithmic scale from the stellar radius (\Rstar) out to the outer envelope radius. 

The effective temperatures, $T_{\rm{eff}}$, and stellar luminosities, $L_{\rm{bol}}$, were taken from \citet{Davies2008ApJ...676.1016D}; see Table~\ref{Tab:outcome}\footnote{We note that those values are slightly different than the ones used in \citet{Beasor2016MNRAS.463.1269B, Beasor2018MNRAS.475...55B, Beasor2020MNRAS.492.5994B}.}.
This translates into stellar radii of $\sim$1100\,--\,1500\,\Rsun\ (see Table~\ref{Tab:outcome}).
Since we only have one rotational CO line, we cannot put constraints on the radial distribution of the kinetic temperature for the individual sources. We therefore assume that the gas kinetic temperature follows a power-law radial profile
\begin{equation}
	T_{\rm{gas}}(r) = T_{\rm{eff}} \left(\frac{\Rstar}{r}\right)^{\epsilon}
\end{equation}
with $\epsilon=0.6$, since such a power law has been shown to be a good representation of the kinetic temperature in circumstellar envelopes \citep[][and references therein]{Decin2006A&A...456..549D}.

For each of the five detected sources, the radial extent of the CO emission zone is larger than the beam size (of $\sim$0\farcs5), but slightly lower than 1\arcsec\ in diameter (see Fig.~\ref{Fig:CO_mom0}). Given a distance of 6\,600\,pc to RSGC1 and stellar radii between 7.55$\times$10$^{13}$\,--\,1.04$\times$10$^{14}$\,cm, this translates into a CO envelope radius of 250$<$\,$R_{\rm{out}}$\,<$\,$650\,\Rstar. While the spectra of F01 and F13 are not spatially resolved for an extraction aperture of 0\farcs75, the spectra of F03 and F04 show the clear absorption depth reminiscent of spatially resolving the CSE. This is in line with the angular sizes estimated from the zeroth moment maps which imply that F03 and F04 have the largest angular extents (Sect.~\ref{Sec:Observations}). These considerations yield a CO envelope radius of $\sim$350\,\Rstar\ for F01 and F13, of $\sim$400\,\Rstar\ for F02, and of $\sim$600\,\Rstar\ for F03 and F04. 
Those values are smaller by a factor of $\sim$1.5\,--\,2 compared to the value used for the proposal preparation, that was based on the $r_{1/2}$-value --- which marks the radius where the CO abundance drops to half of its initial value  --- of \citet{Mamon1988ApJ...328..797M} for a wind mass-loss rate of 1$\times$10$^{-6}$\,\Msun/yr.

To calculate the CO excitation and hence level populations, we account for excitation by the stellar photons,  the microwave background and the dust radiation field  \citep{Decin2006A&A...456..549D}. For the latter, we assume silicates \citep{Decin2006A&A...456..549D} that start condensing at a temperature of $\sim$1750\,K, which translates into a dust condensation radius, $R_{\rm{dust}}$, of 3--4\,\Rstar. To get the local mean radiation field at each radial point in the grid, we calculate the dust radiation field for a canonical gas-to-dust ratio ($r_{\rm{gd}}$) of 200, representative for a Milky Way cluster \citep{Beasor2020MNRAS.492.5994B}. As we discuss in App.~\ref{App:sigmas}, the specific dust-to-gas ratio has only a minimal effect on the CO excitation  by the dust radiation field for these simulations.

%/Users/leen/leda/ALMA/RSGs/programs/compute_mdots_fixed_mass_RSG.pro
\begin{figure*}[htpb]
	\includegraphics[angle=0,height=5.5cm]{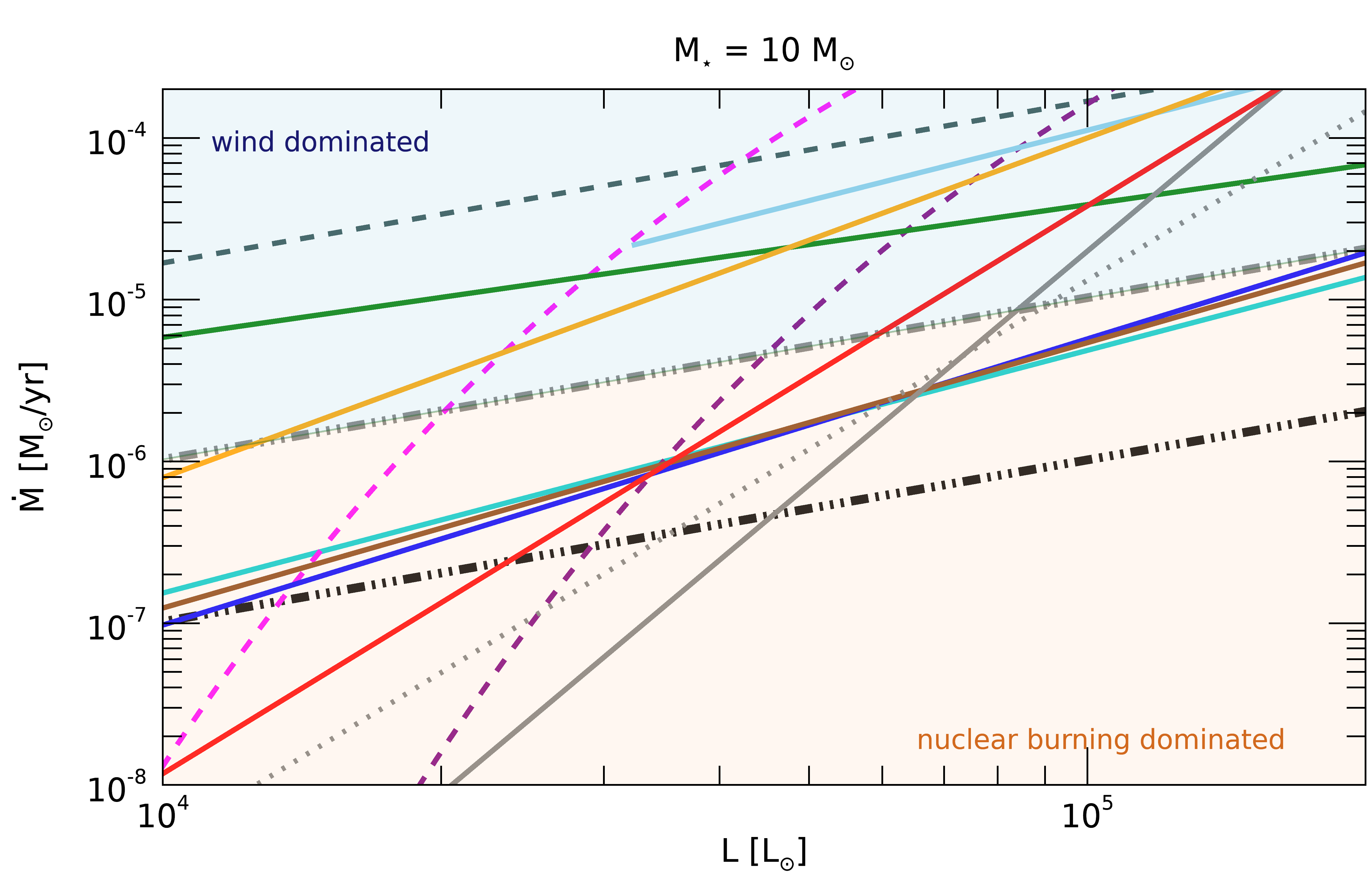}
	\hfill
	\includegraphics[angle=0,height=5.5cm]{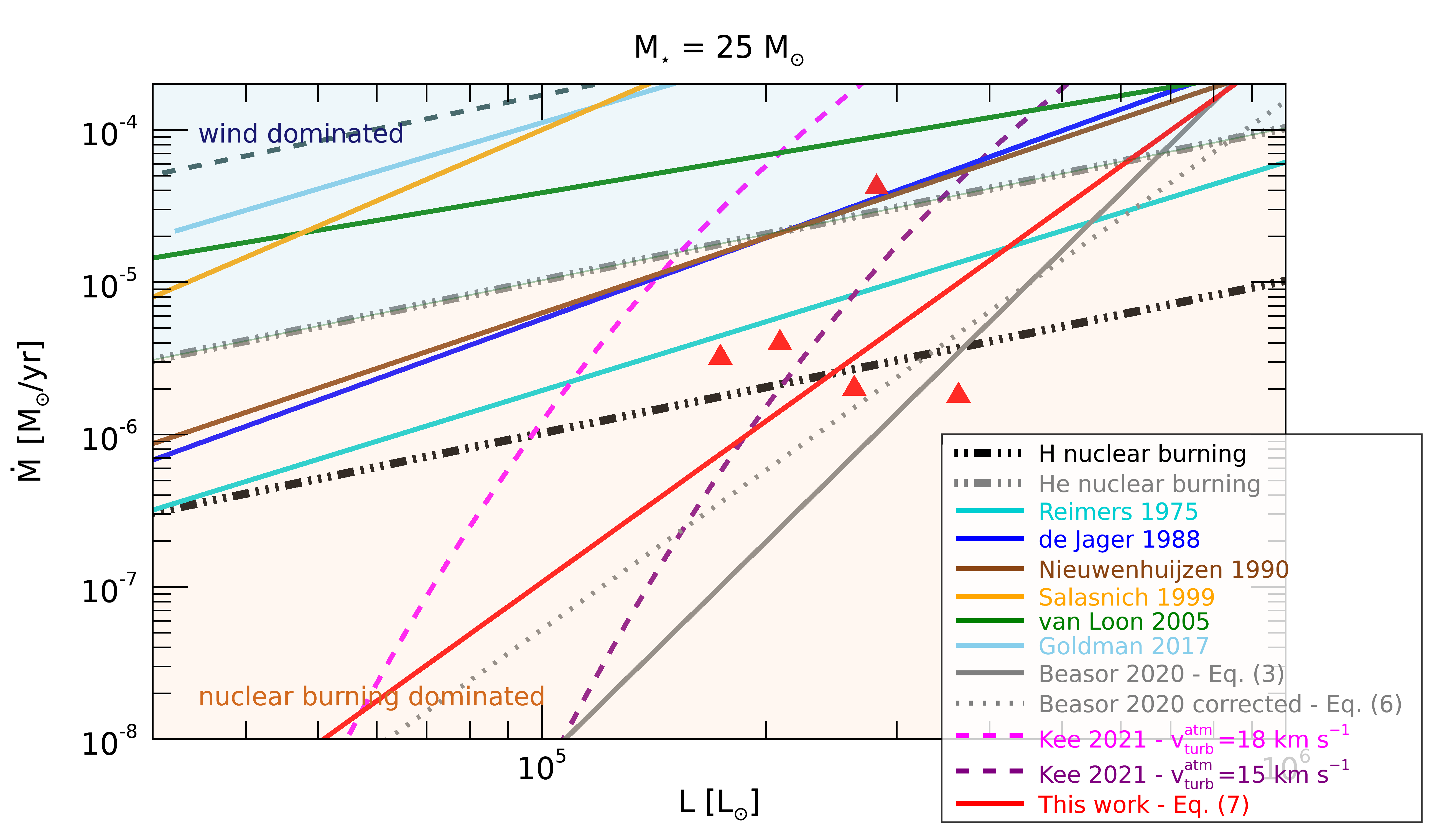}
	\caption{Mass-loss rate as a function of luminosity. Various mass-loss rate relations derived for red supergiants are shown for a fixed stellar mass of 10\,\Msun\ (left panel) and of 25\,\Msun\ (right panel) for an  assumed effective temperature of 3450\,K \citep{Reimers1975MSRSL...8..369R, deJager1988A&AS...72..259D, Nieuwenhuijzen1990A&A...231..134N, Salasnich1999A&A...342..131S, vanLoon2005A&A...438..273V, Beasor2020MNRAS.492.5994B, Goldman2017MNRAS.465..403G, Kee2021A&A...646A.180K}.  Empirical mass-loss rate relations are displayed with a solid line, the theoretical relation of \citet{Kee2021A&A...646A.180K} is shown with a dashed line for 2 different values of the atmospheric turbulent velocity $v_{\rm{turb}}^{\rm{atm}}$. For the empirical relation of \citet{Goldman2017MNRAS.465..403G}, we use the RSG period-luminosity relation as given in Eq.~(C.17) of \citet{DeBeck2010A&A...523A..18D}, which is valid for pulsation periods between $\sim$300\,--\,800 days.
		The corrected \Mdot$_{\rm{SED}}$-relation based on the data of Beasor et al.\ (see Eq.~(\ref{Eq:Mdot_Beasor_corrected})) is shown as dotted grey line; the new  \Mdot$_{\rm{CO}}$-relation derived from the ALMA CO(2-1) measurements of the M-type RSG winds in RSGC1 (Eq.~(\ref{Eq:Mdot_new})) is shown as full red line.	The rate at which hydrogen and helium are consumed by nuclear burning are shown as thick dashed-triple dotted lines; the single-scattering radiation pressure limit for an expansion velocity of 12\,km~s$^{-1}$ is shown as dashed dark grey line. Stellar mass loss rules the evolution of  RSG stars if the wind mass-loss rate exceeds the nuclear burning rate, as indicated by the light-blue region; the nuclear-burning dominated region is indicated by the light-orange region. The red triangles in the right panel indicate the (L, \Mdot)-values as derived in Sect.~\ref{Sec:CO_analysis} from the ALMA CO(2-1) line profiles of F01, F02, F03, F04, and F13.}
	\label{Fig:Mdot}
\end{figure*}

The parametrised $\beta$-type accelerating wind is described by
\begin{equation}
	v_{\rm{gas}}(r)  \simeq v_\infty \left(1-\frac{R_0}{r}\right)^{\beta}\,.
%	v_{\rm{gas}}(r) = v_0 + (v_\infty - v_0) \left(1 - \frac{r_0}{r}\right)^{\beta}
\end{equation}
The terminal wind velocity, $v_\infty$, is deduced from the ALMA CO(2-1) line profiles (see Table~\ref{Tab:outcome}), uncertainty being $\pm$3\,km~s$^{-1}$ (see App.~\ref{App:ALMA}).
As boundary for the velocity structure, we assume that the flow velocity of the gas is equal to the local sound velocity, $v_s$, at $R_0=R_{\rm{dust}}$. For the region between \Rstar\ and $R_{\rm{dust}}$, $\beta$ is assumed to be 1/2 \citep{Decin2006A&A...456..549D}; for the region beyond $R_{\rm{dust}}$ we follow the general conclusions from \citet{Khouri2014A&A...561A...5K}, \citet{Decin2020Sci...369.1497D}, and \citet{Gottlieb2022A&A...660A..94G} that the value of $\beta$ should be larger to represent the slowly accelerating flow in oxygen-rich winds. We here adopt $\beta$\,=\,3. We also include a constant turbulent velocity $v_{\rm{turb}}$ of 3\,km~s$^{-1}$.

To determine the fractional abundance of CO with regard to hydrogen, we assume all photospheric carbon to be locked in CO in the circumstellar envelope (CSE).  For a value of $A({\rm{C)}} = \log({\rm{C}}/{\rm{H}}) + 12$ as derived by \citet{Davies2009ApJ...696.2014D} for RSGC1, this yields a CO fractional abundance of 8.9$\times$10$^{-5}$. 

This set of stellar and circumstellar parameters allows us to retrieve the mass-loss rate, \Mdot$_{\rm{CO}}$, from the ALMA CO(2-1) line profiles for each of the five detected sources; the derived values are listed in Table~\ref{Tab:outcome}, and range between 1.8--42$\times$10$^{-6}$\,\Msun/yr.
The evolution of the four sources with mass-loss rate below $\sim$4$\times$10$^{-6}$\,\Msun/yr (F01, F02, F03, and F04) is dominated by nuclear burning, while in the case of F13 the wind mass-loss rate is currently determining its RSG evolution (see Fig.~\ref{Fig:Mdot}).
 For the RSGC1 red supergiants that remained undetected in the CO(2-1) line, an upper limit on the mass-loss rate of $\sim$7$\times$10$^{-7}$\,\Msun/yr is derived.

Given this set of input parameters, the largest uncertainty in \Mdot$_{\rm{CO}}$ arises from uncertainties in the terminal wind velocity ($\sim$35\%),  and then in the distance, the outer CO envelope size, and the CO fractional abundance (each $\sim$20\%); for additional details, readers can refer to App.~\ref{App:sigmas}. The combined effect of the errors in the individual input parameters on \Mdot$_{\rm{CO}}$ is a factor $\sim$1.4; for additional details, readers can refer to App.~\ref{App:sigmas}. This error on  \Mdot$_{\rm{CO}}$ is lower than the errors on  \Mdot$_{\rm{SED}}$; see Table~\ref{Tab:outcome}. Measurements of dust emission are usually angularly unresolved, hence rely on SEDs, and depend on more factors known to vary such as composition and size of grains,  the stellar contribution to the SED, etc.\ for which reason estimates of the dust-to-gas ratio can differ by an order of magnitude.

We note that terminal wind velocities determined solely from the half line width at zero intensity might be lower limits since it was shown by \citet{Decin2018A&A...615A..28D} that line widths are sensitivity limited, and hence that in some cases higher signal-to-noise data could indicate higher $v_\infty$ values, and hence higher \Mdot$_{\rm{CO}}$ values. However, if the terminal velocity increases, so does the width of the full line profile and of the width between the two horns in spatially resolved, optically thin line profiles. We therefore have used all these characteristics together to determine $v_\infty$. An increase in terminal velocity by 3\,km~s$^{-1}$, would induce an increase in \Mdot$_{\rm{CO}}$ by $\sim$40\% (see Table~\ref{Table:sigma}).

The ALMA data also constrain the CO outer envelope radius (see Table~\ref{Tab:outcome}). In recent studies, \citet{Groenewegen2017A&A...606A..67G} and \citet{Saberi2019A&A...625A..81S} improved  the calculations on CO photodissociation in circumstellar envelopes made by \citet{Mamon1988ApJ...328..797M}. Using the ratio of the maximum flux density over 3 times the noise  as a proxy for the fractional abundance of $f_0/f_{\rm{CO}}(r)$ (with $f_0$ being the initial CO abundance), we can use Eq.~(3) of \citet{Saberi2019A&A...625A..81S} and their values for $r_{1/2}$ and $\alpha$ (their Table~B.1.) to compare the theoretical predicted CO envelope size with the ALMA observations. The observed ALMA CO envelope radius of the four RSGC1 sources F01, F02, F03, and F04 is only lower by a factor  $\sim$1.1\,--\,1.8 compared to what \citet{Saberi2019A&A...625A..81S} predicted, which is a remarkable agreement  given the fact that the values for $r_{1/2}$ and $\alpha$ were calculated for a standard \citet{Draine1978ApJS...36..595D} interstellar radiation field (ISRF). The ISRF in the massive young cluster RSGC1 might actually be higher given the fact that newly formed stars affect the surrounding materials strongly via their UV photons. However, unlike the case of 47~Tucanae \citep{McDonald2015MNRAS.453.4324M}, an estimate of the strength and local variation of the interstellar radiation field in RSGC1 is currently lacking. 
\citet{Groenewegen2017A&A...606A..67G} has shown that an increase in ISRF by a factor $\sim$3 leads to a decrease in photodissociation radius by a factor $\sim$1.5. The notable exception is F13 for which the predicted outer radius is almost an order of magnitude larger than the observed one. F13 stands also out in other aspects; for further discussion, readers are referred to  Sect.~\ref{Sec:Discussion}.

\subsection{Comparison to SED retrievals}\label{SED:analysis}

%%%/Users/leen/leda/ALMA/RSGs/programs/plot_correlations_with_alma_rsgs_paper_revised.pro
%/Users/leen/leda/ALMA/RSGs/programs/plot_correlations_RSGs_paper_with_MC_results
\begin{figure*}[!htpb]
	\includegraphics[angle=0,width=\textwidth]{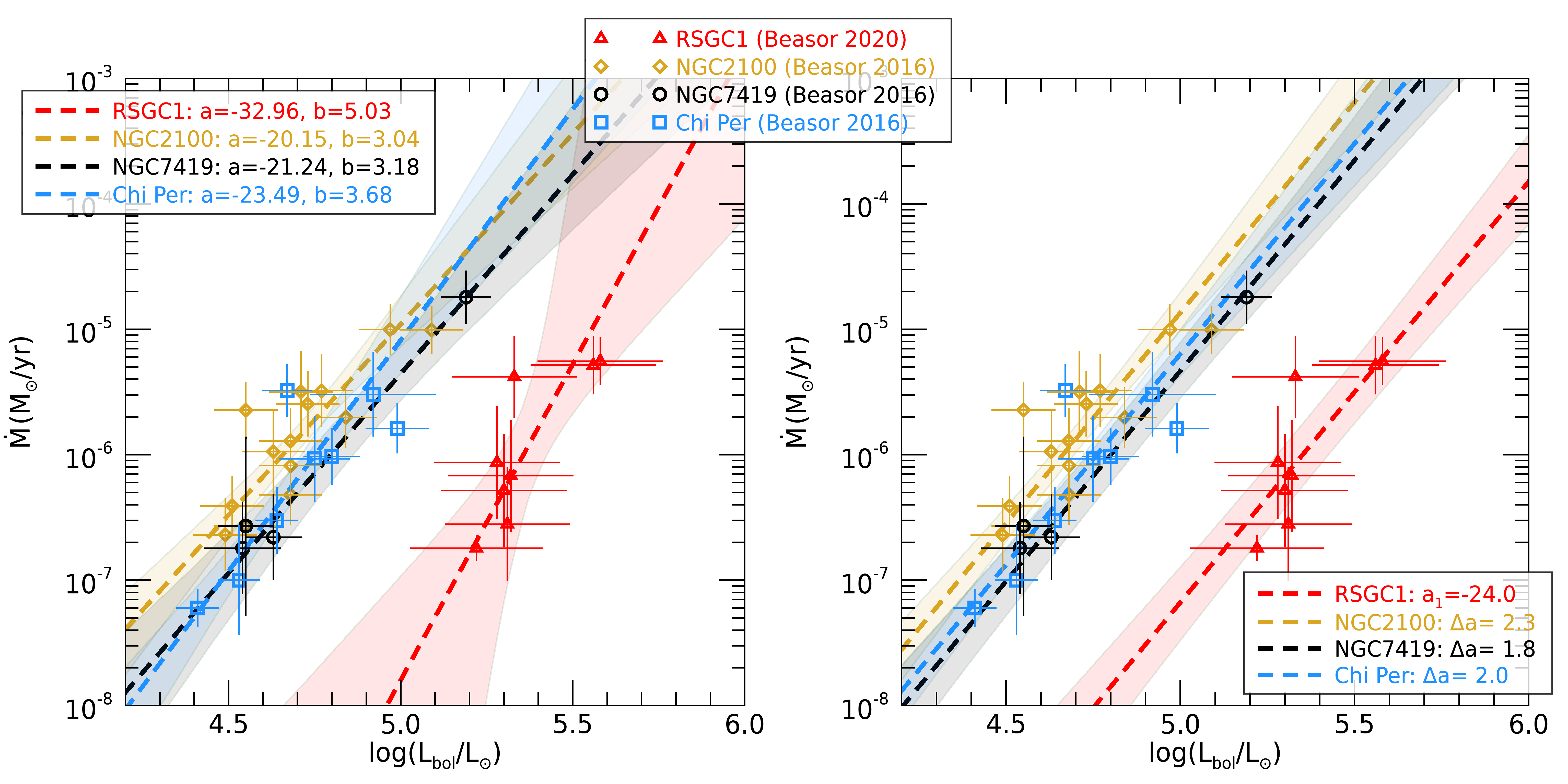}
	\caption{\Mdot$_{\rm{SED}}$ -luminosity relations for the four open clusters RSGC1, NGC\,2100, NGC\,7419, and $\chi$ Per. The coloured open symbols represent the (L$_{\rm{bol}}$, \Mdot$_{\rm{SED}}$)-values for the four clusters studied by \citet{Beasor2016MNRAS.463.1269B} and \citet{Beasor2020MNRAS.492.5994B}; error bars indicate their most conservative error estimates (see App.~\ref{App:Beasor}).
		The dashed lines in the left panel show the individual fits to Eq.~(\ref{Eq:relation_ab}) where the values of both $a$ and $b$ are free parameters in each fit. In the right panel, we force the slope $b$ to be identical for all clusters and fit Eq.~\eqref{Eq:Mdot_for_MC}. The best-fit values are listed in the legend and in Table~\ref{Tab:res_SED}. The shaded areas provide a visualisation of the uncertainties of the fits. They were constructed through MC as the 68\%-locus of all best-fit relations derived from a fit to mock data sets used to assess the errors on the best-fit parameters.
		 } 
	\label{Fig:Beasor_new}
\end{figure*}

In various works, Beasor et al.\ have derived the mass-loss rate for red supergiants in four open clusters, including RSGC1, NGC\,7410, $\chi$~Per, and NGC\,2100 \citep{Beasor2016MNRAS.463.1269B, 
	 Beasor2018MNRAS.475...55B, Beasor2020MNRAS.492.5994B}. The first 3 clusters reside in the Milky Way, while the latter one is an LMC cluster. For each of those clusters, the age and initial mass $M_{\rm{ini}}$ of the RSGs was determined. This yielded values of 12$\pm$2\,Myr and 25$\pm$2\,\Msun\ for RSGC1. For the other 3 clusters, the respective values are 21$\pm$1\,Myr and 10$\pm$1\,\Msun\ (NGC\,2100), 20$\pm$1\,Myr and 11$\pm$1\,\Msun\ (NGC\,7419), and 21$\pm$1\,Myr and 11$\pm$1\,\Msun\ ($\chi$~Per). The aim of their works was to derive a new mass-loss rate prescription for red supergiants that can be used in stellar evolution codes.  \citet{Beasor2020MNRAS.492.5994B} derived a general \Mdot-luminosity relation that is dependent on the initial mass, 
\begin{eqnarray}
	\log(\Mdot_{\rm{SED}}/\Msun\  {\rm{yr^{-1}}}) & = & (-26.4 - 0.23 \times M_{\rm{ini}}/\Msun) \nonumber \\ 
	& & \ \ + 4.8 \log(L_{\rm{bol}}/\Lsun)\,.
	\label{Eq:Beasor}
\end{eqnarray} 
By keeping $M_{\rm{ini}}$ constrained, \cite{Beasor2016MNRAS.463.1269B, Beasor2018MNRAS.475...55B} showed that the \Mdot-luminosity relation has a tighter correlation with a smaller value for the dispersion. The standard deviation for the slope is given to be 4.8$\pm$0.6; the standard deviations on the other numerical values were not listed by \citet{Beasor2020MNRAS.492.5994B}, but have been determined in App.~\ref{App:Beasor}.
 The slope of Eq.~(\ref{Eq:Beasor})  is  steeper than any other mass-loss rate relations derived for red supergiants (see Fig.~\ref{Fig:Mdot}). We here re-examine the results of \citet{Beasor2020MNRAS.492.5994B} and compare the SED mass-loss rate to the ones derived from the ALMA CO(2-1) line in the current work.

\begin{table*}[htp]
	\caption{Best-fit parameters for the \Mdot$_{\rm{SED}}$-luminosity relation for each cluster. }
	\label{Tab:res_SED}
	\centering
	\begin{tabular}{l|c|cc|c}
		\hline
		\xrowht[()]{8pt}Cluster & Pearson correlation & Intercept ($a$) & Slope ($b$) & difference in intercept ($\Delta a_i$)\\
		& coefficient & Eq.~\eqref{Eq:relation_ab} & Eq.~\eqref{Eq:relation_ab} & Eq.~\eqref{Eq:Mdot_for_MC} \\
		\hline
		& & & & \\[-2ex]
		RSGC1 & 0.82 & $-$32.96$\pm$18.43 & 5.03$\pm$3.44 &   0 \\
		NGC\,2100 & 0.85 & $-$20.15$\pm$3.13 & 3.04$\pm$0.64 &  2.31$^{+0.40}_{-0.37}$ \\
		NGC\,7419 & 0.99 & $-$21.24$\pm$3.47 & 3.17$\pm$0.70 &  1.85$^{+0.40}_{-0.38}$\\
		$\chi$~Per & 0.82 &$-$23.49$\pm$4.20 & 3.68$\pm$0.86 & 1.98$^{+0.42}_{-0.38}$ \\
		\hline 
	\end{tabular}
\tablefoot{The second column lists the linear Pearson correlation coefficient. The third and fourth column  list, respectively, the intercept $a$ and slope $b$ with their standard deviation derived by fitting Eq.~\eqref{Eq:relation_ab} to the data. The fifth column lists the fit to Eq.~\ref{Eq:Mdot_for_MC} which yields a best-fit slope $b$\,=\,3.35$^{+0.43}_{-0.37}$ and intercept  $a_1$\,=\,$-$23.96$^{+2.00}_{-2.32}$.}
\end{table*}

In a first step, \citet{Beasor2020MNRAS.492.5994B} has determined a \Mdot-luminosity relation for all clusters in their sample by fitting the relation
\begin{equation}
	\log(\Mdot_{\rm{SED}}/\Msun {\rm{yr^{-1}}}) = a + b \log(L_{\rm{bol}}/\Lsun)
	\label{Eq:relation_ab}
\end{equation}
to their data points (see Table~\ref{Table:all_data_Beasor} and App.~\ref{App:Beasor}). We  repeat the analysis applying the same {\sc{IDL}} routine {\sc{fitexy}}\footnote{\url{https://idlastro.gsfc.nasa.gov/ftp/pro/math/fitexy.pro}} and using the most conservative error estimates for both $L_{\rm{bol}}$ and \Mdot$_{\rm{SED}}$ (see Table~\ref{Table:all_data_Beasor}). The fit to Eq.~(\ref{Eq:relation_ab}) is shown in Fig.~\ref{Fig:Beasor_new}; the values derived for $a$ and $b$ are listed in Table~\ref{Tab:res_SED}, together with the Pearson correlation coefficient. Similar to \citet{Beasor2020MNRAS.492.5994B}, the standard deviation on the intercept $a$ and the slope $b$ is large for the cluster RSGC1.
For all clusters, our values for the intercept are systematically higher than the values listed in Table~4 of \citet{Beasor2020MNRAS.492.5994B} (copied in Table~\ref{Tab:res_SED_Beasor}), and lower for the slopes, although they agree within the 1-sigma uncertainties, the exception being NGC\,2100.
In particular, for RSGC1 we derive {$a$\,=\,$-$32.98$\pm$18.43 and $b$\,=\,5.03$\pm$3.44, as compared to \citet{Beasor2020MNRAS.492.5994B} who listed $a$\,=\,$-$52.0$\pm$51.2 and $b$\,=\,8.8$\pm$9.5 (see left panel in Fig.~\ref{Fig:Beasor} and discussion in App.~\ref{App:Beasor})\footnote{We use 2 decimal points in errors to ensure accurate propagation of uncertainties.}.

The values of the slope $b$  obtained from each cluster  are compatible with one another. This had led \citet{Beasor2020MNRAS.492.5994B} to fix $b$ to $4.8\pm0.6$ (corresponding to the weighted mean of the slope obtained for each cluster) in order to decrease the number of degrees of freedom of the fit from 8 to 5. We follow a similar strategy to limit the number of degrees of freedom, however we do not fix $b$ from the previous exercise. We rather fit a common value of $b$ for all four clusters simultaneously with each individual intercepts ($a_i$) using a multivariate linear fitting method implemented with the MPFIT  Levenberg-Marquardt least-squares solver \citep{ Markwardt2009ASPC..411..251M}. Uncertainties were estimated through Monte Carlo (MC) by generating $10^4$ statistically equivalent data sets randomly drawn from normal distributions centred on the best fit solutions and repeating the fit for each artificial data set. The 68\%-confidence interval on the best-fit parameters are given by the 0.16 and 0.84 percentiles of the distributions of obtained parameters. Fitted at face values, the errors on the individual intercepts $a_i$  are tightly correlated. To prevent this correlation, we adopt a slightly different functional form:
 \begin{equation}
	\log(\Mdot_{\rm{SED}}/\Msun {\rm{yr^{-1}}}) = a_1 +\Delta a_i + b \log(L_{\rm{bol}}/\Lsun)\,,
	\label{Eq:Mdot_for_MC}
 \end{equation}
 	where $a_1$ is the intercept for the cluster RSGC1, and $\Delta a_i$ indicates the difference in intercept of each cluster $i$\,=\,1,2,3,4
 	with regard to the intercept $a_1$ (hence, $\Delta a_1$\,=\,0). That is, each cluster gets its own intercept ($a_i = a_1+\Delta a_i$) and the slope is forced to be identical.
		The MC method yields $b$\,=\,3.34$^{+0.48}_{-0.41}$, $a_1$\,=\,$-$23.83$^{+2.15}_{-2.64}$  with corresponding $\Delta a_i$-values listed in the last column of Table~\ref{Tab:res_SED}; see right panel in Fig.~\ref{Fig:Beasor_new}\footnote{Using instead another cluster for determining $a_i$ yields similar results for the value of $a_j = a_i + \Delta a_j$ (with $j \ne i$) for the 3 other remaining clusters; for additional details, readers can refer to  App.~\ref{App:NGC2100}.}.  While mathematically equivalent to directly fitting the intercept $a_i$ instead of their difference $\Delta a_i$ to a reference intercept (here arbitrarily chosen to be that of RSGC1, but see App.~\ref{App:NGC2100}), the errors on $a_1$,  $\Delta a_2$,  $\Delta a_3$, $\Delta a_4$ are now uncorrelated, which allows for a better sense of whether the intercepts from each cluster vary from one another from a direct comparison of the $\Delta a_i$ values and their errors.

 	To parametrise the mass-loss rate in terms of both the luminosity and the initial mass, we perform a MC fit to all four clusters together using a parametrisation similar to Eq.~\ref{Eq:Beasor}, that is
 	\begin{equation}
 		\log(\Mdot_{\rm{SED}}/\Msun\  {\rm{yr^{-1}}})   =  R  +S M_{\rm{ini}}/\Msun + b \log(L_{\rm{bol}}/\Lsun)\,.
 		\label{Eq:Mdot_Beasor_corrected}
 	\end{equation}
The combined fit yields $R$\,=\,$-$20.63$^{+1.93}_{-2.38}$, $S$\,=\,$-$0.16$^{+0.03}_{-0.04}$, and $b$\,=\,3.47$^{+0.57}_{-0.45}$. This new mass-loss rate relation has a higher constant $R$ and a shallower dependence on  the initial mass and  the luminosity as compared to \citet{Beasor2020MNRAS.492.5994B}; see Eq.~\ref{Eq:Beasor} and Fig.~\ref{Fig:Mdot} (full and dotted grey lines). Except for the mass-dependent intercept, these values  do not agree within their respective 1-sigma uncertainties; see the standard deviations for the parameters of Eq.~\eqref{Eq:Beasor} derived in App.~\ref{App:Beasor}.

However, the uncertainties associated with the intercept $R$ are substantial, spanning two orders of magnitude for the associated mass-loss rate. But  it is crucial to acknowledge that these uncertainties pertaining to the intercept do not reflect the uncertainties on the mass-loss rates within the range of values for the initial mass and luminosities under study. By normalising the mass-loss rate, initial mass and luminosities to representative values, the resulting parameter uncertainties become more readily applicable \citep[e.g.][]{vanLoon2005A&A...438..273V}.  This yields the following analytical relation   
	\begin{eqnarray}
	\log(\Mdot_{\rm{SED}}/10^{-5}\,\Msun\  {\rm{yr^{-1}}})  &  =  & R  +S M_{\rm{ini}}/10\,\Msun \nonumber \\
 & & 	+ b \log(L_{\rm{bol}}/10^5\,\Lsun)\,, 
	\label{Eq:Mdot_SED_scaled}
\end{eqnarray}
with $R$\,=\,1.71$^{+0.54}_{-0.44}$, $S$\,=\,$-1.63^{+0.30}_{-0.36}$, and $b$\,=\,3.47$^{+0.57}_{-0.45}$; the uncertainties on the intercept now being of the order of 30\%.

\section{Discussion}\label{Sec:Discussion}

\subsection{CO-based mass-loss relation for RSGs}

The  CO gas mass-loss rate values of those red supergiants in common with  \citet{Beasor2020MNRAS.492.5994B} are systematically lower, on average by a factor of $\sim$2  (see Table~\ref{Tab:outcome}) --- although we here must provide the caveat of dealing with small number statistics.  CO mass-loss rates, \Mdot$_{\rm{CO}}$, are  not afflicted with uncertain dust extinction corrections, dust-to-gas conversion ratios, and unknown expansion velocities as is the case for \Mdot$_{\rm{SED}}$. The main reason for the difference in \Mdot$_{\rm{SED}}$ and \Mdot$_{\rm{CO}}$ for the 3 RSGC1 sources analysed in both this study and by \citet{Beasor2020MNRAS.492.5994B} is the expansion velocity which was assumed to be 25$\pm$5\,km~s$^{-1}$ by \citet{Beasor2020MNRAS.492.5994B}, but which is lower for all RSGs in which CO(2-1) was detected. Correcting for the terminal velocities as deduced from the ALMA CO(2-1) data, \Mdot$_{\rm{CO}}$ and \Mdot$_{\rm{SED}}$ agree very well for the three sources in common to both studies (F01, F02, and F03) with average difference being a factor $\sim$0.9 and maximum percentage difference of 32\% (see also last two columns in Table~\ref{Tab:outcome}).

A point of concern on the reliability of the mass-loss rates could be binarity. The binary fraction of unevolved massive stars is thought to be above 70\% \citep{Sana2012Sci...337..444S, Moe2017ApJS..230...15M, Lee2022MNRAS.513.5847P}. For a predicted merger fraction of 20--30\%, and a binary interaction fraction of 40--50\%, the total RSG binary fraction is estimated around 20\%  \citep{Patrick2019A&A...624A.129P, Patrick2020A&A...635A..29P, Neugent2020ApJ...900..118N, Sana2022arXiv220316332S}. As discussed by \citet{Decin2021ARA&A..59..337D} and \citet{Gottlieb2022A&A...660A..94G}, a binary system with small orbital distance can develop an equatorial density enhancement (EDE) promoting the formation of dust grains. Hence, it is expected that for those systems mass-loss rate estimates based on dust spectral features in the SED might yield too high an \Mdot-estimates, and that mass-loss rate estimates based on a CO-analysis should be preferred \citep{Decin2019NatAs...3..408D}. The sensitivity of the CO(2-1) integrated line flux to binary-induced morphologies has been shown to be less than a factor of $\sim$2, for both spiral structures \citep[see Fig.~16 in][]{Homan2015A&A...579A.118H} and equatorial density enhancements \citep{Decin2019NatAs...3..408D}, the latter morphologies being much better traced in other molecular diagnostics, such as the SiO v=0 J=8-7 transition \citep{Kervella2016A&A...596A..92K, Decin2020Sci...369.1497D}.

Despite of  (i)~the caveat of dealing with small-number statistics, and (ii)~the fact that we are dealing with the 3 RSGs in RSGC1 that have the largest \Mdot$_{\rm{SED}}$-values from the study of \citet{Beasor2020MNRAS.492.5994B}, we can try to improve on the mass-loss rate prescription derived in Eq.~(\ref{Eq:Mdot_Beasor_corrected}). Similar to the analysis of \citet{Beasor2020MNRAS.492.5994B}, we exclude F13 from the regression analysis since it stands out in the sample of 14 RSGs in RSGC1: F13 is unusually red, has an \Mdot$_{\rm{CO}}$ being an order of magnitude larger than the other four RSGs detected in CO(2-1), and is the only source for which the observed CO envelope size is much lower than any prediction for CO photodissociation in a circumstellar envelope (see Sect.~\ref{Sec:CO_analysis}). This tentatively suggests that another mass-loss mechanism is active  (see Sect.~\ref{Sec:phase_change} and Sect.~\ref{Sec:evolution}). 
The limited and biased sample led us decide to fit both  the ($L_{\rm{bol}}$, \Mdot$_{\rm{CO}}$) and ($L_{\rm{bol}}$, \Mdot$_{\rm{SED}}$) measurements together with a four parameter set of equations:
	\begin{eqnarray}
	 		\log(\Mdot_{\rm{SED}}/10^{-5}\,\Msun\  {\rm{yr^{-1}}})   & =  &  R_{\rm{SED}}  +S M_{\rm{ini}}/10\,\Msun \nonumber\\
	 		& & + b \log(L_{\rm{bol}}/10^5\,\Lsun) \nonumber\\
	 		\log(\Mdot_{\rm{CO}}/10^{-5}\,\Msun\  {\rm{yr^{-1}}})   & =  &  (R_{\rm{SED}} + \Delta R)  \nonumber \\
	 		& & + S M_{\rm{ini}}/10\,\Msun  \nonumber\\
	 		 & & + b \log(L_{\rm{bol}}/10^5\,\Lsun).
	\label{Eq:Mdot_new}	
	\end{eqnarray}
with best-fit values being $R_{\rm{SED}}$\,=\,$1.77^{+0.58}_{-0.46}$, $S$\,=\,$-1.68^{+0.31}_{-0.40}$, $b$\,=\,3.50$^{+0.60}_{-0.46}$, and $\Delta R$\,=\,0.49$^{+0.44}_{-0.41}$; see Fig.~\ref{Fig:Beasor_Mini_new}.
The \Mdot$_{\rm{CO}}$ measurements have an intercept that differs by $0.5\pm0.4$ compared to the \Mdot$_{\rm{SED}}$ measurements; the likelihood that such a difference occurs by chance is only $\sim$12\%.  This new \Mdot-luminosity relation derived for M-type supergiants is plotted as full red line in Fig.~\ref{Fig:Mdot}, and is clearly different from the \Mdot-relation derived by \citet{Beasor2020MNRAS.492.5994B} (full grey line in Fig.~\ref{Fig:Mdot}).  

%  email from Hugues Sana on 17-18/11/2022
%email from Hugues Sana on 19/09/2023
\begin{figure*}[!htpb]
	\includegraphics[angle=0,width=\textwidth]{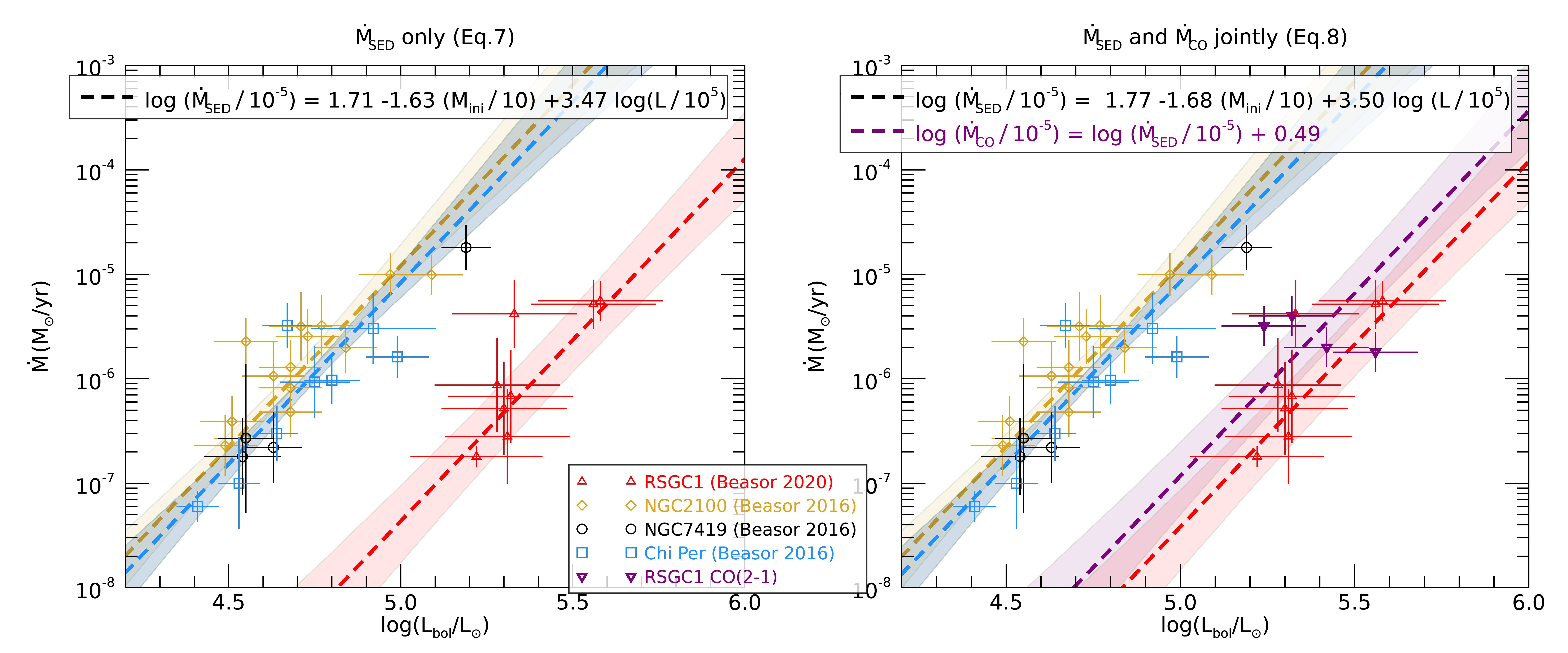}
	\caption{\Mdot-luminosity relations parametrised in terms of $\log L_\mathrm{bol}$ and $M_\mathrm{ini}$ and fitted jointly for the four open clusters RSGC1, NGC\,2100, NGC\,7410, and $\chi$ Per. Left panel presents the best-fit to the \Mdot$_{\rm{SED}}$ measurements of the four clusters (Eq.~\ref{Eq:Mdot_SED_scaled}) while the right panel includes in the fit the \Mdot$_{\rm{CO}}$-values of four stars in RSGC1 (purple downward triangles) using Eq.~\eqref{Eq:Mdot_new}.  Symbols, colours and shades have the same meaning as in Fig.~\ref{Fig:Beasor_new}, with the exception of the addition of the \Mdot$_{\rm{CO}}$ measurements derived in Sect.~\ref{Sec:CO_analysis} using the $L_{\rm{bol}}$ values of \citet{Davies2008ApJ...676.1016D}.}
	\label{Fig:Beasor_Mini_new}
\end{figure*}

 To assess the predictive power of Eq.~(\ref{Eq:Mdot_new}), we have compared its predicted \Mdot-values with \Mdot$_{\rm{CO}}$-values derived for some well-known Galactic M-type supergiants towards which several rotational CO lines have been observed ($\alpha$~Ori, $\mu$~Cep, VX~Sgr, and VY~CMa; see App.~\ref{App:other_rsgs}). The analysis presented in App.~\ref{App:other_rsgs} proves that Eq.~(\ref{Eq:Mdot_new}) predicts the gas mass-loss rate for M-type red supergiants with effective temperature between $\sim$3200\,--\,3800\,K with good accuracy, the average difference only being $\sim$30\% (see Table~\ref{Tab:outcome_other_RSGs}). 
 
Moreover, we can use the derived \Mdot$_{\rm{CO}}$ values to derive the gas-to-dust ratio in the winds of individual sources. This ratio can be determined by comparing \Mdot$_{\rm{SED}}$-values with  \Mdot$_{\rm{CO}}$-values (adjusted for differences in wind speed and distance, as used by different authors). Our analysis reveals a gas-to-dust ratio of 235 $\pm$71 for the RSGC1 sources. For the Galactic sources, the gas-to-dust ratio ranges between $\sim$200\,--\,550. The exception is the extreme RSG VY CMa for which a low ratio of $\sim$20 is derived, but there are indications that that ratio is changing throughout its mass-loss history  (see App.~\ref{App:other_rsgs}).

Remarkably, the new \Mdot$_{\rm{CO}}$ values for the M-type red supergiants F01, F02, F03, and F04 are smaller than predicted by all empirical mass-loss rate prescriptions derived prior to 2020 and used in stellar evolution codes (see right panel in  Fig.~\ref{Fig:Mdot}). Again here, (part of) the explanation is based on the fact that all empirical \Mdot-relations shown in Fig.~\ref{Fig:Mdot} are based on an SED analysis that are prone to uncertain gas-to-dust ratios and expansion velocities, and that samples of stars will be flawed by a large fraction of stars that experience binary interaction. In addition, those previous studies were often biased towards samples with high mass-loss rate objects for which the infrared excess is easier to detect and model, as acknowledged by those authors; see, for example, the discussion in \citet{vanLoon2005A&A...438..273V}. Moreover, distances --- and hence corresponding luminosities ---  are very uncertain for Galactic samples. This latter caveat was avoided in the studies by Beasor et al.\ who focused on open clusters, which eventually led in 2020 to the RSG mass-loss prescription given in Eq.~(\ref{Eq:Beasor}) \citep{Beasor2020MNRAS.492.5994B} and shown as full grey line in Fig.~\ref{Fig:Mdot}.
The only theoretical mass-loss rate prescription for RSGs is derived by \citet{Kee2021A&A...646A.180K}. That relation can fit the RSGC1 data under condition of an atmospheric turbulent velocity of $\sim$15$\pm$1\,km~s$^{-1}$, which is lower than the values quoted in Table~2 of \citet{Kee2021A&A...646A.180K} for stars with mass around 25\,\Msun. 

In that regard, it is also worth turning our attention to F13 and to 
	the mass-loss rate prescriptions from \citet{vanLoon2005A&A...438..273V} and \citet{Goldman2017MNRAS.465..403G} which are derived for dusty RSGs, some of which displaying clear OH maser action. The relation of \citet{Goldman2017MNRAS.465..403G} yields a mass-loss rate that is a factor $\sim$10 higher than what we derive, although we need to remark that the pulsation period estimated from the period-luminosity relation given by \citet{DeBeck2010A&A...523A..18D} is very uncertain. 
	The \citet{vanLoon2005A&A...438..273V} relation predicts a mass-loss rate for F13 of 5.8$\times$10$^{-5}$\,\Msun/yr, only a factor 1.4 larger than our derived \Mdot$_{\rm{CO}}$ (see Fig.~\ref{Fig:Mdot}).   \citet{vanLoon2005A&A...438..273V} discussed that their recipe overestimates mass-loss rates for Galactic `optical' RSGs, on average by a factor $\sim$2.8 \citep{Mauron2011A&A...526A.156M} with standard deviation on that value being $\sim$2.5, hence in line with our result for F13. 

\subsection{Phase change in mass loss}\label{Sec:phase_change}

The amount of mass lost during the RSG phase, its speed, and how soon before core collapse the material is removed can have a dramatic effect on the resulting supernova light curve and spectrum \citep{Smith2009AJ....137.3558S}.
For luminosities above 10$^5$\,\Lsun, a wind mass-loss rate $\le$2$\times$10$^{-5}$\,\Msun/yr implies that the nuclear burning rate exceeds the wind mass-loss rate (see right panel in Fig.~\ref{Fig:Mdot}), and hence that core-He burning is dominating the evolution of F01, F02, F03, and F04 (see Fig.~\ref{Fig:Mdot}). Accounting for the uncertainty in $L_{\rm{bol}}$ (Table~\ref{Tab:outcome}) and \Mdot$_{\rm{CO}}$, only F13 is  above the boundary where the wind mass-loss rate dominates the star's evolution. This outcome is in line with the results from \citet{vanLoon1999A&A...351..559V} and \citet{Javadi2013MNRAS.432.2824J}, who suggested that red supergiants come in two flavours, those dominated by nuclear burning which takes about 75\% of the RSG lifetime, and those dominated by intense mass loss taking $\sim$25\% of the RSG lifetimes.

The measured wind speed of the four RSGC1 sources F01, F02, F03, and F04, whose evolution is dominated by nuclear burning (see Fig.~\ref{Fig:Mdot}), is 11$\pm$3 km/s. However, in the case of source F13,  the wind speed is significantly higher, at $\sim$22\,km/s. The measured wind speeds conform to the wind speed-luminosity relation  ($v_\infty \propto Z L^{0.4}$) established by \citet{Goldman2017MNRAS.465..403G} for OH/IR stars in the Large Magellanic Cloud (LMC). However, they are noticeably lower compared to the Galactic samples \citep[see Fig.~17 in][]{Goldman2017MNRAS.465..403G}. This finding aligns with the outcomes reported by \citet{Davies2009ApJ...696.2014D}, who identified a subsolar iron content in the RSGC1 sources and indicated a metallicity between Z = 0.008 and Z = 0.02. This preliminary alignment with the relation proposed by \citet{Goldman2017MNRAS.465..403G} for lower metallicities tentatively corroborates  the findings of \citet{Goldman2017MNRAS.465..403G} of expansion velocities being consistent with the predictions of  dust-driven wind theory \citep{vanLoon2000A&A...354..125V}.

One advantage of our study is the ability to observe essentially the `same' star at different stages of post-main sequence evolution within a single cluster. When considering the changes in mass-loss rate and wind speed, this leads to the hypothesis that RSGs may undergo one or multiple phase changes in mass loss during their RSG evolution. It is conjectured that there could be intense and potentially eruptive mass loss occurring for a shorter period of the RSG lifetime, which disrupts the otherwise more tranquil mass-loss process that takes place over a larger portion of the RSG branch. For further discussion on this topic, we refer to Sect.~\ref{Sec:evolution}.

\subsection{Implications for stellar evolution}\label{Sec:evolution}

Accurate stellar mass-loss predictions are fundamental for stellar evolutionary models, in particular for the prediction of the nature of the end-products. The outcome of this study impacts the formation frequency of core-collapse supernovae of type IIP,  black holes and neutron stars, and hence the frequency of gravitational wave events that can be detected with current (and to be developed) detectors. For massive stars with initial mass $\la$30\,\Msun, winds during the main-sequence phase will only remove $\le$0.8\,\Msun \citep{Beasor2021ApJ...922...55B}. Hence, the only evolutionary phase during which these massive stars can potentially loose a significant amount of mass is during the cool red supergiant phase, which for a star of $\sim$25\,\Msun\ lasts $\sim$10$^{5.5}$\,yr \citep{Meynet2015A&A...575A..60M}.

 The {\sc mesa} evolutionary code \citep{Paxton2019ApJS..243...10P} was used to compute the evolution of a 25\,\Msun\ star until core carbon depletion\footnote{The files to reproduce the simulations are available at \url{https://zenodo.org/doi/10.5281/zenodo.10020628}.}. Following \citet{Brott2011A&A...530A.115B}, overshooting was modelled by extending the convective region by $0.335$ pressure scale heights, while winds during the main sequence and the Wolf-Rayet phase are modelled using the prescriptions of \citet{Vink2001A&A...369..574V} and \citet{Hamann1995A&A...299..151H}. For temperatures below $10^4$\,K we switch to either the prescription of \citet{Nieuwenhuijzen1990A&A...231..134N} or the newly derived one from Eq.~(\ref{Eq:Mdot_new}). Composition and opacities are determined from solar metallicity and metal fractions as given by \citet{Asplund2009ARA&A..47..481A}.

Both simulations evolve to very different endpoints, with the one using the \citet{Nieuwenhuijzen1990A&A...231..134N} wind prescription managing to strip its outer stellar envelope and evolve to become a hot Wolf-Rayet star. Using the new prescription of Eq.~(\ref{Eq:Mdot_new}), only 1.93\,\Msun\ of the hydrogen-rich stellar envelope is lost (out of a total envelope mass of 11.97\,\Msun) implying that such massive stars would explode as RSG upon core-collapse (see Fig.~\ref{Fig:evol_track}). Following \citet{Smith2009AJ....137.3558S}, this leads to the suggestion that if F01, F02, F03, or F04 were to explode in their current RSG phase, this would produce a Type II SN with limited level of interaction with the circumstellar material, without enough inertia to substantially decelerate the blast wave and with no substantial narrow H$\alpha$ emission from the post-shock gas.  Mass-loss rates of order $\sim$11.97/1.93\,=\,6.20 times higher would be needed to successfully strip the H-rich stellar envelope.

\begin{figure}[htp]
	\centering
	\includegraphics[width=0.48\textwidth]{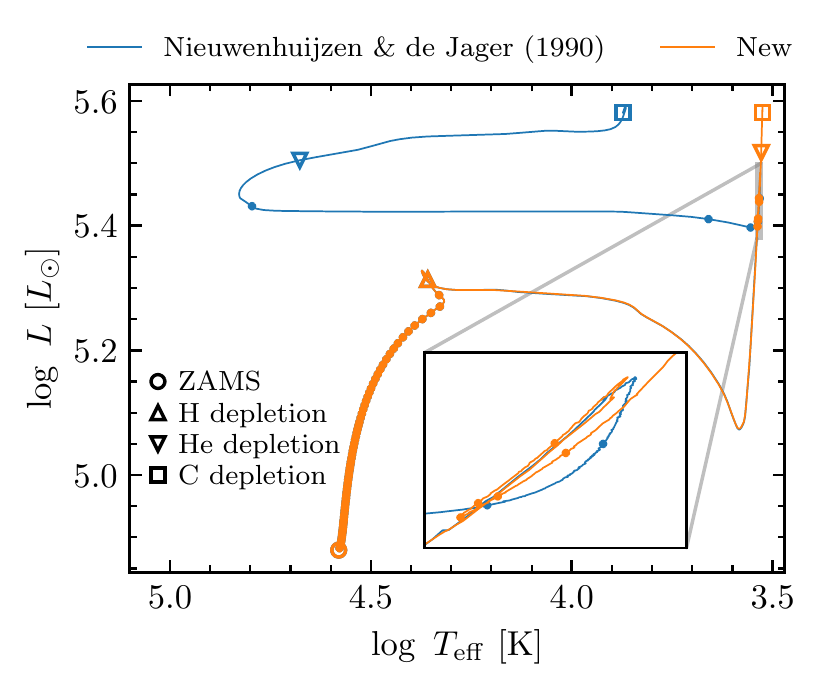}
	\caption{Evolutionary track as computed with the {\sc{mesa}} evolutionary code for a star of  initial mass 25\,\Msun\ applying the RSG mass-loss rate prescription from \citet{Nieuwenhuijzen1990A&A...231..134N} (blue) and from Eq.~(\ref{Eq:Mdot_new}) (orange). To discern long-lived versus short thermal evolution phases, dots have been added every $10^5$ years; tracks go till carbon depletion indicated by the open square. The inset at the RSG, for a range in $\log T_{\rm{eff}}$ of 0.01\,dex, shows how both tracks digress there.}
	\label{Fig:evol_track}
\end{figure}

In that sense, F13 might be an important outlier mass-loss wise, since its \Mdot$_{\rm{CO}}$ is more than an order of magnitude larger than for the other four RSGs. The retrieved \Mdot$_{\rm{CO}}$ value of  F13 does not fit  Eq.~(\ref{Eq:Mdot_new}) which is $\sim$10 times above the prescription. Both the strong CO(2-1) line  of F13 and its high near-infrared extinction \citep{Davies2008ApJ...676.1016D} indicate that F13 is surrounded by a lot of circumstellar material produced by a wind with mass-loss rate higher than  the other four RSGC1 sources.  It might be that the luminosity of F13 is underestimated given the high \Mdot$_{\rm{CO}}$ and the fact that it is anomalously red compared to the other RSGs in RSGC1. For fitting the relation Eq.~(\ref{Eq:Mdot_new}), $\log(L_{\rm{bol}}/L_\odot)$ should be 5.74, which would imply that it would be brighter than any known RSG and is far above the Humphreys-Davidson limit. Another possibility is an evolutionary phase change. \citet{Davies2008ApJ...676.1016D} has shown that F13 is spatially coincident with H$_2$O, OH, and SiO maser emission\footnote{SiO masers are also detected associated with F01 and F02 \citep{Davies2008ApJ...676.1016D}.}. Such maser emission is often associated with evolved stars that are long-period variables (periods of 300\,--\,500\,days) and have winds with a very high mass-loss rate.  
This induces the suggestion that another mass-loss rate mode has become active in F13. Among a coeval sample of RSGs one may expect to see enhanced mass loss, and hence masers, in those objects furthest along their evolution, that have high $L_\star$/$M_\star$ ratios --- close to the (modified) Eddington limit --- so that only small changes in the atmospheric structure --- for example caused by pulsations or changes in the high opacity due to variations in the hydrogen ionisation beneath the stellar surface  --- makes these stars unstable to more furious episodic mass ejections. Eq.~(\ref{Eq:Mdot_new}) would then represent the more quiescent mass-loss process, while F13 is an example RSG that undergoes stronger, potentially eruptive, mass loss. 
 
  In that sense, F13 shares its status as extreme RSG with VY~CMa (see App.~\ref{App:other_rsgs}), the only Galactic RSG for which the \Mdot$_{\rm{CO}}$ prediction using Eq.~(\ref{Eq:Mdot_new}) is a factor $\sim$6.6 too low. VY~CMa is one of a small class of evolved massive stars characterised by extensive asymmetric ejections and multiple high mass-loss events lasting several hundred years \citep{Decin2006A&A...456..549D, OGorman2012AJ....144...36O, Kaminski2019A&A...627A.114K, Humphreys2021AJ....161...98H} attributed to large-scale surface and magnetic activity. 
 Both  F13 and VY~CMa are indicative of a stronger, potentially eruptive,  mass-loss process that breaks from the prescription given in Eq.~(\ref{Eq:Mdot_new}); they demonstrate that one should be careful about applying any \Mdot-prescription, including Eq.~(\ref{Eq:Mdot_new}), more globally in stellar evolution models as they will not reproduce the behaviour of those extreme RSGs; see also the recent discussion by \citet{Massey2022arXiv221114147M}. 
 
 If we consider $F$ to be the fraction of time spent on RSG phase with stronger mass loss, and $B$ the enhancement of the mass-loss rate during that stage as compared to the more quiescent mass-loss rate (here $B$\,=\,10.43), we can estimate that the enhancement of a lifetime averaged mass-loss rate would be equal to $(1-F)+B \times F$. For an enhancement of 6.20 and $B$\,=\,10.43, we need $F=0.55$, so 55\% of the lifetime on the strong mass-loss rate RSG phase to successfully strip the H-rich stellar envelope. However, this derived fraction of $F=55$\% is much higher than what one derives from  using number statistics to determine the fraction of time spent in the strong mass-loss rate stage. I.e.\ out of 14 RSGs in RSGC1, only one of them is observed in the strong mass-loss rate stage. To assess the likelihood of observing such a ratio (1 out of 14), a binomial distribution can be employed. By conducting a Bayesian analysis with a flat prior for $F$, the posterior distribution of $F$ can be determined, which corresponds to a beta-distribution with shape parameters $\alpha=2$ (representing the number of stars in high mass-loss rate phase + 1) and $\beta=14$ (representing the number of stars in the quiescent RSG mass-loss rate stage + 1). Consequently, a 90\% credible interval for $F$ is obtained as $10.9_{-8.5}^{+17}$\%. This interval represents the median value along with the range between the 5th and 95th percentiles.
 
This tension between the outcome of number statistics and the fraction of time needed to strip the H-envelope during the RSG phase when considering both the quiescent and (potentially eruptive) high mass-loss rate phase induces the suggestion that the RSGs in RSGC1 will not be able to strip their entire H-envelope and will explode as RSG upon core-collapse. For the full H-rich envelope to be stripped,  mass-loss rates much stronger than observed for F13 would need to be invoked. A potential mechanism thereof has been explored by \citet{Heger1997A&A...327..224H}, who suggested that sufficiently evolved stars become dynamically unstable and exhibit large-amplitude pulsations with periods of the order of the Kelvin-Helmholtz time scale, which eventually become strong enough to dynamically eject shells of matter from the stellar surface, implying the loss of (most of) the H-rich envelope. Simulations for episodic mass ejections from common-envelope binaries yield a similar outcome of dynamical unstable mass ejections with periods in the range of a few years to a few decades, leading to a time-averaged mass-loss rate of the order of 10$^{-3}$\,\Msun/yr \citep{Clayton2017MNRAS.470.1788C}. 	The probability of observing RSGs in such a stage of large amplitude pulsation and associated strong mass loss ($\ga$10$^{-4}$\Mdot/yr)  is not very large; however such events  have marked consequence on the appearance of the supernova explosion.

% If the phase of evolution of F13 represents $\sim$10\,--\,15\% of the total RSG lifetime, then it would suffice to fully strip H-rich stellar envelope.

Similarly, F15 is an important outlier, as being the only post-RSG in the RSGC1 sample of \citet{Davies2008ApJ...676.1016D}. F15 matches the picture of being a post-RSG star, with its luminosity of $\log(L_{\rm{bol}}/L_\odot)$\,$\sim$\,5.36 matching that of the brightest RSGs in RSGC1. Even though no mass-loss constraint could be made, F15 could be indicative of the mass-loss process operating in the RSG shutting off as the star becomes hotter ($T_{\rm{eff}}$\,$\sim$\,6850\,K)  and transitions to a radiative envelope. If F15 is indeed a post-RSG, it indicates that the RSG mass loss managed to strip the H-rich stellar envelope to the point that it would evolve to the blue.

\section{Conclusions}\label{Sec:Conclusions}

The ALMA detection of CO(2-1) emission towards RSGs residing in the open cluster RSGC1  provides us with a powerful diagnostic to derive the gas mass-loss rates of those RSGs. Of importance is that the RSG cluster stars are co-eval, which allows for stars to be studied with the same initial conditions --- mass, metallicity, local environment, etc. Since the cluster stars all have roughly the same initial masses (of $\sim$25\,\Msun, within a few tenths of a solar mass), the evolutionary path should be the same, allowing for the luminosity to be used as a proxy for evolution. Based on the CO(2-1) detections, we propose a new mass-loss rate relation for M-type RSGs with effective temperatures between $\sim$3200\,--\,3800\,K, that scales with luminosity and mass.  The new \Mdot-luminosity relation proposed in Eq.~(\ref{Eq:Mdot_new}) is validated against some other well-known Galactic RSGs towards which multiple CO rotational lines have been observed. 

The gas mass-loss rates derived from CO diagnostics are systematically lower than the values retrieved from an SED analysis on which current stellar evolution codes are based (\Mdot$_{\rm{CO}}$$<$\Mdot$_{\rm{SED}}$). Implementing our new mass-loss rate relation will impact the frequency rate of type IIP SNe, neutron stars, and black holes. In particular, models suggest that the RSG mass loss would not allow single massive stars to evolve back to the blue and explode as a H-poor SN. However, the mass-loss rate of both  the RSG  F13 in RSGC1 and the well-known Galactic extreme RSG VY~CMa are almost an order of magnitude higher than predicted by Eq.~(\ref{Eq:Mdot_new}), which is indicative of a stronger mass-loss process different than captured by Eq.~(\ref{Eq:Mdot_new}). Statistical reasoning implies that the RSGs in RSGC1 will not be able to strip their entire H-rich envelope and will explode as a RSG upon core collapse.

Only five RSGs in RSGC1 were detected during the current observation run. A completion of this RSGC1 study with ALMA should allow a more accurate mass-loss rate relation to be derived, that can be checked against lower mass-loss rate RSG stars with lower luminosities. Given the fact that ALMA is now 50\,--\,100\% more sensitive than in 2015, such deeper observations are now well feasible and would allow the observation of large samples that are, as much as possible, uniform and unbiased. In addition, we intend to return to observe other clusters with slightly
different ages, such as  RSGC2 \citep{Davies2007ApJ...671..781D}, which will then allow us to repeat this study but for RSGs with slightly different masses. Ultimately, we will be able to provide the
time-averaged mass-loss rates and total mass lost during the RSG phase as a function of initial mass, crucial inputs for the theory of stellar evolution, and SN progenitors.

%-------------------------------------------------------------------
\begin{acknowledgements}
	The authors acknowledge the referees for their constructive feedback. The authors thank Emma Beasor and Benjamin Davies for fruitful discussions.
	LD acknowledges support from the European Research Council (ERC) under the
	European Union's Horizon 2020 research and innovation programme (grant
	agreements No.\ 646758: AEROSOL with PI L.\ Decin) and from the FWO grant G099720N. 
	LD, HS, and PM acknowledge support from the KU Leuven C1 excellence grant MAESTRO C16/17/007 (PI L.\ Decin). PM acknowledges support from the FWO junior postdoctoral fellowship No. 12ZY520N.
	This paper makes use of the following ALMA data: ADS/JAO.ALMA2013.1.01200.S. ALMA is a partnership of ESO (representing 
	its member states), NSF (USA) and NINS (Japan), together with NRC 
	(Canada) and NSC and ASIAA (Taiwan), in cooperation with the Republic of 
	Chile. The Joint ALMA Observatory is operated by ESO, AUI/NRAO and NAOJ.
	The authors acknowledge the UK Science and Technology Facilities Council (SFTC) IRIS for provision of high-performance computing facilities. This work was partly performed using the Cambridge Service for Data Driven Discovery (CSD3), part of which is operated by the University of Cambridge Research Computing on behalf of the STFC DiRAC HPC Facility. The DiRAC component of CSD3 was funded by BEIS capital funding via STFC capital grants ST/P002307/1 and ST/R002452/1 and STFC operations grant ST/R00689X/1. 
\end{acknowledgements}

%-------------------------------------------------------------------
%bibliography
\bibliographystyle{aa}
\bibliography{RSGC1}

 %\Online
 \newpage
  \onecolumn
 
 \begin{appendix}
 	
 	\section{ALMA observation}\label{App:ALMA}
%such as coordinates etc?

%get also info from script scriptForImaging.py

For each star, the input for the ALMA observations is listed in Table~\ref{Tab:targets}. For the five sources detected in the ALMA band 6 continuum data and in CO(2-1) line emission, Table~\ref{Tab:targets}  lists the coordinates of the peak of the continuum emission, the peak intensity, and the rms noise of the continuum  and  CO(2-1) observations. 

\medskip
For each source, \citet{Davies2008ApJ...676.1016D} has estimated the local standard of rest velocity, $v_{\rm{LSR}}$, based on high spectral-resolution observations of the 2.293\,$\mu$m CO band head (see Table~\ref{Tab:targets}), uncertainty being $\pm$4\,km~s$^{-1}$. These $v_{\rm{LSR}}$  values were used as input for the ALMA observations. For sources F01, F02 , F04,  and F13, \citet{Nakashima2006ApJ...647L.139N} also measured $v_{\rm{LSR}}$ using SiO maser data, with  an uncertainty of $\pm$2\,km~s$^{-1}$ (see last column in Table~\ref{Tab:targets}). Using the ALMA data, we have re-assessed the $v_{\rm{LSR}}$  values based on the considerations that  (i)~the CO line profile is symmetric around the $v_{\rm{LSR}}$ value (see also footnote~\ref{footnote_vel}), and (ii)~for spatially resolved sources for which the CO optical thickness is not too high (typically, a maximum $\tau_{\nu}$ of $\sim$2), the CO profile is two-horn like (see Fig.~\ref{Fig:CO}). Those values are listed in Table~\ref{Tab:outcome} and in Table~\ref{Tab:targets}. Given the spectral resolution of the data, the uncertainty on the $v_{\rm{LSR}}$ values is $\pm$3\,km~s$^{-1}$.
 The newly derived $v_{\rm{LSR}}$ values for all sources agree with the values from \citet{Nakashima2006ApJ...647L.139N}.  For the sources F03, F04, and F13, the  $v_{\rm{LSR}}$ values also agree with \citet{Davies2008ApJ...676.1016D}, F02 is just within the uncertainty range of both studies, but the value is off for F01. In particular, in the case of F01 \citet{Davies2008ApJ...676.1016D} lists a value of 129.5\,km~s$^{-1}$, but the ALMA CO and SiO maser data of \citet{Nakashima2006ApJ...647L.139N} indicate a value around 117.5\,km~s$^{-1}$.

%with input from Anita in file conts.txt

\begin{table*}[!htpb]
	\caption{Data for the stars observed within the ALMA proposal 2013.1.01200.S.}
		\label{Tab:targets}
	\vspace*{-1.5ex}
	\setlength{\tabcolsep}{1.mm}
	\begin{tabular}{l|ccc|cccc|cc|c}
		\hline
     	& & & & & & & & && \\[-2ex]
		\multicolumn{1}{c|}{} & \multicolumn{3}{c|}{Input ALMA observations$^{b}$} &  \multicolumn{4}{c|}{ALMA continuum} &  \multicolumn{2}{c|}{ALMA CO(2-1)} & \\ 
		\multicolumn{1}{c|}{ID$^{a}$}	& $\alpha$(J2000.0) & $\delta$(J2000.0) & $v_{\rm{LSR}}^{\rm{input}}$  &
		$\alpha$ (peak flux) & $\delta$ (peak flux) & Peak intensity & $\sigma_{\rm{rms}}$ & $\sigma_{\rm{rms}}$ & $v_{\rm{LSR}}$ & $v_{\rm{LSR}}$ $^{(c)}$\\
		& & & & & & [mJy/beam] & [mJy/beam] &  [mJy/beam]  & [km~s$^{-1}$]  & [km~s$^{-1}$]\\
		\hline
		& & & & & & & &&  \\[-2ex]
		F01 & 18 37 56.29  & $-$06 52 32.2 & 129.5 & 18 37 56.31& $-$06 52 32.25 & 0.420 & 0.051 &  1.7 &  117.5  & 117.7\\
		F02 & 18 37 55.28 & $-$06 52 48.4 & 114.2 & 18 37 55.29 & $-$06 52 48.50 & 0.359 & 0.051 &  1.9 &  120.5 & 119.7\\
		F03 & 18 37 59.72 & $-$06 53 49.4 & 127.2 & 18 37 59.74 & $-$06 53 49.50 & 0.160 & 0.053 &  2.1  &  123.5 & $-$ \\
		F04 & 18 37 50.90 & $-$06 53 38.2 & 121.2 & 18 37 50.89 & $-$06 53 38.45 &  0.097 & 0.055 &  2.0 &  123.5 & 124.3  \\
		F05 & 18 37 55.50 & $-$06 52 12.2 & 124.8 & & & & 0.051 & 2.0 & & $-$\\
		F06 & 18 37 57.45 & $-$06 53 25.3 & 120.7 & & & &0.051& 2.0  & & $-$ \\
		F07 & 18 37 54.31 & $-$06 52 34.7 & 121.6 & & & &0.055& 2.2  & & $-$ \\
		F08 & 18 37 55.19 & $-$06 52 10.7 & 128.2& & & & 0.050& 2.0  & & $-$\\
		F09 & 18 37 57.77 & $-$06 52 22.2 & 121.6& & & & 0.053& 2.4  & & $-$\\
		F10 & 18 37 59.53 & $-$06 53 31.9 & 122.0& & & & 0.051& 2.0 & & $-$\\
		F11 & 18 37 51.72 & $-$06 51 49.9 & 124.1 & & & & 0.052& 2.0 & & $-$\\
		F12 & 18 38 03.30 & $-$06 52 45.1 & $-$ & & & &0.051& 1.8  & & $-$\\
		F13 & 18 37 58.90 & $-$06 52 32.1 & 125.4 & 18 37 58.90 & $-$06 52 32.05 & 0.148 & 0.051 &  1.8 & 123.5 & 120.5 \\
		F14 & 18 37 47.63 & $-$06 53 02.3 & 122.0 & & & & 0.052& 1.7 & & $-$\\
		F15 & 18 37 57.78 & $-$06 52 32.0 & 120.8 & & & &0.051&1.9 & & $-$ \\
		\hline
	\end{tabular}
	\tablefoot{First part of the table lists the target identifier, the input coordinates for the ALMA observations, and the $v_{\rm{LSR}}$ as derived by \citet{Davies2008ApJ...676.1016D}. The second part of the table lists for the four sources detected in the ALMA band~6 continuum data and in CO(2-1) the coordinates of the peak of the continuum emission, the peak flux, and the rms noise of the continuum observations. The third part lists the CO(2-1) line rms noise and the $v_{\rm{LSR}}$ deduced from the ALMA CO(2-1) data. The last part list the $v_{\rm{LSR}}$ values deduced by \citet{Nakashima2006ApJ...647L.139N} from the analysis of SiO masers. \\
		\tablefoottext{a}{The stellar IDs from \citet{Figer2006ApJ...643.1166F}.}\\
		\tablefoottext{b}{From \citet{Davies2008ApJ...676.1016D}.}\\
		\tablefoottext{c}{From \citet{Nakashima2006ApJ...647L.139N}.}
	}
\end{table*}

 	\newpage
 	
 	\section{ALMA SiO and continuum images}\label{App:continuum}
 	
 	For the five sources in RSGC1 detected in the continuum and in CO v=0 J=2-1 emission,  bright SiO v=3 J=5-4 line emission (at 212.582\,GHz) is present as well. No other sources showed either line or had significant continuum detections.
 	
 	 Given the weak continuum flux for the five detected  sources, we have used the bright SiO emission for locating the source and its continuum emission. 	 Fig.~\ref{Fig:SiO} shows the SiO v=3 J=5-4 zeroth moment (i.e., total intensity) maps, and Fig.~\ref{Fig:cont} the continuum emission. The location of the maximum of the continuum emission and the maximum continuum flux are listed in Table~\ref{Tab:targets}. The continuum rms noise is listed in Table~\ref{Tab:targets}, and is $\sim$0.05\,mJy/beam. The rms noise in the SiO zeroth moment maps is 28, 34, 29, 34 and 35 mJy/beam km/s for F01, F02, F03, F04, and F13, respectively.

\begin{figure*}[htp]
	\centering
	\begin{minipage}[t]{0.85\textwidth}
		\includegraphics[angle=0, width=0.47\textwidth]{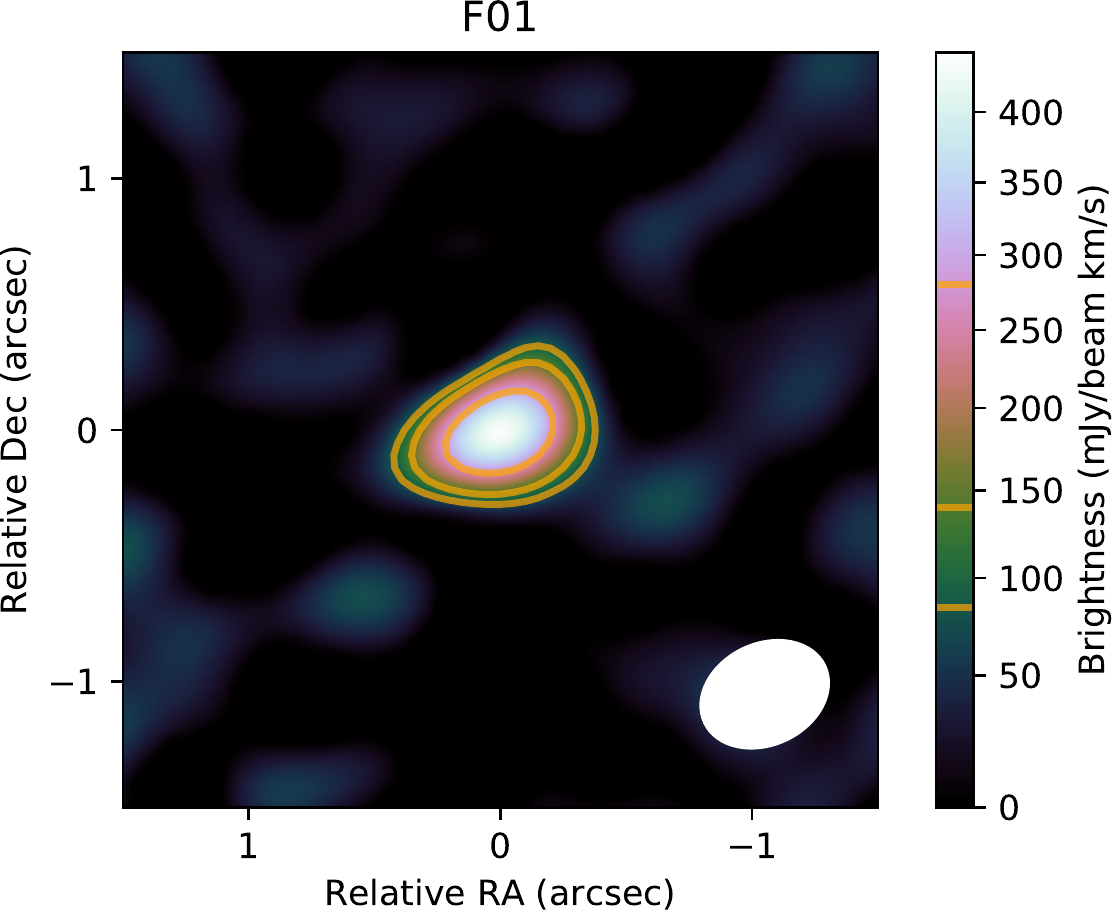}
		\hfill
		\includegraphics[angle=0, width=0.47\textwidth]{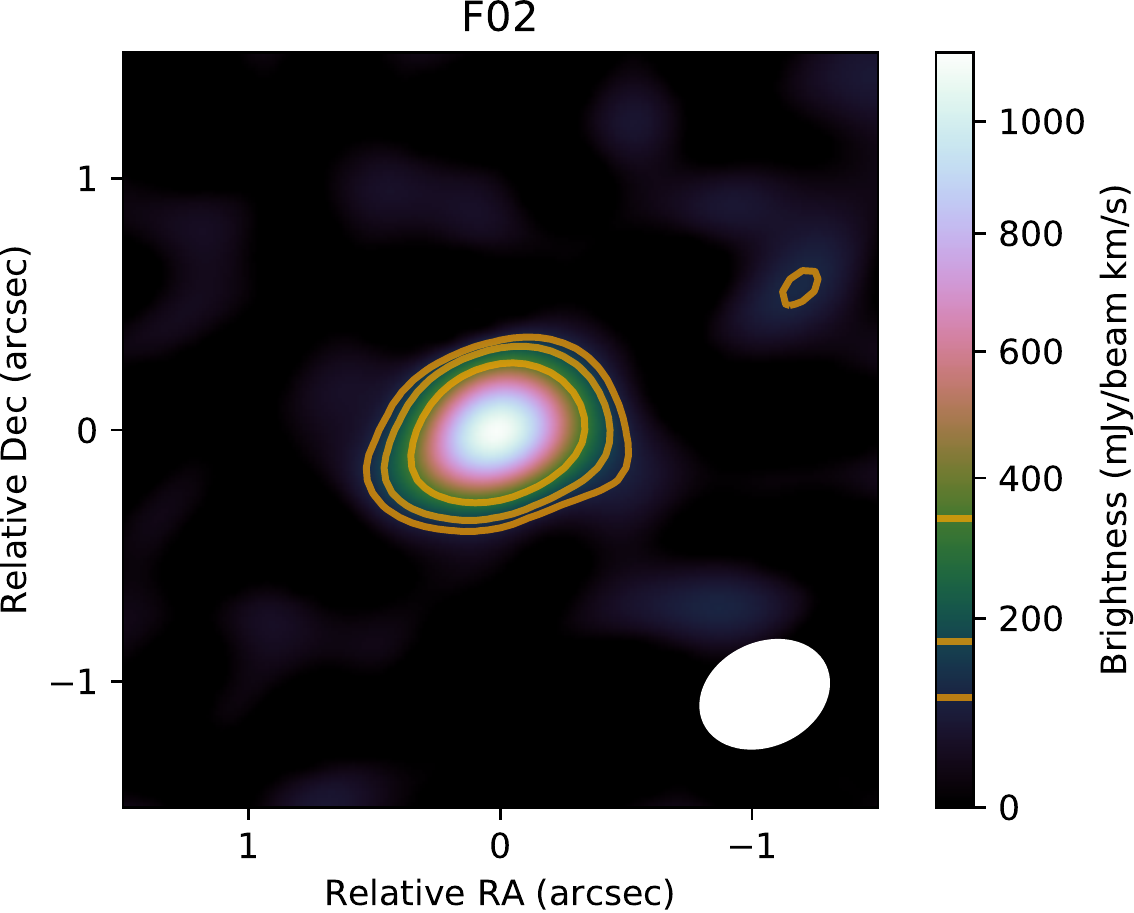}
	\end{minipage}\\[2ex]
	\begin{minipage}[t]{0.85\textwidth}
		\includegraphics[angle=0, width=0.47\textwidth]{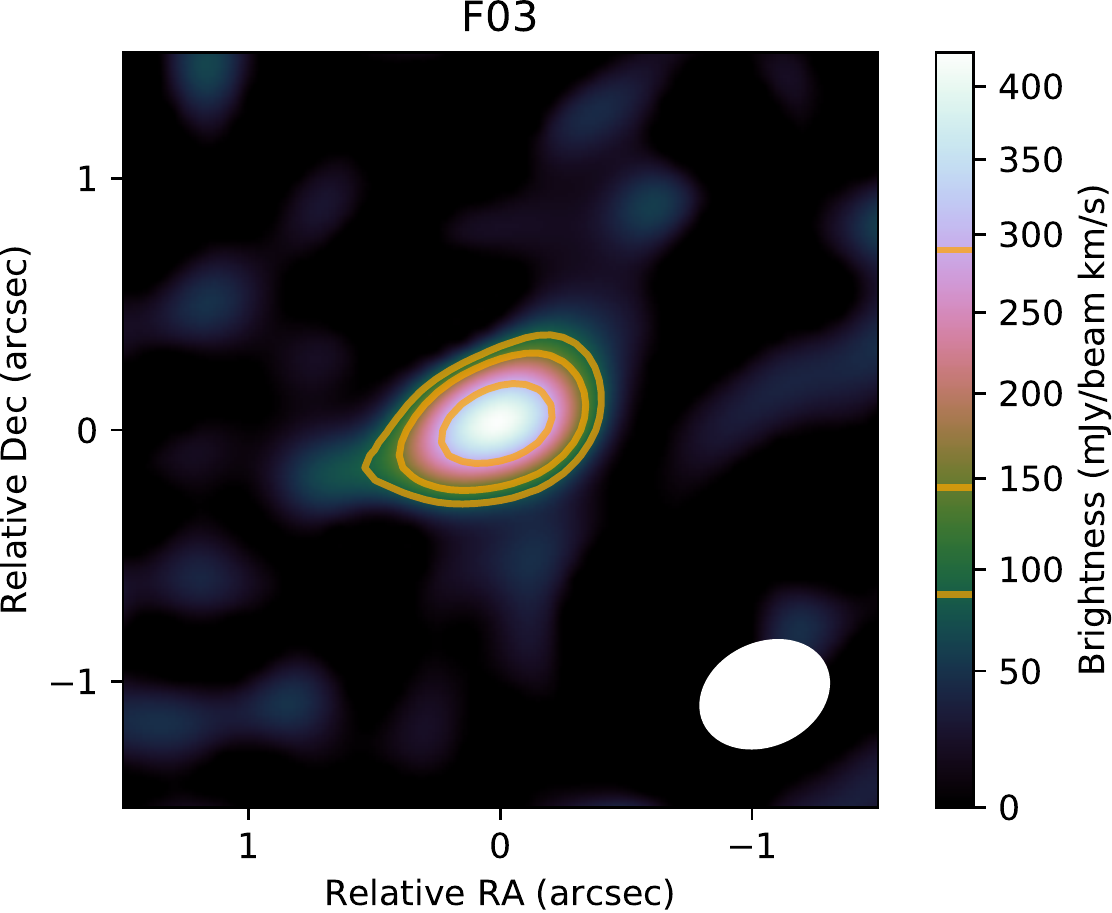}
		\hfill
		\includegraphics[angle=0, width=0.47\textwidth]{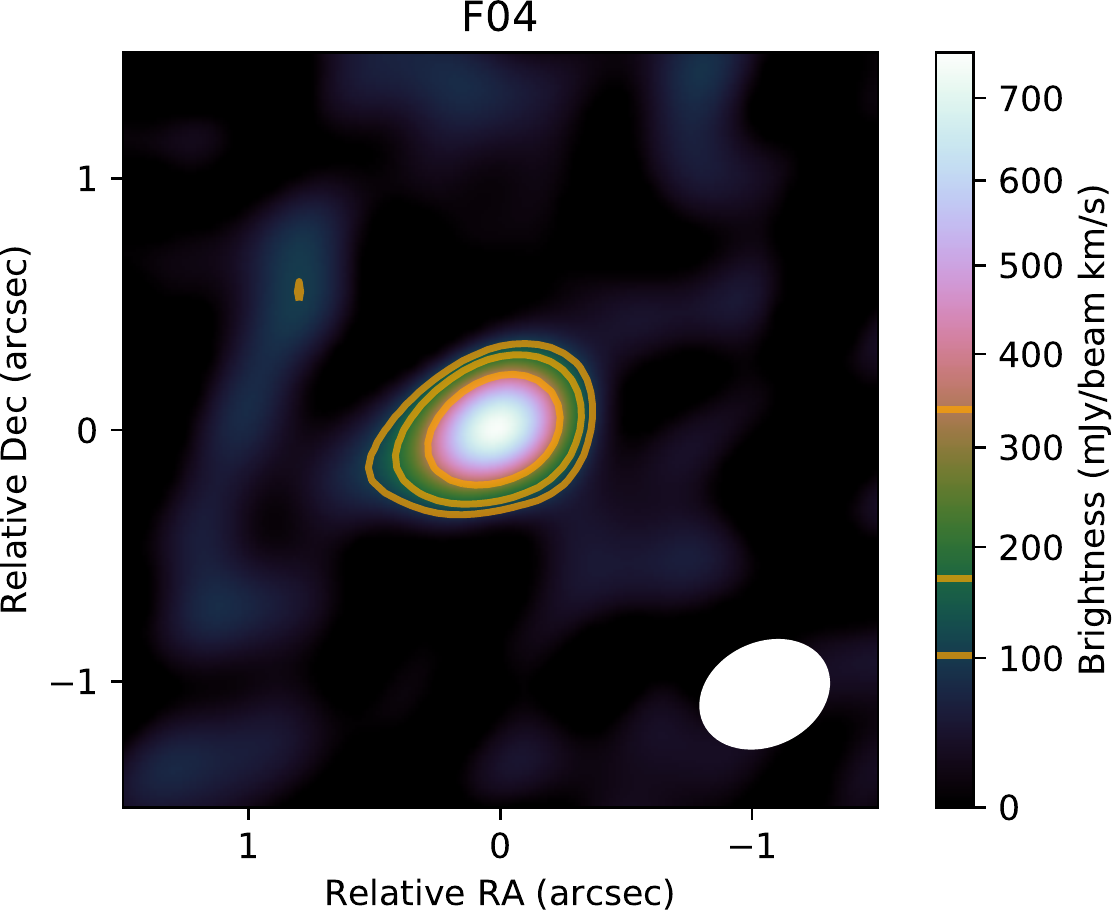}
	\end{minipage}\\[2ex]
	\begin{minipage}[t]{0.85\textwidth}
		\includegraphics[angle=0, width=0.47\textwidth]{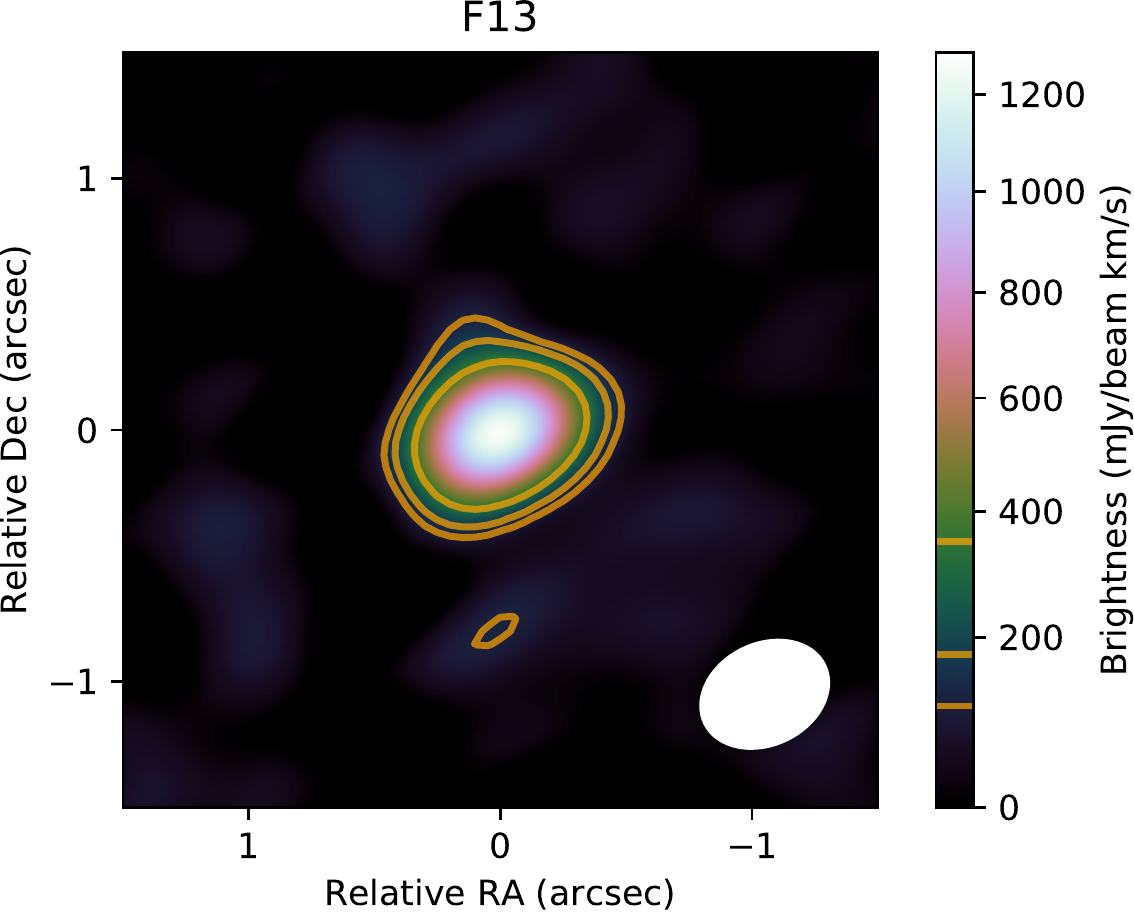}
		\hfill
	\end{minipage}
	\caption{ALMA band~6 SiO v=3 J=5-4 zeroth moment maps of five red supergiants in RSGC1: F01 (upper left), F02 (upper right), F03 (middle left), F04 (middle right), and F13 (bottom left). 
			The ordinate and co-ordinate axis give the offset of the right ascension and declination, respectively, with respect to the input coordinates for the ALMA observations (see columns~2 and 3 in Table~\ref{Tab:targets}).
		The ALMA synthesised beam is shown as a white ellipse in the lower right corner of each panel. The contours  (in orange) show the SiO emission at [3, 5, 10] times the rms value.}
	\label{Fig:SiO}
\end{figure*}

 \begin{figure*}[htp]
 	\centering
 	\begin{minipage}[t]{\textwidth}
 	\includegraphics[angle=0, width=0.45\textwidth]{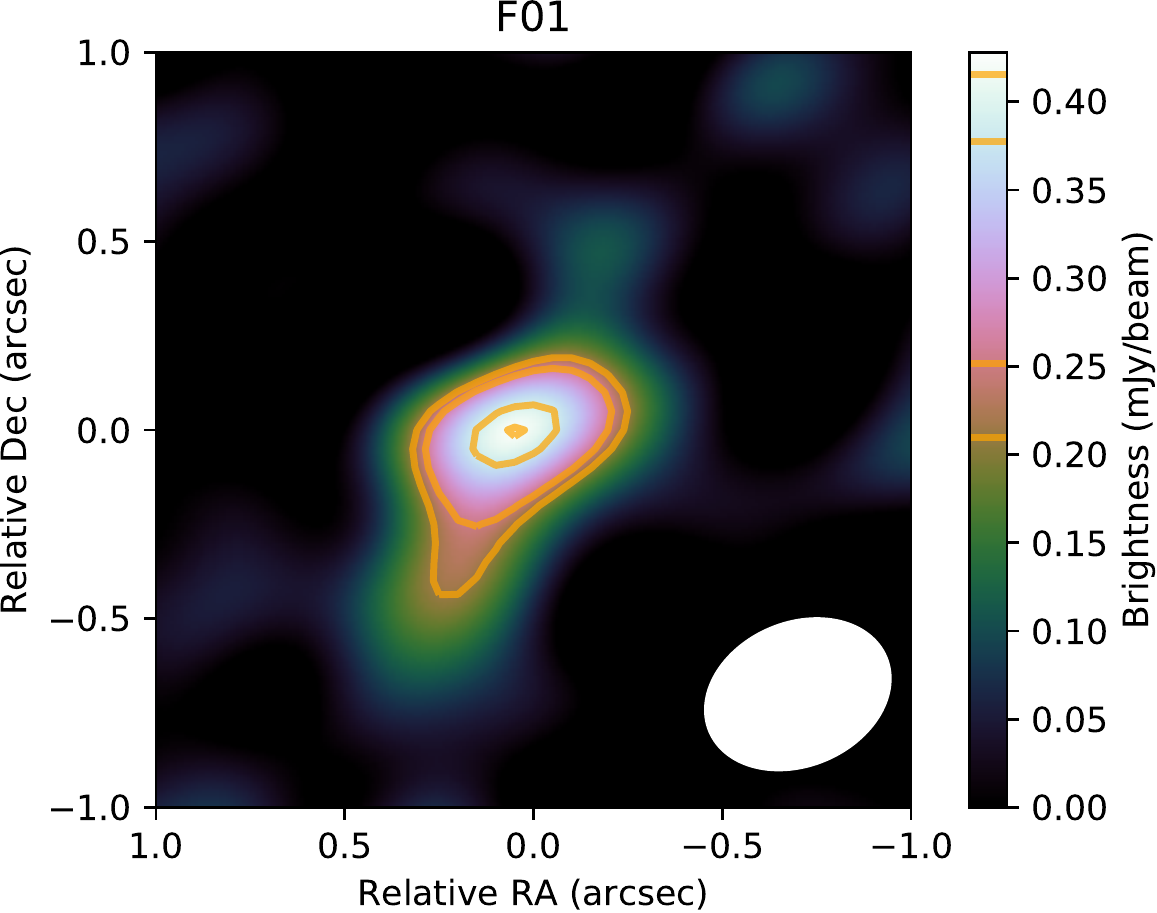}
 	\hfill
 	\includegraphics[angle=0, width=0.45\textwidth]{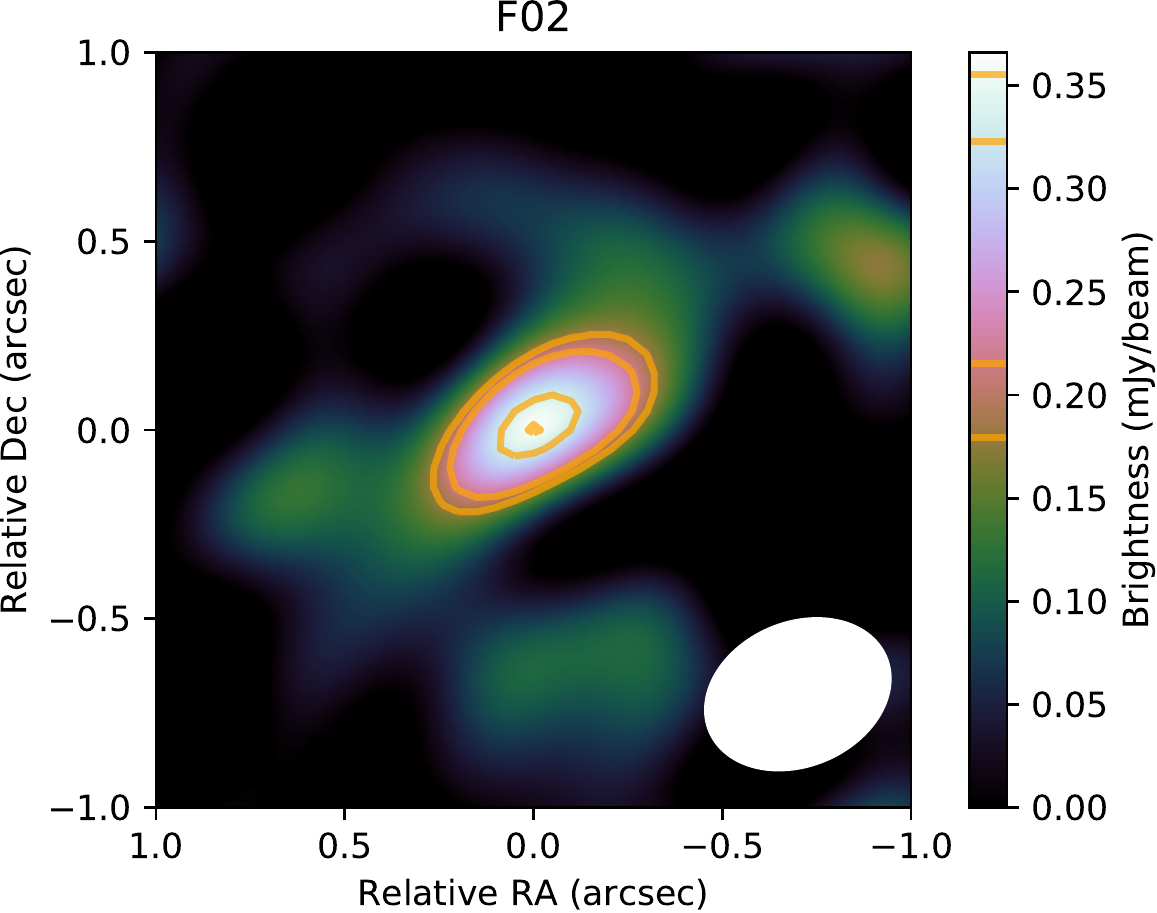}
 	\end{minipage}\\[2ex]
 	\begin{minipage}[t]{\textwidth}
  	\includegraphics[angle=0, width=0.45\textwidth]{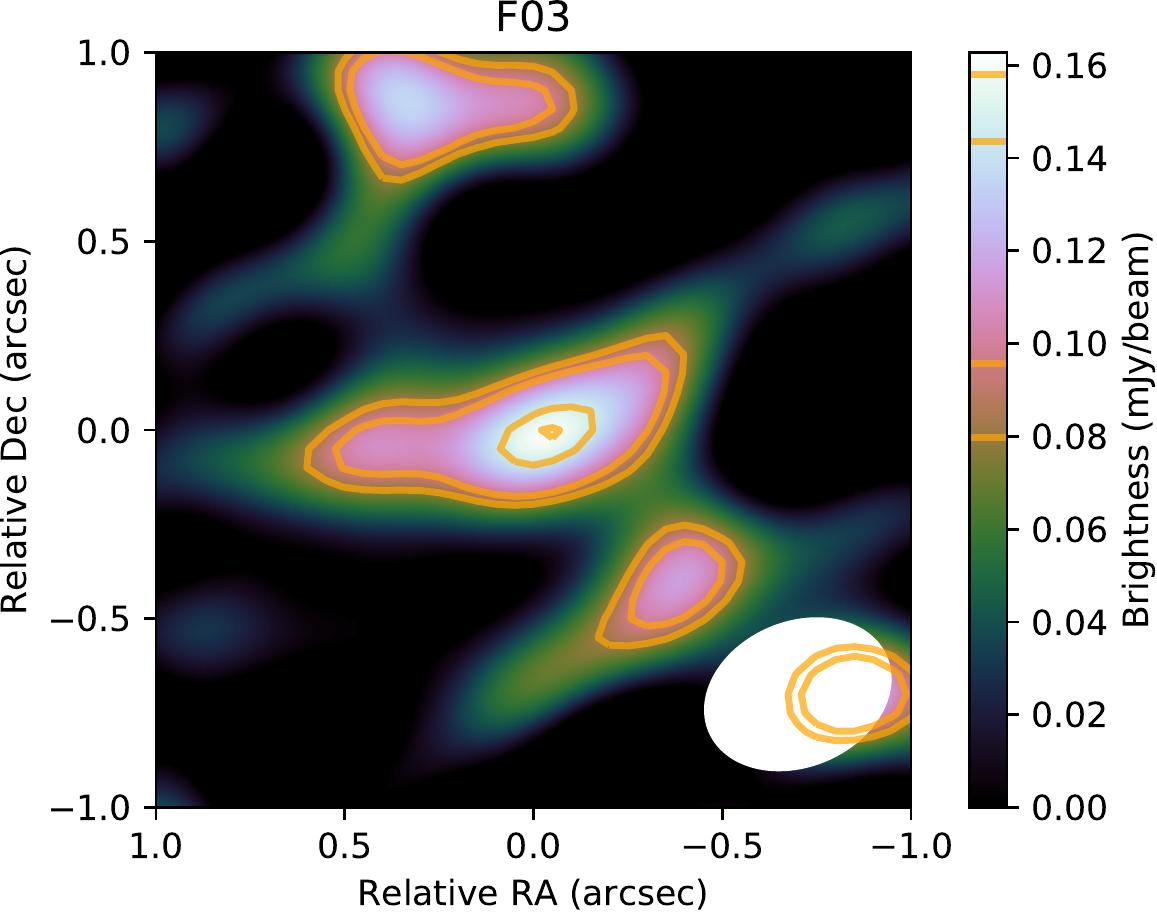}
  	\hfill
  	\includegraphics[angle=0, width=0.45\textwidth]{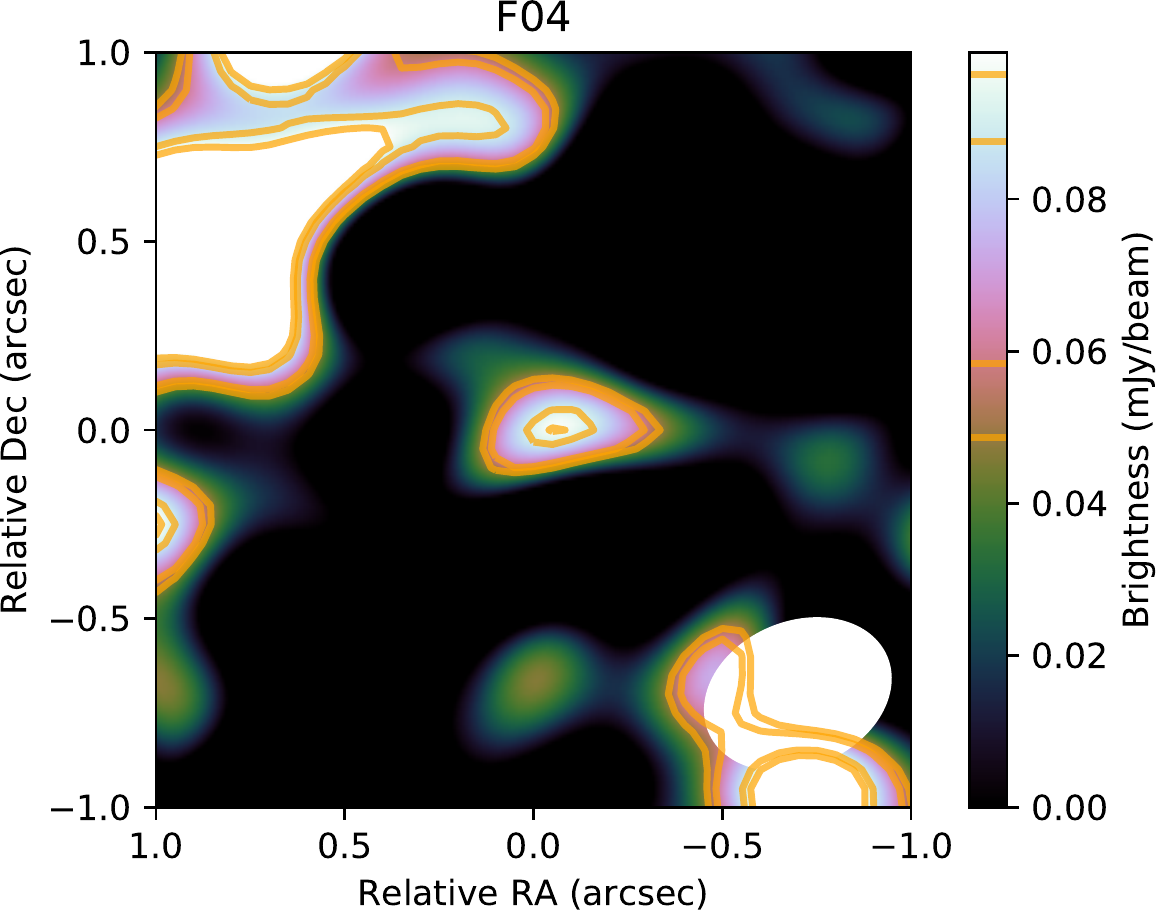}
  	\end{minipage}\\[2ex]
  \begin{minipage}[t]{\textwidth}
  	\includegraphics[angle=0, width=0.45\textwidth]{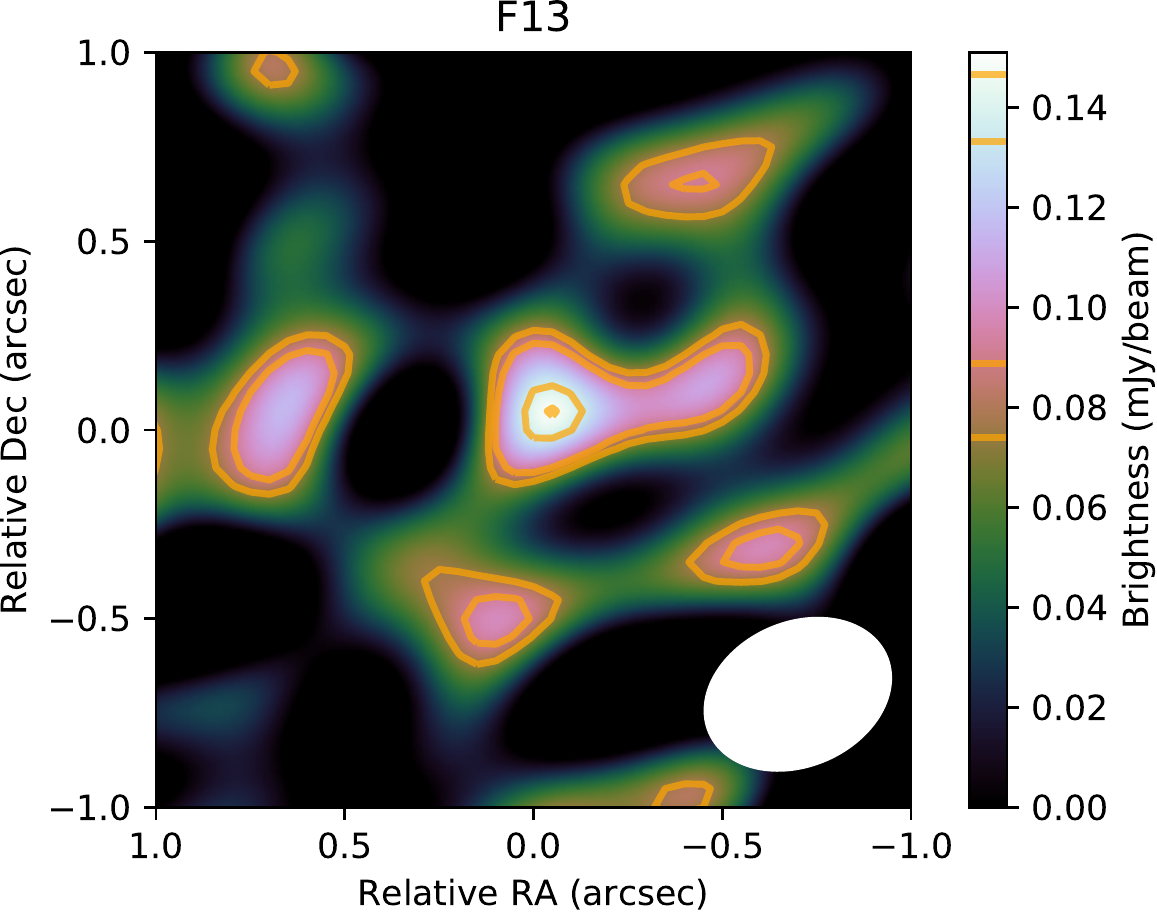}
  	\hfill
  	\end{minipage}
  		\caption{ALMA band~6 continuum maps of five red supergiants in RSGC1: F01 (upper left), F02 (upper right), F03 (middle left), F04 (middle right), and F13 (bottom left). 
  		Contours (in orange) are indicated at [50, 60, 90, and 99]\% of the peak continuum emission (see Table~\ref{Tab:targets}). 
  		The ordinate and co-ordinate axis give the offset of the right ascension and declination, respectively, with respect to the peak of the SiO v=3 J=5-4 zeroth moment map.
  		The ALMA synthesised beam is shown as a white ellipse in the lower right corner of each panel.}
 	\label{Fig:cont}
 \end{figure*}
 
 \afterpage{\clearpage}
 
 %remove if split in above plots
 \newpage
\section{ALMA CO v=0 J=2-1 total intensity and channel maps}\label{App:CO}

In this section, we first present the ALMA CO v=0 J=2-1 total intensity (zeroth moment) and channel maps for all sources that were detected in CO(2-1); i.e.\ F01, F02, F03, F04, and F13. Thereafter, we discuss for each source individually the spectral regions contaminated by emission from the ISM.

For all sources, the continuum was less than the spectral channel rms so no continuum subtraction  was performed. Fig.~\ref{Fig:CO_mom0} shows the CO v=0 J=2-1 total intensity map of F01, F02, F03, F04, and F13. The rms in each image varies due to different amounts of ISM contamination as well as velocity widths averaged, with values being, respectively, 26, 21, 25, 23, and 29 mJy/beam km/s. The peak signal-to-noise ratio ranges between $\sim$12\,--\,34.
	 
%/Users/leen/leda/ALMA/RSGs/programs/plot_CO_mom0.py
\begin{figure*}[htp]
	\centering
	\begin{minipage}[t]{0.85\textwidth}
		\includegraphics[angle=0, width=0.49\textwidth]{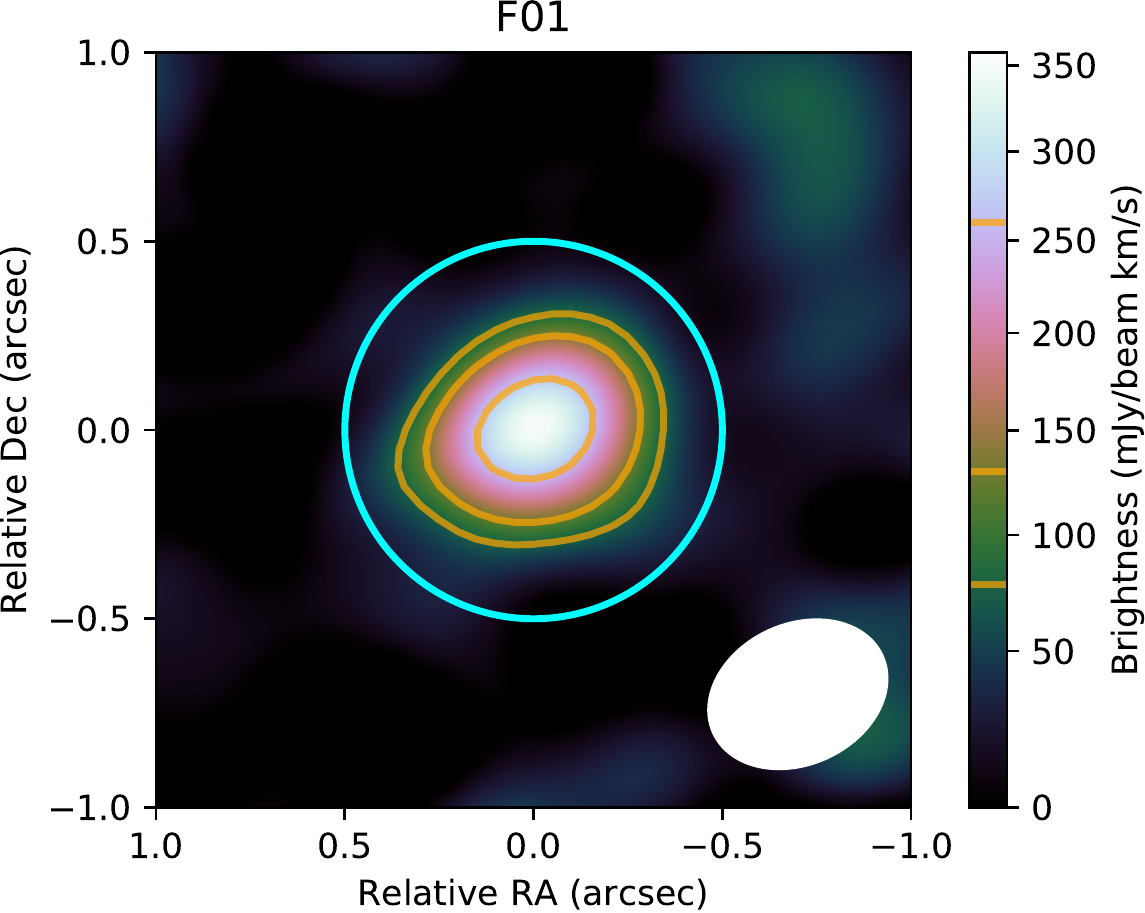}
		\hfill
		\includegraphics[angle=0, width=0.49\textwidth]{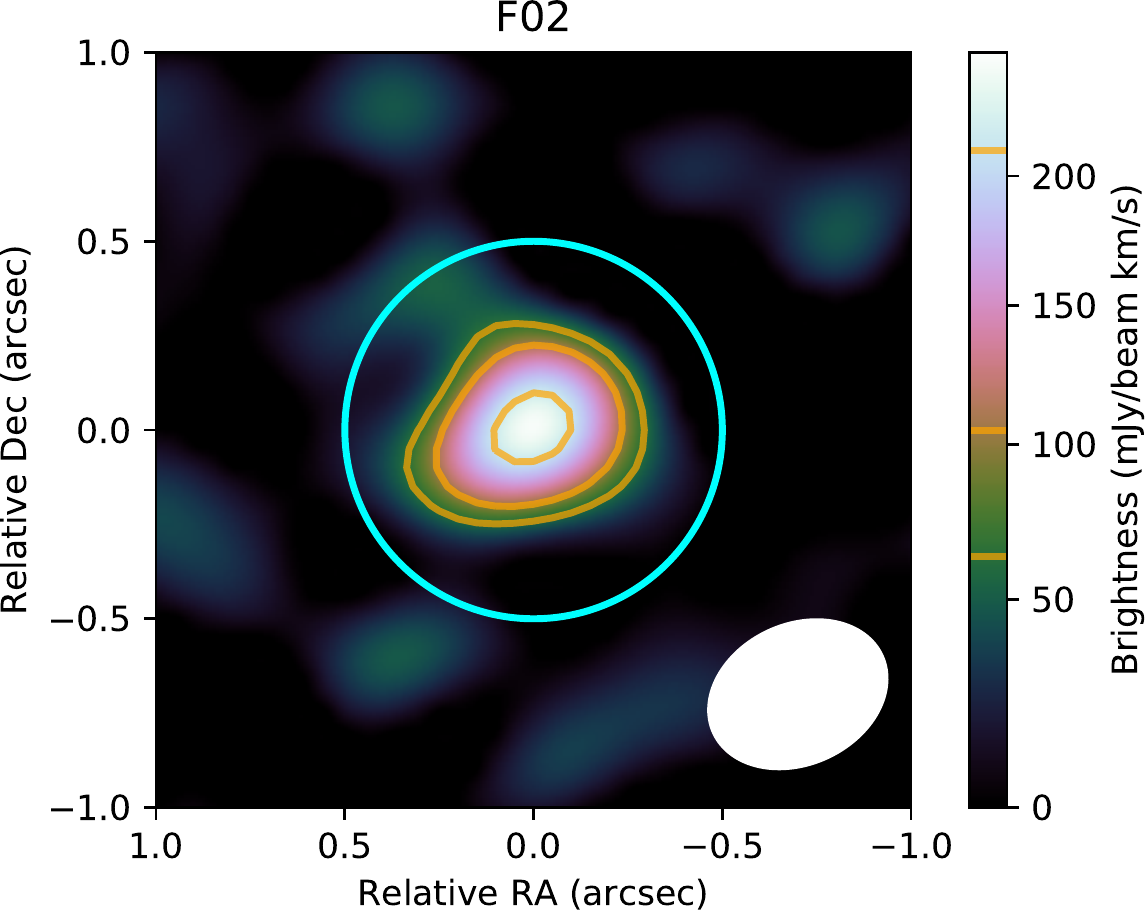}
	\end{minipage}\\[2ex]
	\begin{minipage}[t]{0.85\textwidth}
		\includegraphics[angle=0, width=0.49\textwidth]{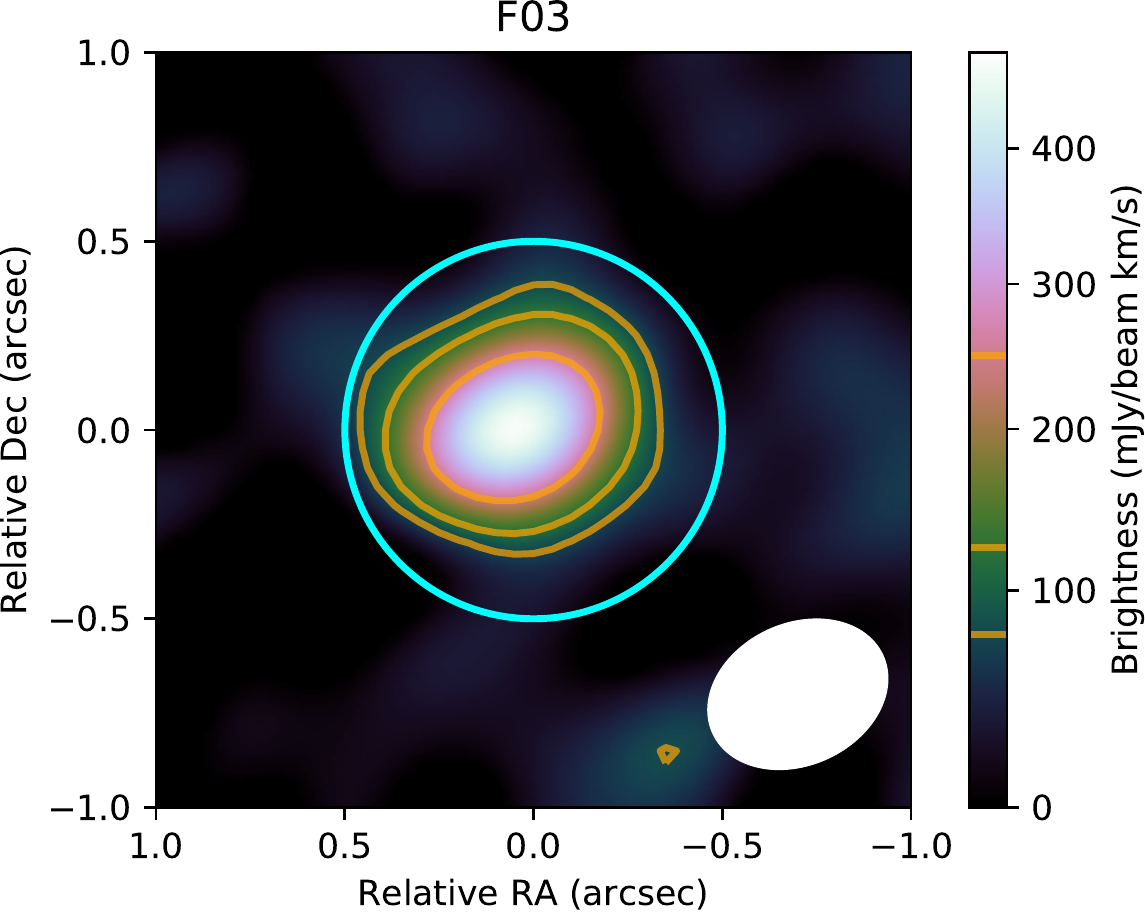}
		\hfill
		\includegraphics[angle=0, width=0.49\textwidth]{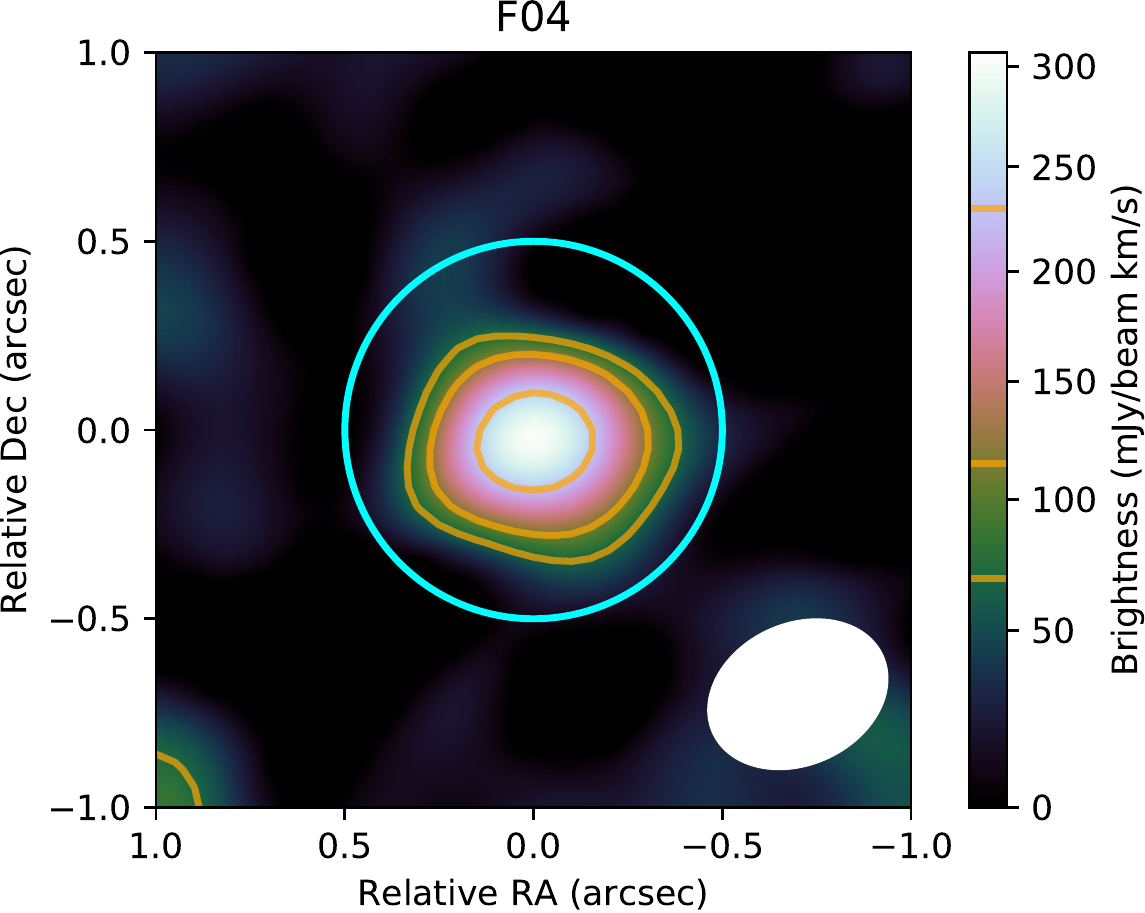}
	\end{minipage}\\[2ex]
	\begin{minipage}[t]{0.85\textwidth}
		\includegraphics[angle=0, width=0.49\textwidth]{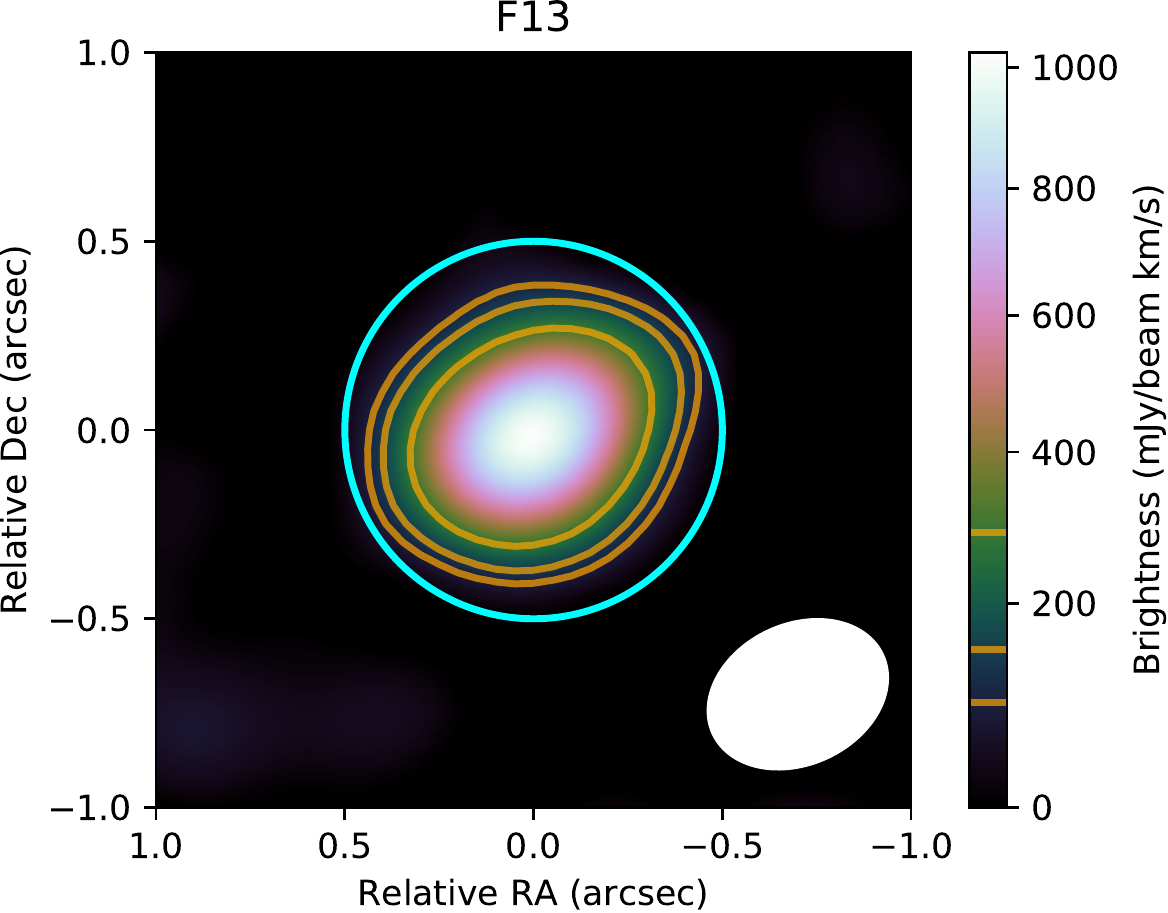}
		\hfill
	\end{minipage}
	\caption{ALMA band~6 CO v=0 J=2-1 zeroth moment maps of five red supergiants in RSGC1: F01 (upper left), F02 (upper right), F03 (middle left), F04 (middle right), and F13 (bottom left). 
		The ordinate and co-ordinate axis give the offset of the right ascension and declination, respectively, with respect to the input coordinates for the ALMA observations (see columns~2 and 3 in Table~\ref{Tab:targets}).
		The ALMA synthesised beam is shown as a white ellipse in the lower right corner of each panel. The contours  (in orange) show the CO emission at [3, 5, 10] times the rms value. The cyan circle has a diameter of  size 1\arcsec\ and is centred on the peak of the continuum emission.}
	\label{Fig:CO_mom0}
\end{figure*}

\begin{figure}[htp]
	\centering
	\includegraphics[width=0.5\textwidth]{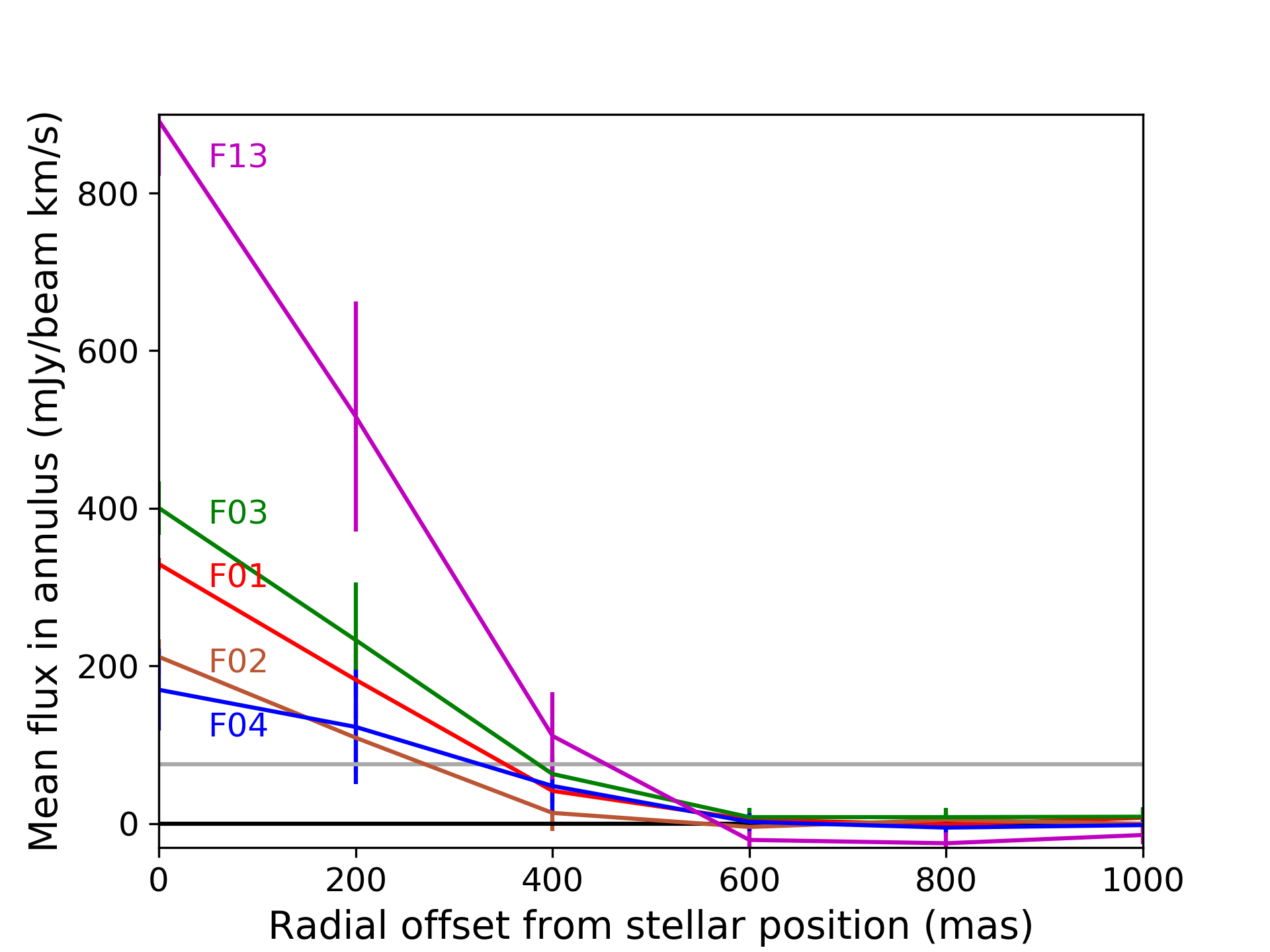}
	\caption{Measurements of the azimuthally averaged flux in annuli of 200-mas thick. The grey line indicates the 3$\times$25\,=\,75 mJy beam$^{-1}$ km s$^{-1}$ cutoff.
	}
	\label{Fig:CO_annuli}
\end{figure}

\bigskip

For each source, two channel maps are presented, first 
 a zoom onto the central 2\farcs2 region and then a larger map of size 8\arcsec\   (Figs.~\ref{Fig:CO_F01} -- \ref{Fig:CO_F13}). The combined use of these maps is needed to assess the ISM contamination. 

%from /Users/leen/leda/ALMA/RSGs/programs/make_channel_maps_paper_new_data.py
\begin{figure}[htp]
	\centering\centering\includegraphics[angle=0, width=0.72\textwidth]{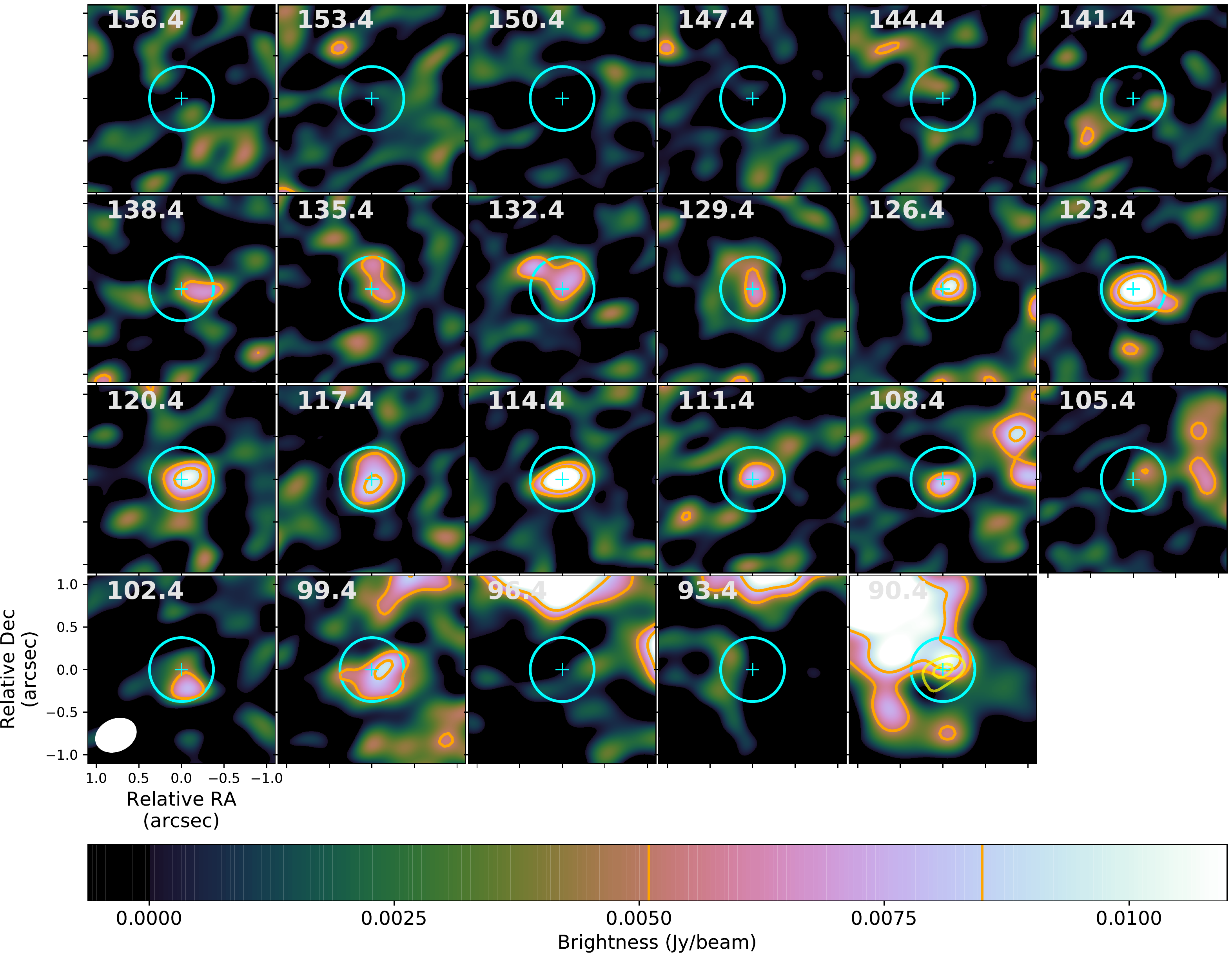}
	\centering\centering\includegraphics[angle=0, width=0.72\textwidth]{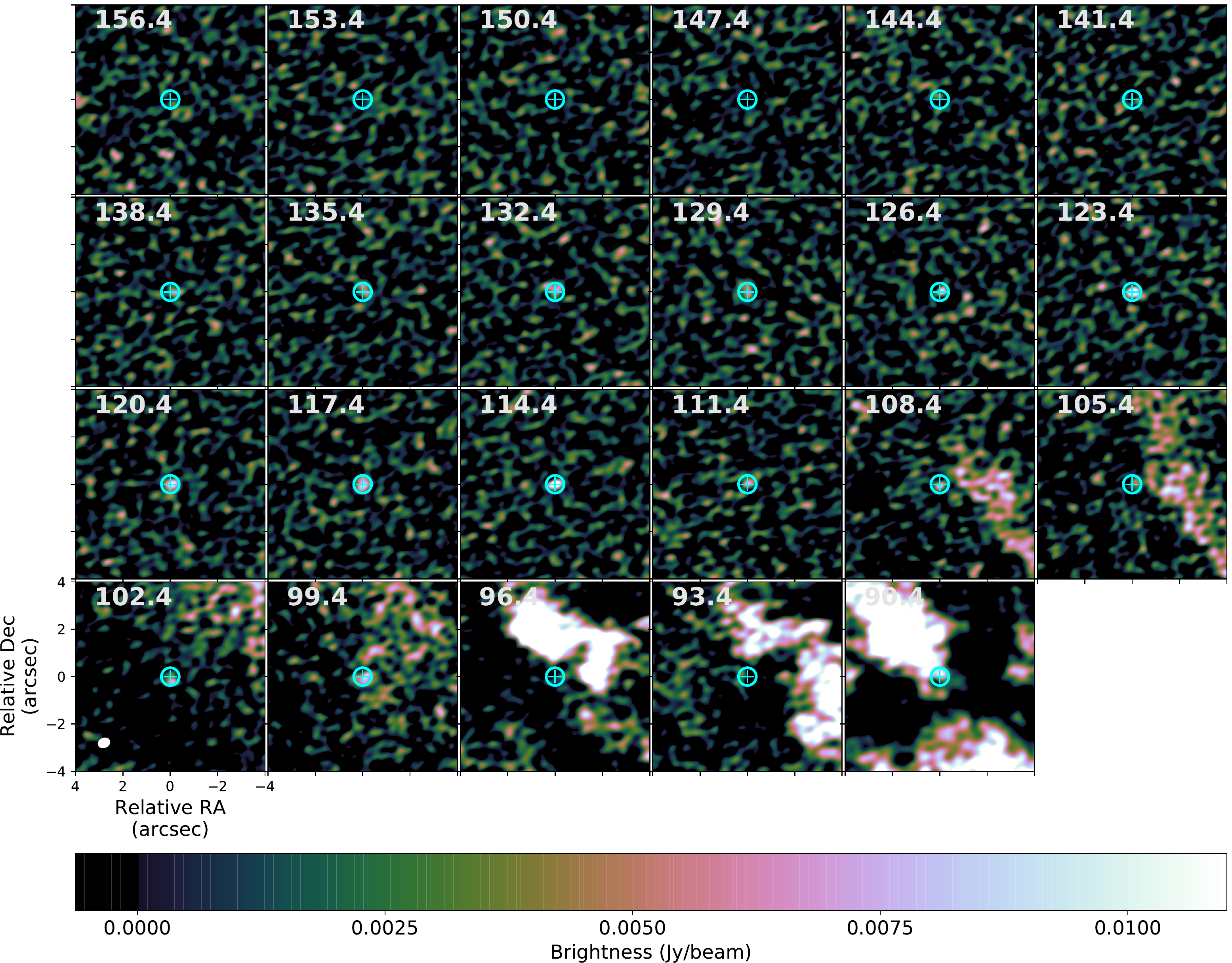}
\caption{
\textbf{$^{12}$CO v=0 J=2-1 channel map of F01.}
The ordinate and co-ordinate axis give the offset of the right ascension and declination, respectively, with respect to the peak of the continuum emission (indicated by the cyan cross in the upper panels). The velocity is given in the upper left corner of each panel (in units of km~s$^{-1}$). The cyan circle  indicates the circular aperture of 0\farcs75 (diameter) centred on the peak of the continuum emission that was used for extracting the CO(2-1) line profile. The ALMA beam is shown as a white ellipse in the bottom left corner of the bottom left panel. 
Upper plot: The contours (in orange) show the CO emission at [3, 5]$\times$$\sigma_{\rm{rms}}$ (see Table~\ref{Tab:targets}). The contours (in yellow) in the last panel show the continuum emission at [60, 90, and 99]\% of the peak continuum emission.
 }
  \label{Fig:CO_F01}
\end{figure}

\clearpage

 \begin{figure}[htp]
 	\centering\centering\includegraphics[angle=0, width=0.8\textwidth]{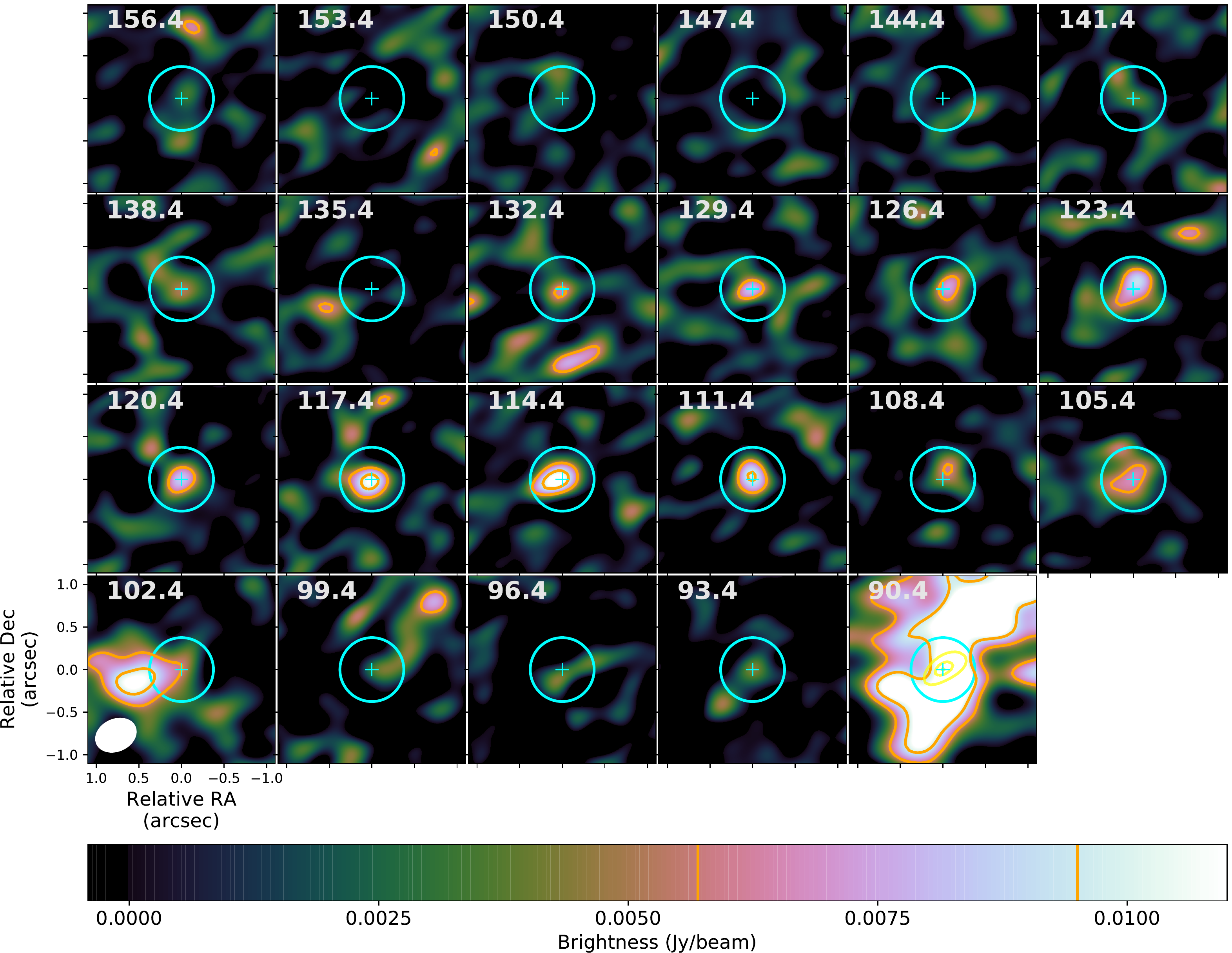}
 	\centering\centering\includegraphics[angle=0, width=0.8\textwidth]{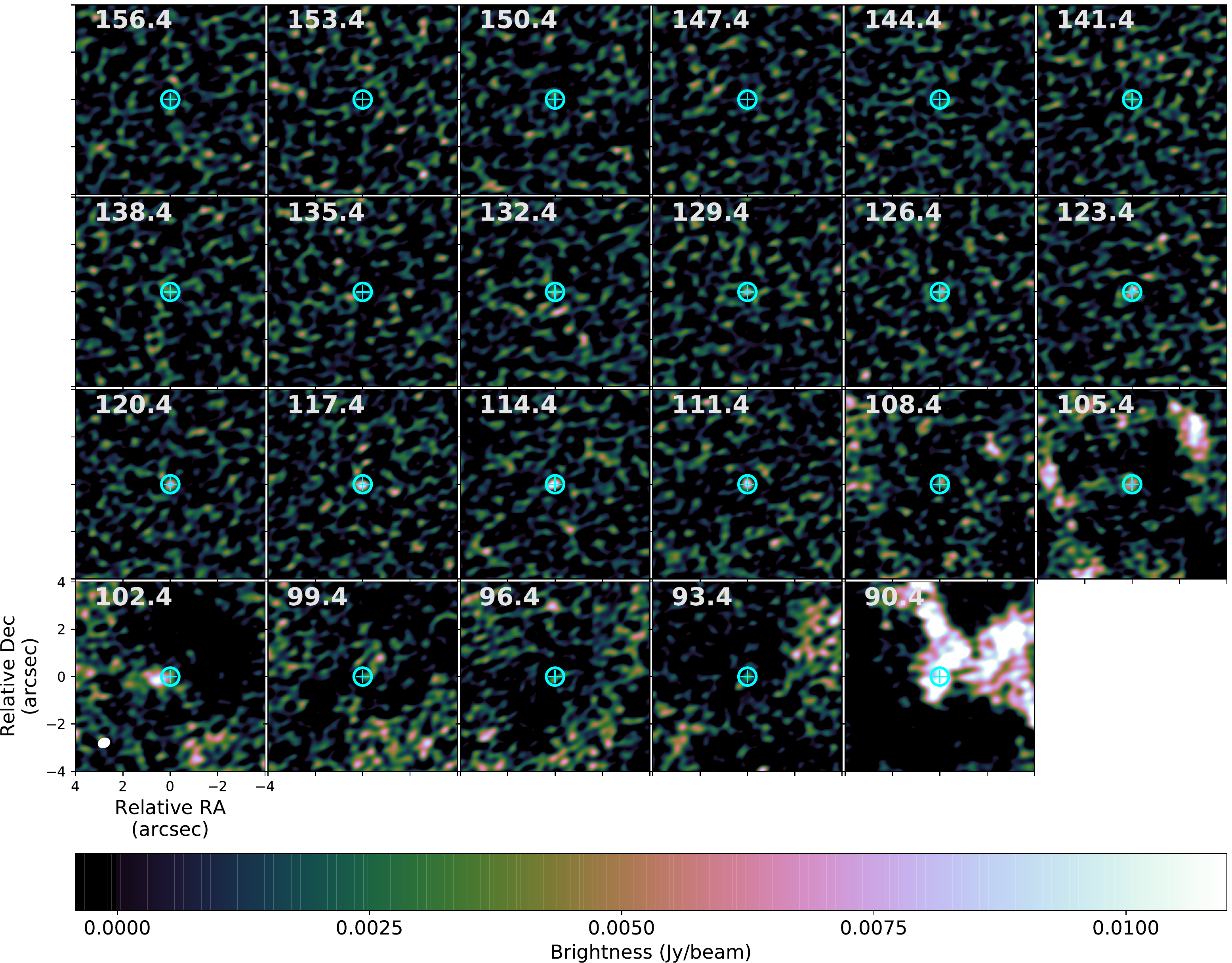}
 	\caption{
 		\textbf{$^{12}$CO v=0 J=2-1 channel map of F02.} Same as Fig.~\ref{Fig:CO_F01}, but for F02.}
 	\label{Fig:CO_F02}
 \end{figure}
 \clearpage
 
 \begin{figure}[htp]
	\centering\centering\includegraphics[angle=0, width=0.8\textwidth]{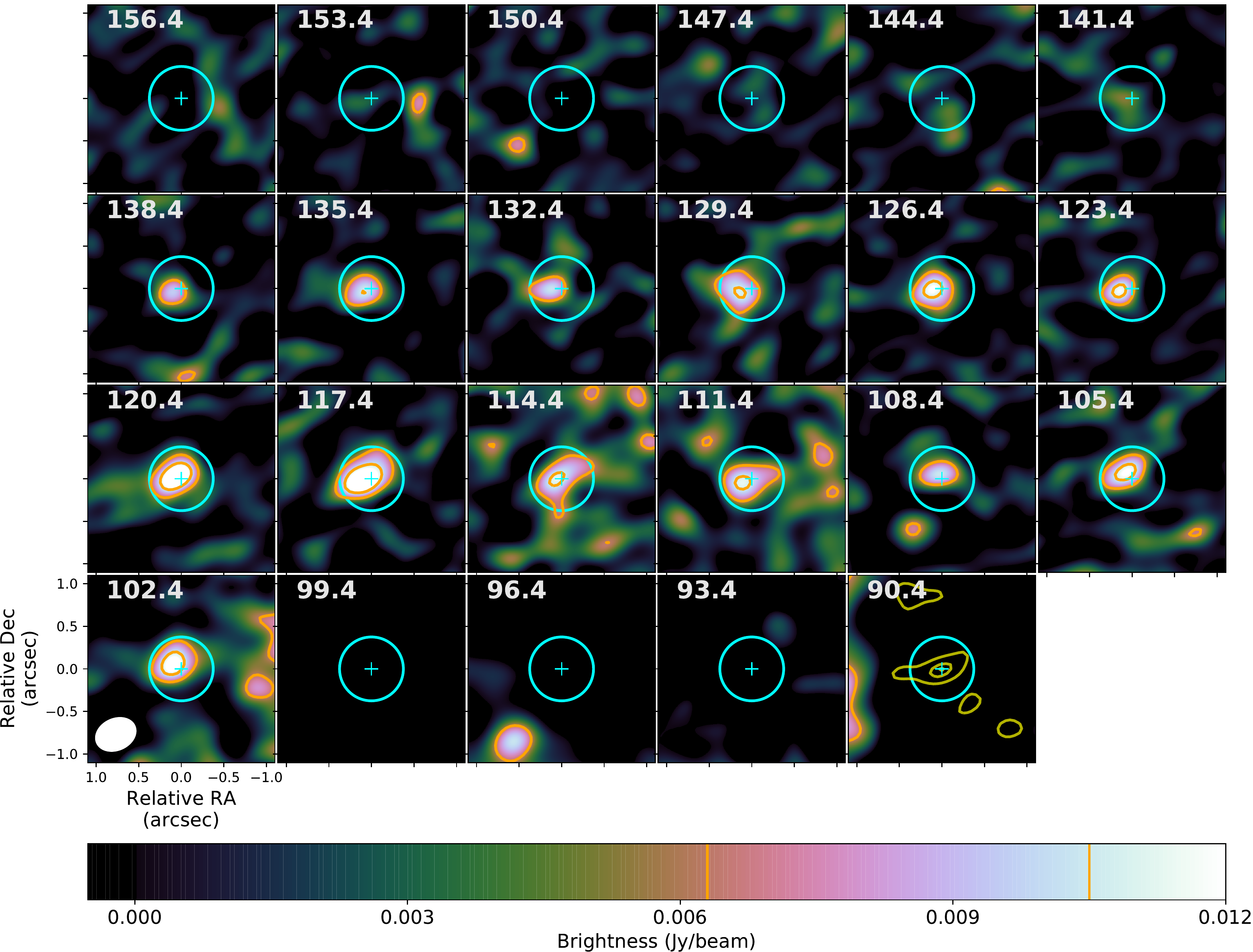}
	\centering\centering\includegraphics[angle=0, width=0.8\textwidth]{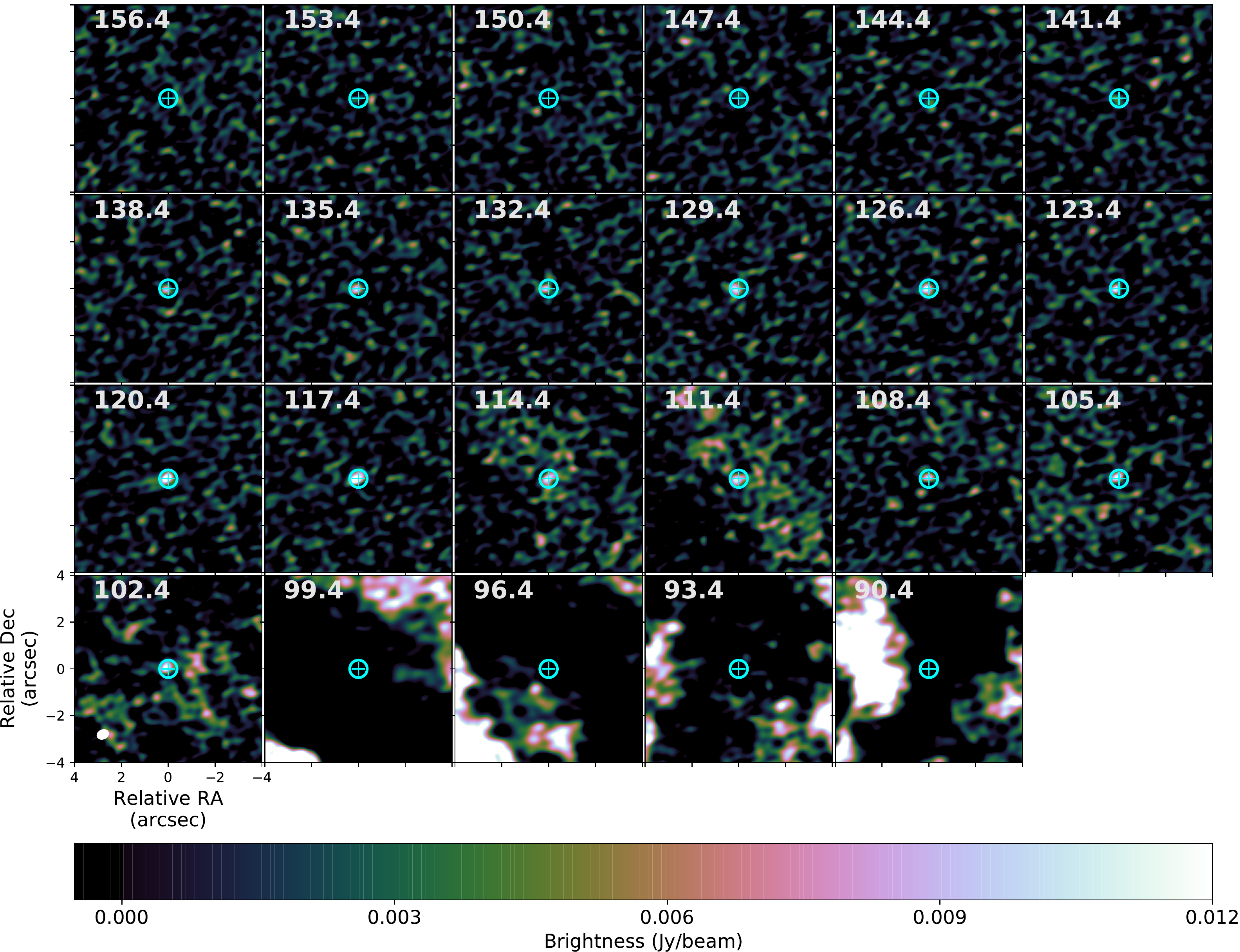}
	\caption{
		\textbf{$^{12}$CO v=0 J=2-1 channel map of F03.} Same as Fig.~\ref{Fig:CO_F01}, but for F03.}
	\label{Fig:CO_F03}
\end{figure}
\clearpage

 \begin{figure}[htp]
	\centering\centering\includegraphics[angle=0, width=0.8\textwidth]{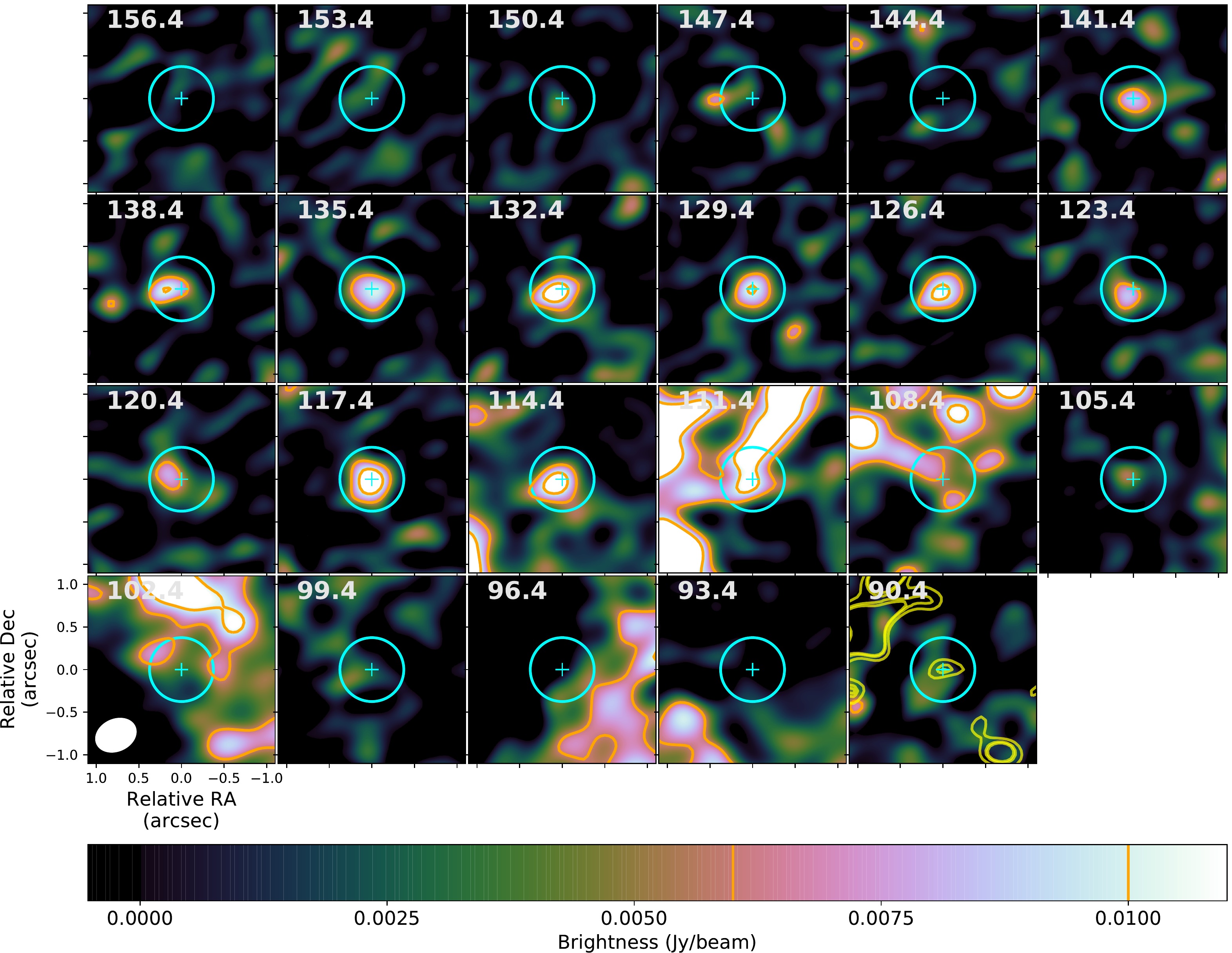}
	\centering\centering\includegraphics[angle=0, width=0.8\textwidth]{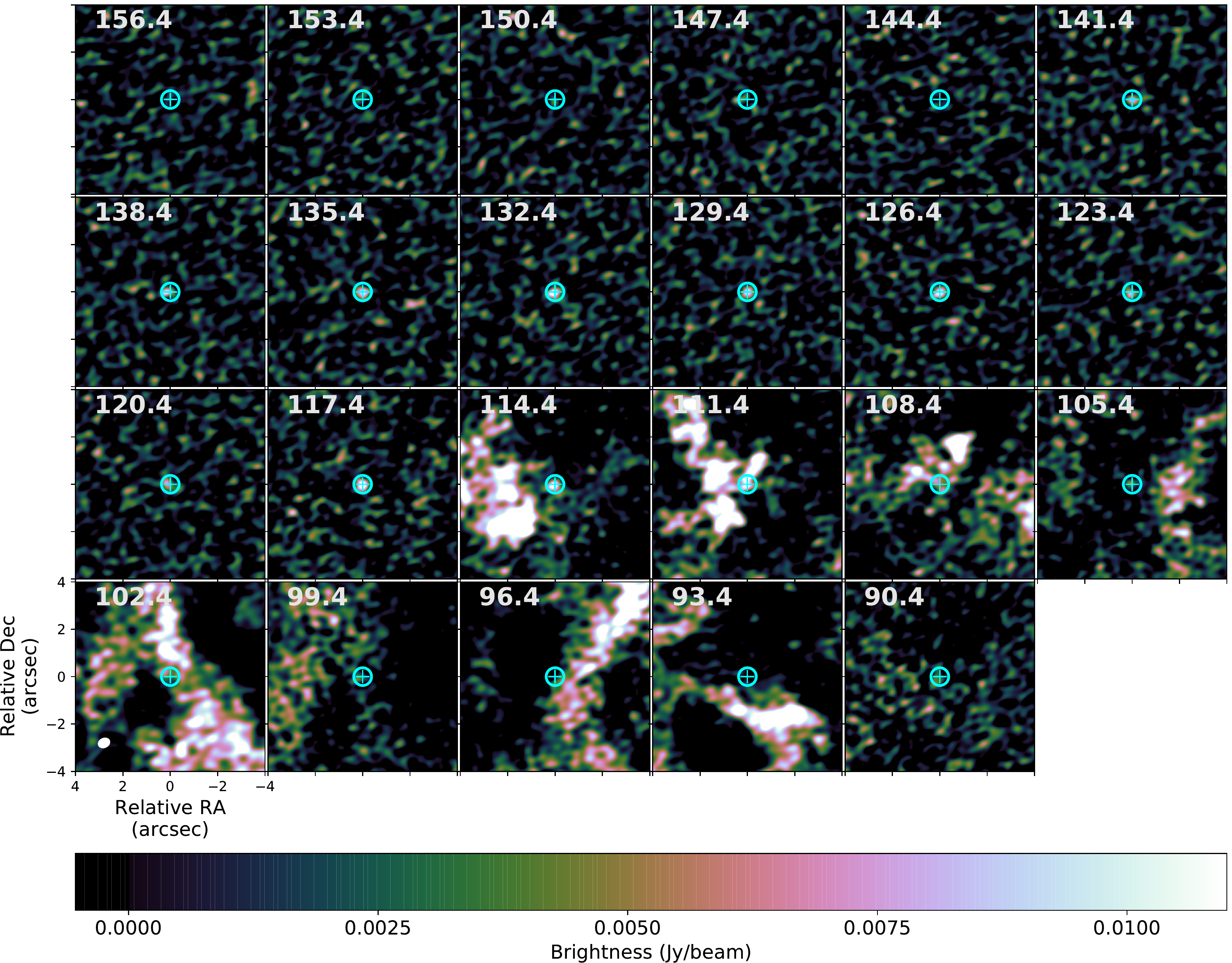}
	\caption{
		\textbf{$^{12}$CO v=0 J=2-1 channel map of F04.} Same as Fig.~\ref{Fig:CO_F01}, but for F04.}
	\label{Fig:CO_F04}
\end{figure}
\clearpage

 \begin{figure}[htp]
	\centering\centering\includegraphics[angle=0, width=0.8\textwidth]{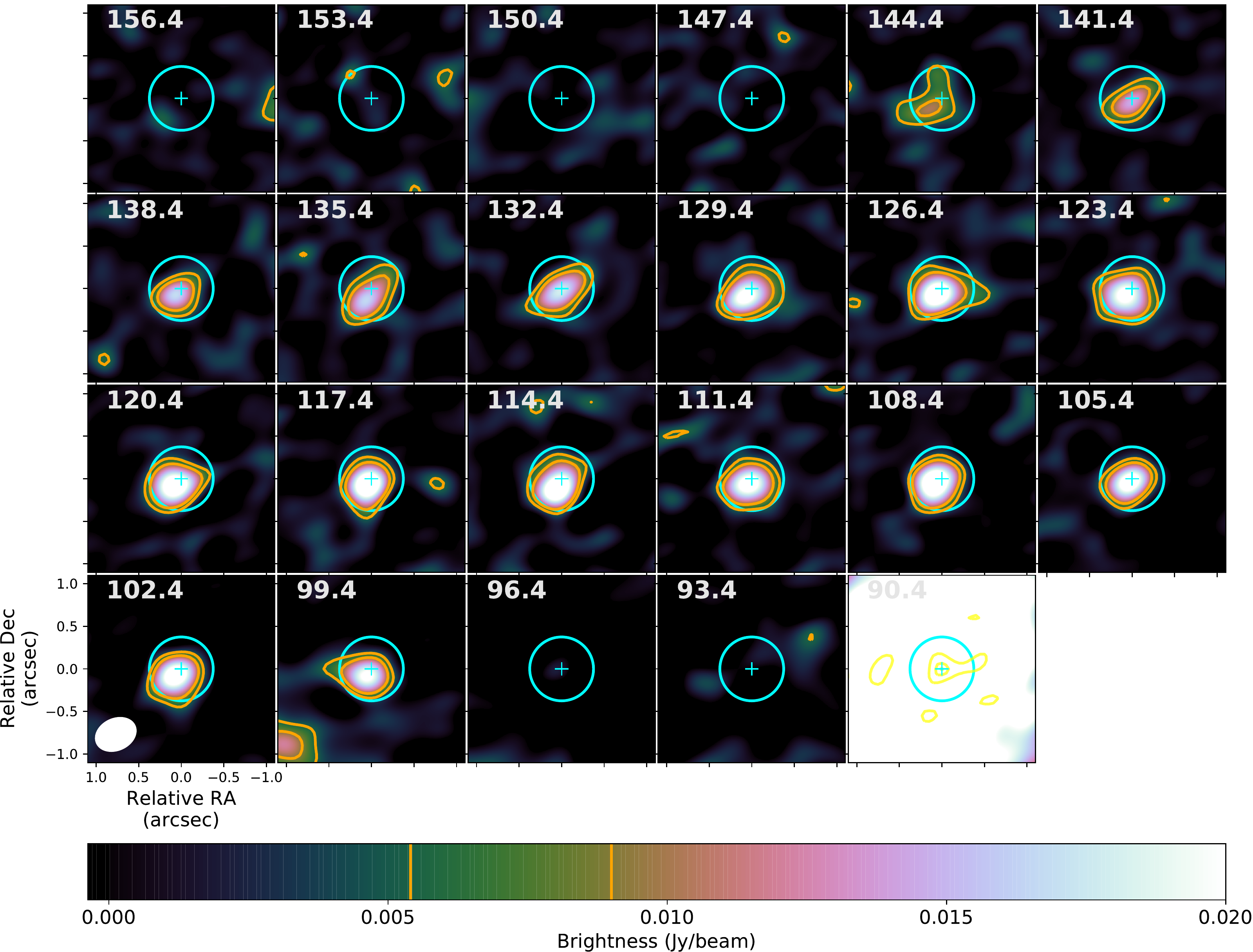}
	\centering\centering\includegraphics[angle=0, width=0.8\textwidth]{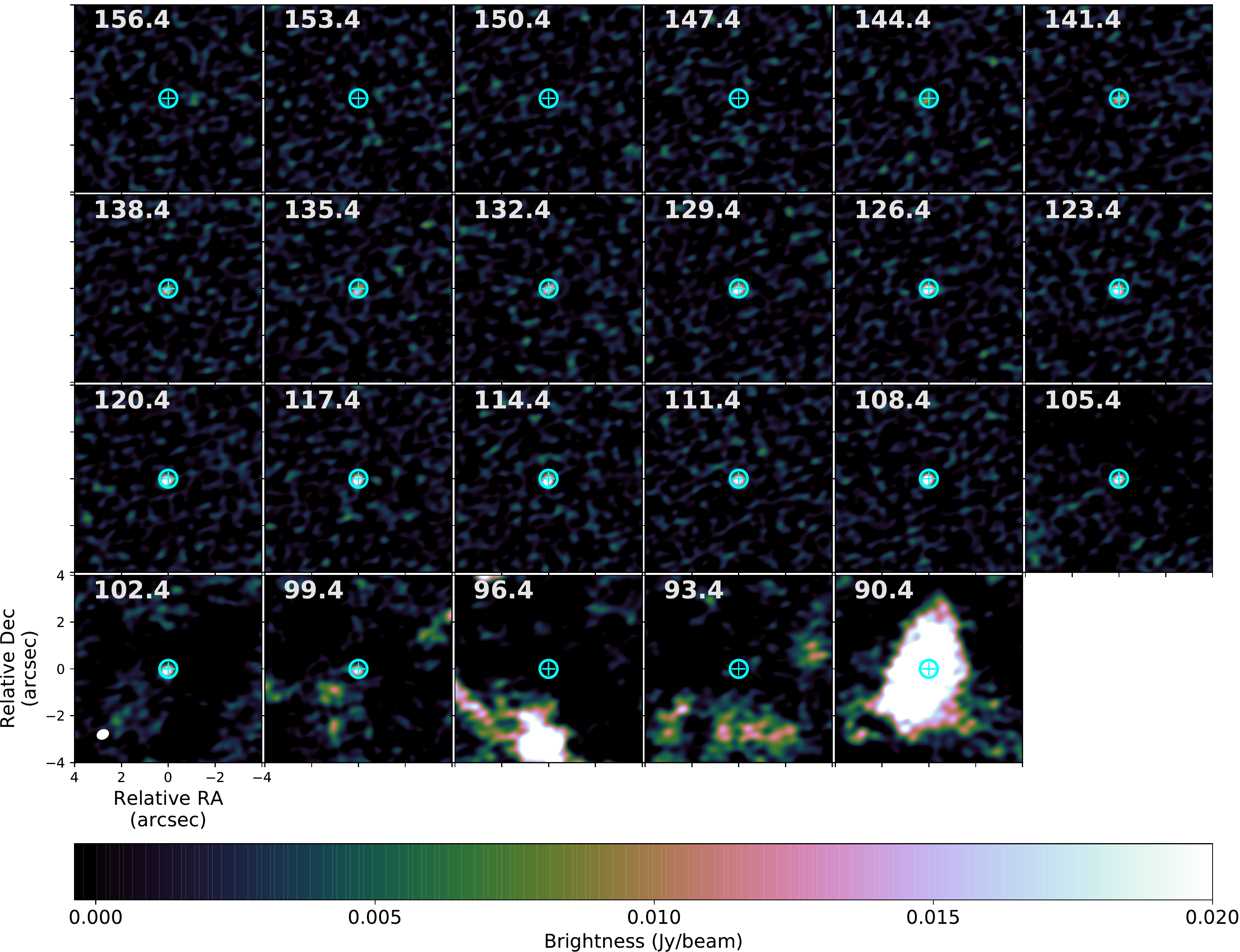}
	\caption{
		\textbf{$^{12}$CO v=0 J=2-1 channel map of F13.} Same as Fig.~\ref{Fig:CO_F01}, but for F13.}
	\label{Fig:CO_F13}
\end{figure}

\afterpage{\clearpage} 
\newpage
\mbox{}

For all sources, there is much less (or no) ISM contamination visible on the high velocity side, although there might be a noise background. In particular:
\vspace*{-2ex}
\paragraph{F01:} At velocities lower than $\sim$105\,km~s$^{-1}$ the ISM emission starts contributing. In general, the angular extent of a spherically symmetric source with monotonically increasing velocity will be larger around zero rest velocity, and will decrease when going to redder and bluer velocities. For the source F01, we notice extra emission in the channels from 129.4--135.4\,km~s$^{-1}$ offset from the centre to the north-east, that is also visible as increase in flux density in Fig.~\ref{Fig:CO}. Either this is noise (the blob is at $<$4$\times$$\sigma_{\rm{rms}}$) or it is a genuine source (as, for example, another cluster member, a companion, ...). When fitting the CO(2-1) line profile for retrieving the gas mass-loss rate, this extra emission is neglected. In the case that F01 would still contribute significant flux in the 129.4-135.3\,km~s$^{-1}$ channels, the terminal velocity would be underestimated. Under the rough assumption that the extra blob to the north-east would contribute $\sim$2.8\,mJy in the 129.4\,--\,135.5\,km~s$^{-1}$ velocity range, the terminal velocity could reach up to $\sim$16\,km~s$^{-1}$. A good prediction of the (flux-corrected) red wing and peak flux around 135\,km~s$^{-1}$ (with peak flux then being around 5 mJy) is achieved for a mass-loss rate of 2.2$\times$10$^{-6}$\,\Msun/yr and $v_{\rm{LSR}}$ around 127\,km~s$^{-1}$, but the peak around 115\,--\,125\,km~s$^{-1}$ is then underpredicted by a factor of $\sim$2 (see dotted black line in upper left panel in Fig.~\ref{Fig:CO}). That value for \Mdot$_{\rm{CO}}$ is only $\sim$10\% higher than the value listed in Table~\ref{Tab:outcome}.

\vspace*{-2ex}
\paragraph{F02:} The CO(2-1) spectrum of F02 is quite clean, with the exception of the ISM contamination at velocities lower than $\sim$105\,km~s$^{-1}$.
\vspace*{-2ex}
\paragraph{F03:} For velocities lower than $\sim$114.4\,km~s$^{-1}$, there is a strong contribution from the ISM, which then vanishes at $\sim$99.4\,km~s$^{-1}$. The theoretical line profile for a terminal wind velocity of 13\,km~s$^{-1}$ and \Mdot$_{\rm{CO}}$ of 3.2$\times$10$^{-6}$\,\Msun/yr is slightly underpredicting the red wing  of the ALMA CO(2-1) data (Fig.~\ref{Fig:CO}). A terminal wind velocity of 17\,km~s$^{-1}$ and \Mdot$_{\rm{CO}}$ of 3.8$\times$10$^{-6}$\,\Msun/yr gives a better prediction of the red wing for flux values below $\sim$4\,mJy, but then the typical two-horn profile is predicted too wide. That difference of 15\% in \Mdot$_{\rm{CO}}$ owing to an uncertainty in $v_\infty$ is within the quoted uncertainty on the retrieved \Mdot$_{\rm{CO}}$-values, see Table~\ref{Table:sigma}.

\vspace*{-2ex}
\paragraph{F04:} Similar to F03, the ISM is contaminating the channel maps and spectrum for velocities $\la$114.4\,km~s$^{-1}$.
\vspace*{-2ex}
\paragraph{F13:} F13 is the source for which the CO(2-1) emission is the least contaminated by the ISM. Only for velocities $\la$99.4\,km~s$^{-1}$, the line profile might be contaminated.

\section{Uncertainty estimate on CO(2-1) line profile predictions}\label{App:sigmas}

The radiative transfer analysis of the CO(2-1) emission is based on a number of input parameters. To assess the impact of the model predictions to uncertainties on the different derived or assumed input parameters, we varied the  input parameters independently and calculated the percentage impact on the line profile and the associated uncertainty on the mass-loss rate \Mdot$_{\rm{CO}}$ (see Table~\ref{Table:sigma}).
Uncertainties on the effective temperature (of $\pm$130\,K), luminosity and hence stellar radius \Rstar\ ($\pm$215\,\Rsun), and distance $D$ ($\pm$890\,pc) are taken from \citet{Davies2008ApJ...676.1016D}, on the fractional carbon abundance and hence on [CO/H] (of $\pm$1.4$\times$10$^{-5}$) from \citet{Davies2009ApJ...696.2014D}, on the terminal wind velocity (of $\pm$3\,km/s) and the outer radius (of $\pm$50\,\Rstar) from our ALMA data, and on the coefficients for the wind acceleration profile $\beta$ ($\pm$2) and for the temperature profile $\epsilon$ ($\pm$1) from the analysis presented in \citet{Decin2006A&A...456..549D} and \citet{DeBeck2010A&A...523A..18D}.  The canonical gas-to-dust ratio of 200 was taken representative for a Milky Way cluster \citep{Beasor2020MNRAS.492.5994B}, but is quite ill-constrained. We therefore let it range between 100\,--\,1000.	The largest uncertainty in \Mdot$_{\rm{CO}}$ arises from uncertainties in the terminal wind velocity ($\sim$35\%), and then in the distance, the outer CO envelope size, and the CO fractional abundance (each $\sim$20\%).  Uncertainties on all other parameters only have a minor effect. We note that the uncertainty estimates in Table~\ref{Table:sigma} are conservative values since only based on reproducing the peak flux, but not accounting for the full shape of the line profile which carries important diagnostics on, for example, the optical thickness of a specific line.
	
	%We note though that the uncertainty induced by the terminal wind velocity is a conservative estimate:  if the terminal velocity increases, not only is the peak flux decreasing but the width of the full line profile and the width between the two horns in spatially resolved, optically thin line profiles are increasing. It is the ensemble of those diagnostics 

To calculate the combined effect of the errors in the individual input parameters on \Mdot$_{\rm{CO}}$, one should  perform a Monte Carlo Markov Chain (MCMC) analysis. However, this is not feasible since non-LTE radiative tranfser calculations are too computer-intensive to be embedded in a MCMC algorithm. We therefore need to resort to a simpler approach similar to, for example, \citet{DiazLuis2019A&A...629A..94D} who have used an analytical prescription derived by \citet{Knapp1985ApJ...292..640K} to calculate the mass-loss rate given some individual parameters such as the expansion velocity, distance, integrated line intensity, and [CO/H] ratio \citep[see, for example, Eq.~3 in][]{DiazLuis2019A&A...629A..94D}. However, we here use our model grid and results presented in \citet{DeBeck2010A&A...523A..18D} since based on the same {\sc{GASTRoNOoM}} models, in particular we use Eq.~9 of \citet{DeBeck2010A&A...523A..18D}. The CO envelope size, $R_{\rm out}$, is not an individual input value of Eq.~9 of \citet{DeBeck2010A&A...523A..18D}, but is inherently captured in Eq.~9 since the [CO/H] ratio determines the photodissociation radius of the circumstellar envelope.
	Contrary to \citet{DiazLuis2019A&A...629A..94D}, we do not add the individual uncertainties in (unweighted) quadrature, but use a proper error propagation based on Eq.~9 of \citet{DeBeck2010A&A...523A..18D} which weighs the sensitivity of each individual uncertainty. 
	Hence, we derive that the total uncertainty on the estimated \Mdot$_{\rm{CO}}$-values is a factor of $\sim$1.4. In that, the uncertainty in terminal wind velocity contributes some 50\% of the total uncertainty factor and the distance $\sim$30\%.

	It might seem surprising that this total uncertainty on \Mdot$_{\rm{CO}}$ is similar to the highest uncertainty raised by an individual parameter (of $\sim$42\%) listed in Table~\ref{Table:sigma}. The reason thereof is that Table~\ref{Table:sigma} uses the sensitivity of the intensity peak to determine the associated change in \Mdot$_{\rm{CO}}$ by using a full non-LTE radiative transfer routine, while Eq.~9 of \citet{DeBeck2010A&A...523A..18D} is an analytic relation based on the integrated line intensity with various weightings (in terms of exponential factors) for the input parameters. Those different approaches hence rule out a one-to-one comparison of both uncertainty values.

% see /Users/leen/leda/ALMA/RSGs/programs/plot_lineprofiles.pro

\renewcommand{\arraystretch}{1.2}
\begin{table}[htp]
	\caption{Uncertainty on the model predictions.}
	\label{Table:sigma}
	\centering
	\begin{tabular}{c|c|c|c}
		\hline
		& & \\[-2ex]
		Parameter & 1$\sigma$ uncertainty parameter & Uncertainty peak intensity & Uncertainty \Mdot$_{\rm{CO}}$\\
		& & \\[-2ex]
		\hline
		\multirow{2}{*}{\centering $D$\,=\,6600\,pc} 	& $+$890\,pc & $-23$\% & $+$19\%\\
		 &$-$890\,pc & $+$32\% & $-$19\%\\
		\hline
		& & \\[-2ex]
		\multirow{1}{*}{\centering $T_{\rm{eff}}$\,=\,3450\,K} 	&  $\pm$130\,K & $\mp$4\% & $\pm$1\%\\
		\hline		
		\multirow{2}{*}{\centering $R_\star$\,=\,1450\,\Rsun} 	& +215\,\Rsun & +10\% & $-$8\%\\
		& $-$215\,\Rsun & $-$15\% & $+$15\%\\
		\hline		
		\multirow{2}{*}{\centering $v_\infty$\,=\,11\,km/s} 	& +3\,km/s &  $-$39\% & +35\%\\
		& $-$3\,km/s & $+$60\% & $-$34\%\\
		\hline
		\multirow{1}{*}{\centering $\beta$\,=\,3.0} 	& $\pm$2 & $\pm$2\% & $\mp$4\%\\
		\hline		
		\multirow{1}{*}{\centering $\epsilon$\,=\,0.6} 	& $\pm$0.1 & $\mp$12\% &  $\pm$4\%\\
		\hline		
		\multirow{2}{*}{\centering $R_{\rm out}$\,=\,400\,\Rstar} 	& +50\,\Rstar & $+$21\% & $-$12\%\\
		& $-$50\,\Rstar & $-$21\% & $+$23\%\\
		\hline		
		\multirow{2}{*}{\centering [CO/H]\,=\,8.9$\times$10$^{-5}$} 	& +1.4$\times$10$^{-5}$ & $+$20\% & $-$12\%\\
		& $-$1.4$\times$10$^{-5}$ & $-$22\% & $+$23\%\\
		\hline		
		\multirow{2}{*}{\centering $r_{\rm{gd}}$\,=\,200} 	& $r_{\rm{gd}}$\,=\,100 & $-$5\% & $+$4\%\\
		& $r_{\rm{gd}}$\,=\,1\,000 &$+$4\% & $-$4\%\\
		\hline		
	\end{tabular}
\tablefoot{First column lists the input parameter, second column 	the 1$\sigma$ uncertainty of the input parameter (unless for the gas-to-dust ratio which lists its range in potential values), third column the associated change in peak intensity, and last column the associated uncertainty on the mass-loss rate \Mdot$_{\rm{CO}}$. Non-symmetric changes in  the associated peak strength and \Mdot$_{\rm{CO}}$ are listed separately.}
\end{table}

\renewcommand{\arraystretch}{1}

\section{\Mdot-luminosity relations from \citet{Beasor2020MNRAS.492.5994B}}\label{App:Beasor}

\begin{table*}[htp]
	\caption{Values for bolometric luminosity and \Mdot$_{\rm{SED}}$ for $\chi$~Per, NGC\,7419, NGC\,2100 and RSGC1.}
	\label{Table:all_data_Beasor}
	\begin{tabular}{lcc|lcc}
		\hline
		\multicolumn{3}{c|}{}	& \multicolumn{3}{|c}{} \\[-2ex]	
	    \multicolumn{3}{c|}{$\chi$ Per} & \multicolumn{3}{|c}{NGC\,7419} \\
		\hline
		\xrowht[()]{8pt}Star & $\log$(L$_{\rm{bol}}$/L$_\odot$) & \Mdot$_{\rm{SED}}$ (10$^{-6}$\,M$_\odot$ yr$^{-1}$) & 	Star & $\log$(L$_{\rm{bol}}$/L$_\odot$) & \Mdot$_{\rm{SED}}$ (10$^{-6}$\,M$_\odot$ yr$^{-1}$) \\
		\hline
		\xrowht[()]{8pt}FZ Per & 4.64$^{+0.06}_{-0.05}$ & 0.30$^{+0.18}_{-0.07}$ & MY CEP & 5.19$\pm$0.07 & 18.04$^{+7.15}_{-8.54}$ \\
		RS Per & 4.92$^{+0.18}_{-0.07}$ & 3.03$^{+2.31}_{-0.94}$ & BMD 139 & 4.55$\pm$0.08 & 0.27$^{+0.44}_{-0.05}$ \\
		AD Per & 4.80$^{+0.08}_{-0.05}$ & 0.97$^{+0.33}_{-0.50}$ &  BMD 696 & 4.63$\pm$0.08 & 0.22$^{+0.17}_{-0.04}$ \\
		V439 Per & 4.53$^{+0.06}_{-0.05}$ & 0.10$^{+0.10}_{-0.01}$ & BMD 435 & 4.54$\pm$0.11 & 0.18$^{+0.15}_{-0.04}$ \\
		V403 Per & 4.41$^{+0.06}_{-0.05}$ & 0.06$^{+0.02}_{-0.02}$ &  & & \\
		V441 Per & 4.75$^{+0.10}_{-0.06}$ & 0.93$^{+0.72}_{-0.31}$ &  & & \\
		SU Per & 4.99$^{+0.09}_{-0.05}$ & 1.62$^{+0.72}_{-0.63}$ &  & & \\
		BU Per & 4.67$^{+0.07}_{-0.05}$ & 3.24$^{+1.53}_{-1.28}$ &  & & \\
		\hline
		\hline
	%	\multicolumn{6}{c} \\
		\multicolumn{3}{c|}{}	& \multicolumn{3}{|c}{} \\[-2ex]	
		\multicolumn{3}{c|}{RSGC1} & \multicolumn{3}{|c|}{NGC\,2100} \\
		\hline
		\xrowht[()]{8pt}Star & $\log$(L$_{\rm{bol}}$/L$_\odot$) & \Mdot$_{\rm{SED}}$ (10$^{-6}$\,M$_\odot$ yr$^{-1}$) & 	Star & $\log$(L$_{\rm{bol}}$/L$_\odot$) & \Mdot$_{\rm{SED}}$ (10$^{-6}$\,M$_\odot$ yr$^{-1}$) \\
		\hline
     	\xrowht[()]{8pt}F01 & 5.58$\pm$0.18 & 5.57$^{+2.37}_{-2.17}$ &  1 & 5.09$\pm$0.09& 9.89$^{+4.20}_{-3.17}$ \\
		F02 & 5.56$\pm$0.18 & 5.18$^{+2.72}_{-1.75}$ &  2 & 4.97$\pm$0.09& 9.97$^{+4.52}_{-3.19}$ \\
		F03 & 5.33$\pm$0.18 & 4.18$^{+3.08}_{-0.84}$ &  3 & 4.84$\pm$0.09& 1.98$^{+1.07}_{-0.74}$ \\
		F06 & 5.32$\pm$0.18 & 0.68$^{+0.69}_{-0.14}$ &  4 & 4.71$\pm$0.09& 3.17$^{+2.34}_{-1.29}$ \\
		F07 & 5.31$\pm$0.18 & 0.28$^{+0.29}_{-0.06}$ &  5 & 4.73$\pm$0.09& 2.54$^{+1.49}_{-0.86}$ \\
		F09 & 5.30$\pm$0.18 & 0.52$^{+0.53}_{-0.10}$ &  6 & 4.77$\pm$0.09& 3.25$^{+2.12}_{-0.72}$ \\
		F10 & 5.28$\pm$0.18 & 0.87$^{+0.89}_{-0.17}$ &  7 & 4.68$\pm$0.09& 0.82$^{+0.27}_{-0.46}$ \\
		F12 & 5.22$\pm$0.19 & 0.18$^{+0.04}_{-0.04}$ &  8 & 4.68$\pm$0.09& 1.29$^{+0.76}_{-0.51}$ \\
		 & & & 9 & 4.68$\pm$0.09& 0.48$^{+0.26}_{-0.18}$ \\
		 & & & 10 & 4.63$\pm$0.09& 1.06$^{+0.80}_{-0.36}$ \\
		 & & & 11 & 4.55$\pm$0.09& 0.28$^{+0.14}_{-0.16}$ \\
		 & & & 12 & 4.51$\pm$0.09& 0.39$^{+0.21}_{-0.21}$ \\
		 & & & 16 & 4.55$\pm$0.09& 2.27$^{+1.14}_{-0.57}$ \\
		 & & & 17 & 4.49$\pm$0.09& 0.23$^{+0.15}_{-0.15}$ \\
		 \hline
	\end{tabular}
\tablefoot{Values used for Fig.~\ref{Fig:Beasor_new} and Fig.~\ref{Fig:Beasor}, and reproduced from Table~3 of \citet{Beasor2016MNRAS.463.1269B}, Table~2 of \citet{Beasor2018MNRAS.475...55B}, and 
	Table~2 of \citet{Beasor2020MNRAS.492.5994B}.}
\end{table*}

In a first step, \citet{Beasor2020MNRAS.492.5994B} has determined a \Mdot-luminosity relation for all clusters in their sample by fitting the relation given in Eq.~(\ref{Eq:relation_ab}) to their data points taken from \citet{Beasor2016MNRAS.463.1269B}, \citet{Beasor2018MNRAS.475...55B} and \citet{Beasor2020MNRAS.492.5994B} (see Table~\ref{Table:all_data_Beasor}). They therefore have used a linear least-square approximation that accounts for both errors on the ordinate and coordinate values. However, plotting the \Mdot-luminosity relation using the values for the intercept and slope  as listed in Table~4 of \citet{Beasor2020MNRAS.492.5994B} (repeated here in Table~\ref{Tab:res_SED_Beasor}) reveals a problem with those values (see left panel in Fig.~\ref{Fig:Beasor}): (1)~A typo occurred in Table~4 of \citet{Beasor2020MNRAS.492.5994B}: the intercept $a$ for RSGC1 should not be $-52$, but $-53$ (Beasor, priv.\ comm.), which indeed yields a much better fit to the data (see dotted line in left panel of Fig.~\ref{Fig:Beasor}); and (2) comparing Fig.~\ref{Fig:Beasor_new} and Fig.~\ref{Fig:Beasor} with Fig.~2 in \citet{Beasor2020MNRAS.492.5994B} shows a problem with the data points plotted  in \citet{Beasor2020MNRAS.492.5994B}, for example the maximum  \Mdot$_{\rm{SED}}$ value for NGC\,7419 is 1.8$\times$10$^{-5}$\,\Msun/yr (see Table~\ref{Table:all_data_Beasor}) but is plotted around 7$\times$10$^{-5}$\,\Msun/yr in Fig.~2 of \citet{Beasor2020MNRAS.492.5994B}; similar problems occur for other data points. 
For that reason we continue with the data listed in Table~\ref{Table:all_data_Beasor}; a corrigendum is currently being prepared by E.~Beasor et al.

\begin{table*}[htp]
	\caption{Best-fit parameters for the \Mdot$_{\rm{SED}}$-luminosity relation for each cluster derived by fitting Eq.~(\ref{Eq:relation_ab}) to the data.}
	\label{Tab:res_SED_Beasor}
	\centering
	\begin{tabular}{l|cc|c}
		\hline
		\xrowht[()]{8pt}Cluster & Intercept ($a$) & Slope ($b$) & Intercept ($a$) \\
		& & & for fixed $b$\,=\,4.8$\pm$0.6 \\
		\hline
		& & &  \\[-2ex]
		RSGC1 & $-$52.0$\pm$51.2 & 8.8$\pm$9.5 & $-$31.90$\pm$1.16 \\
		NGC\,2100 & $-$30.9$\pm$3.6 & 5.3$\pm$0.8 & $-$28.59$\pm$0.75\\
		NGC\,7419 & $-$22.9$\pm$4.9 & 3.6$\pm$1.7 & $-$29.01$\pm$1.41\\
		$\chi$~Per & $-$27.06$\pm$4.90 & 4.5$\pm$1.0 &$-$28.97$\pm$0.99\\
		\hline 
	\end{tabular}
\tablefoot{The second and third column  list, respectively, the intercept $a$ and slope $b$ with their standard deviation as derived by \citet{Beasor2020MNRAS.492.5994B}. The fourth column lists the intercept $a$ and its standard deviation for a slope fixed to $b$\,=\,4.8$\pm$0.6.}
\end{table*}

%/Users/leen/leda/ALMA/RSGs/programs/plot_correlations_with_alma_rsgs_paper.pro
\begin{figure*}[htp]
	\includegraphics[angle=0,width=0.95\textwidth]{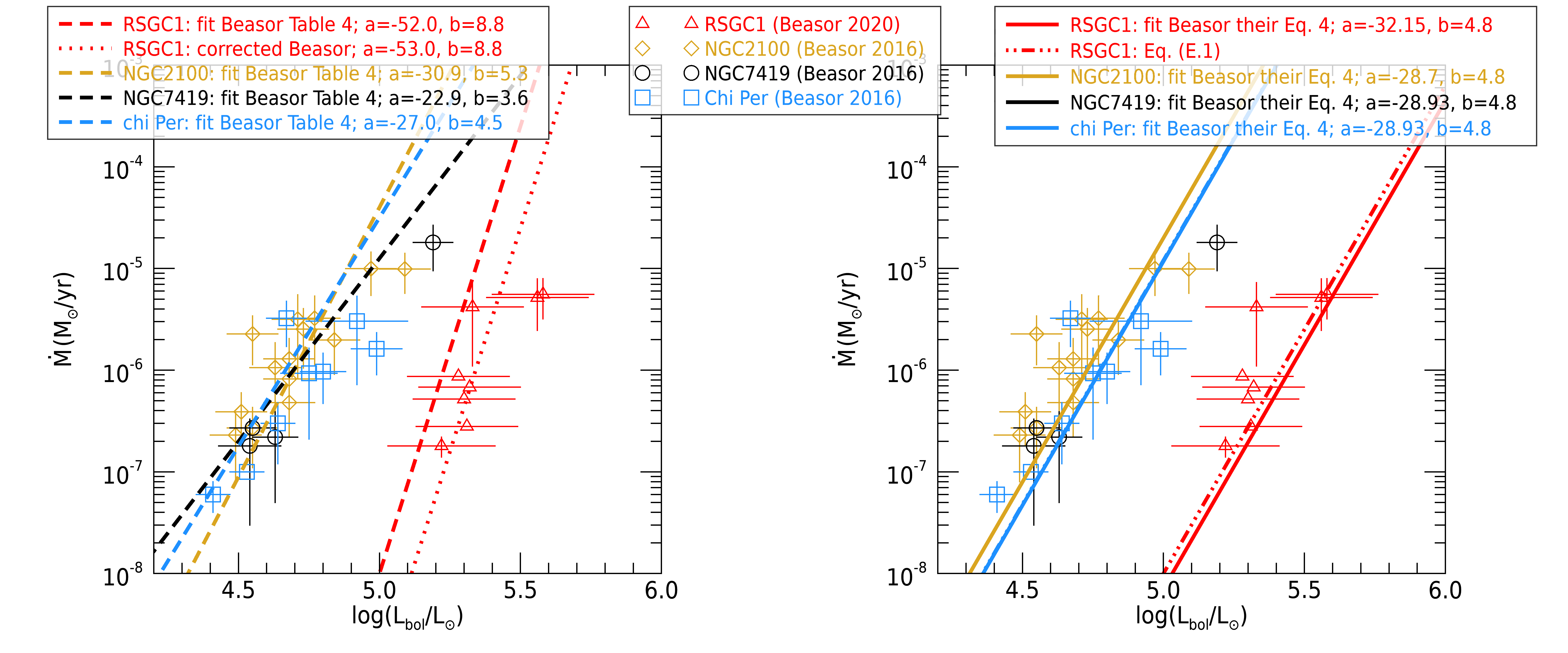}
	\caption{\Mdot-luminosity relations as determined by \citet{Beasor2020MNRAS.492.5994B}. The coloured symbols represent the (L$_{\rm{bol}}$, \Mdot$_{\rm{SED}}$)-values for four open clusters as derived by \citet{Beasor2016MNRAS.463.1269B, Beasor2020MNRAS.492.5994B}; error bars indicate their most conservative error estimates.
		The straight lines in both panels show the individual straight-line fits to each relation $\log(\Mdot_{\rm{SED}}/\Msun {\rm{yr^{-1}}}) = a + b \log(L_{\rm{bol}}/\Lsun)$ for all clusters in the sample. In the left panel, the dashed lines show the \Mdot-$L_{\rm{bol}}$ relation using the values for the offset and slope as given in Table~4 of \citet{Beasor2020MNRAS.492.5994B}; the full lines in the right panel show the fits to the \Mdot-$L_{\rm{bol}}$ relation once the gradient is fixed to $b$\,=\,4.8, following Eq.~4 of \citet{Beasor2020MNRAS.492.5994B} and using the values for the initial mass as given in Table~1 of \citet{Beasor2020MNRAS.492.5994B}.
		The red dotted line in the left panel shows the fit to RSGC1 for an intercept $a$\,=\,-53; see footnote~7.
			The red dashed-triple dotted lin in the right panel shows the fit to RSGC1 following Eq.~\eqref{Eq:Mdot_Beasor_repeated}.	
	}
	\label{Fig:Beasor}
\end{figure*}

In a next step, \citet{Beasor2020MNRAS.492.5994B} chose to fix the gradient in order to reduce the number of degrees of freedom in the fit.  Accounting for the standard deviation on the derived $b$-values, one gets  $b$=4.8$\pm$0.6. The latter value was then used for deriving the intercept values in their more general mass-dependent \Mdot-luminosity relation given in Eq.~(\ref{Eq:Beasor}) \citep[which is Eq.~(4) in][]{Beasor2020MNRAS.492.5994B}. But for RGSC1, the \Mdot-luminosity relation from their Eq.~(4) seems at odd with the empirical (L$_{\rm{bol}}$, \Mdot$_{\rm{SED}}$) values (see right panel in Fig.~\ref{Fig:Beasor}), with the straight-line fit being systematically lower than the empirical values. We therefore have repeated their analysis for a slope of $b$\,=4.8$\pm$0.6. Fixing the slope, only the intercept has to be determined; the derived values with their standard deviation are listed in the fourth column of Table~\ref{Tab:res_SED_Beasor}. From the analysis of the four clusters, we can determine the mass-dependent intercept, yielding the following mass-dependent  \Mdot$_{\rm{SED}}$-luminosity relation
	\begin{eqnarray}
		\log(\Mdot_{\rm{SED}}/\Msun\  {\rm{yr^{-1}}})  & = & (-26.5 - 0.22 \times M_{\rm{ini}}/\Msun)  + 4.8 \log(L_{\rm{bol}}/\Lsun)
		\label{Eq:Mdot_Beasor_repeated}
	\end{eqnarray}
	with standard deviations on the best-fit parameters being  $-$26.50$\pm$1.33,  $-$0.22$\pm$0.10, and  4.8$\pm$0.6; which is similar to Eq.~\eqref{Eq:Beasor} within some round-off errors. Eq.~\eqref{Eq:Mdot_Beasor_repeated} implies a slight shift upwards for RSGC1 as compared to Eq.~\eqref{Eq:Beasor} (see red dashed-triple dotted line in the right panel of Fig.~\ref{Fig:Beasor}), but that shift is not visible on the scale of the plot for the 3 other clusters. 

\afterpage{\FloatBarrier}
%\clearpage

%\afterpage{\clearpage}
%\newpage

\section{Sensitivity to intercept $a_i$}\label{App:NGC2100}

Fig.~\ref{Fig:NGC2100} shows the fit to Eq.~\eqref{Eq:Mdot_for_MC}, but instead of using RSGC1 for determining $a_1$, we use NGC\,2100 for determining the intercept $a_2$ and  the difference in intercept of the other 3 clusters, $\Delta a_i$ ($i$\,=\,1,3,4). We fit both the  ($L_{\rm{bol}}$, \Mdot$_{\rm{SED}}$) and ($L_{\rm{bol}}$, \Mdot$_{\rm{CO}}$) together; corresponding values are listed in Table~\ref{Tab:res_SED_NGC2100}.

%  email from Hugues Sana on 15/03/2023
\begin{figure}[htp]
	\centering
	\includegraphics[angle=0,width=.5\textwidth]{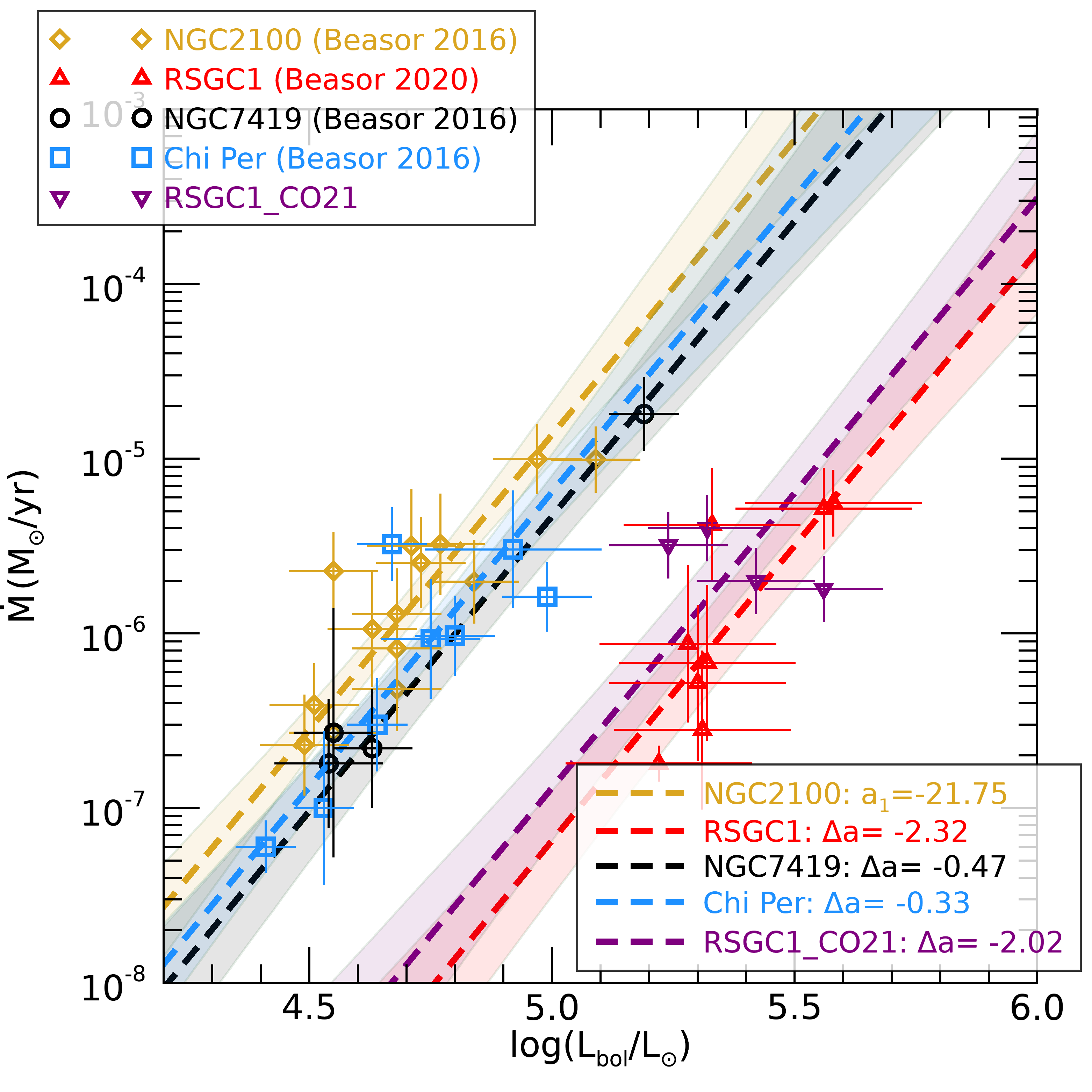}
	\caption{\Mdot-luminosity relations for the four open clusters RSGC1, NGC\,2100, NGC\,7410, and $\chi$ Per. Dashed gold, red, black, and blue lines present the best-fit of Eq.~\eqref{Eq:Mdot_for_MC} to the ($L_{\rm{bol}}$, \Mdot$_{\rm{SED}}$) measurements of the four clusters (shown as open symbols with corresponding colour), and the dashed purple line the best-fit to the ($L_{\rm{bol}}$, \Mdot$_{\rm{CO}}$) values of four stars in RSGC1 (filled purple downward triangles).}
	\label{Fig:NGC2100}
\end{figure}

\begin{table}[htp]
	\caption{($L_{\rm{bol}}$, \Mdot)-relation for each cluster derived when fitting Eq.~\ref{Eq:Mdot_for_MC} to the ($L_{\rm{bol}}$, \Mdot$_{\rm{SED}}$) measurements of the four clusters and the ($L_{\rm{bol}}$, \Mdot$_{\rm{CO}}$) values for RSGC1. The MC method yields  a best-fit slope $b$\,=\,3.38$^{+0.52}_{-0.44}$ and intercept  $a_2$ for NGC\,2100 of $-$21.75$^{+2.05}_{-2.46}$.}
	\label{Tab:res_SED_NGC2100}
	\centering
	\begin{tabular}{l|c}
		\hline
		\xrowht[()]{8pt}Cluster & difference in intercept ($\Delta a_i$)\\
	    & Eq.~\eqref{Eq:Mdot_for_MC} \\
		\hline
		&  \\[-2ex]
		RSGC1 &    $-$2.32$^{+0.38}_{-0.44}$ \\
		NGC\,2100 &   0 \\
		NGC\,7419 &   $-$0.47$^{+0.24}_{-0.25}$\\
		$\chi$~Per & $-$0.33$^{+0.16}_{-0.16}$ \\
		\hline 
		&  \\[-2ex]
		RSGC1 (CO) & $-$2.02$^{+0.38}_{-0.43}$ \\
		\hline
	\end{tabular}
\end{table}

 \section{Other Galactic red supergiants}\label{App:other_rsgs}
 
 In this section, we derive \Mdot$_{\rm{CO}}$ for some well-known Galactic M-type red supergiants following the same modelling strategy as in Sect.~\ref{Sec:CO_analysis}. The aim is to compare the retrieved \Mdot$_{\rm{CO}}$ with the values predicted using Eq.~(\ref{Eq:Mdot_new}). We therefore perform a radiative transfer analysis of the CO rotational line transitions of the red supergiants published by \citet{DeBeck2010A&A...523A..18D}. In particular, we analyse the James Clerk Maxwell Telescope (JCMT) data of $\alpha$~Ori, $\mu$~Cep, and VX~Sgr. The M2 red supergiant VY CMa is excluded from the current analysis, since the CO rotational line data have already been analysed by \citet{Decin2006A&A...456..549D} and we will resort to their outcomes.  We also exclude the hypergiant NML~Cyg, since its estimated initial mass of $\sim$40\,\Msun\ \citep{Schuster2009ApJ...699.1423S} puts its above the mass range of 8$\la$$M_\star$$\la$30\,\Msun\ of interest to this study. %We also note that the CO rotational lines of NML~Cyg are contaminated with interstellar CO \citep{DeBeck2010A&A...523A..18D}. 
 The JCMT data of $\alpha$~Ori, $\mu$~Cep, and VX~Sgr are plotted in Fig.~\ref{Fig:other_RSGs}. The intensity is plotted as main beam temperature $T_{\rm{mb}}$, in units of K, which  can be converted to flux density $F_\nu$ or brightness $I_\nu$ by following formula
\begin{equation}
	T_{\rm{mb}} \,=\, \frac{\lambda^2 \, F_\nu}{2\,k\,\Omega} \,=\,1.36  \left(\frac{\lambda_{\rm{cm}}}{\theta\arcsec}\right)^2 I_\nu,
\end{equation} 
with $\lambda$ the wavelength, $k$ the Boltzmann constant, $\Omega$ the beam solid angle,  $\lambda_{\rm{cm}}$ the wavelength in units of cm, $\theta$ the beam size in units of arcsec, and $I_\nu$ the brightness in units of mJy/beam. The beam size of the JCMT CO(2-1), CO(3-2), and CO(4-3) data are 19\farcs7, 13\farcs2, and 10\farcs8, respectively.

 Similar to the analysis of the five red supergiants in RSGC1 (Sect.~\ref{Sec:CO_analysis}), we assume that the main wind component can be represented by a spherically symmetric envelope. As has been discussed in detail in the literature, high spatial and spectral resolution observations of those Galactic RSGs prove that those winds are composed of a smooth outflow and individual mass ejections in the form of gas and dust clumps \citep{Decin2006A&A...456..549D, Montarges2019MNRAS.485.2417M, Montarges2021Natur.594..365M, Humphreys2022AJ....163..103H}. The smooth outflow contains by far most of the mass and its mass-loss rate is a factor of a few up to almost 2 orders of magnitude higher than that traced by the individual clumps \citep{ Montarges2019MNRAS.485.2417M, Montarges2021Natur.594..365M, Humphreys2022AJ....163..103H}. With the aim of (i)~deriving the overall (average) mass-loss rate \Mdot$_{\rm{CO}}$, and (ii)~comparing \Mdot$_{\rm{CO}}$ with the predictions from Eq.~(\ref{Eq:Mdot_new}), we only focus on modelling the dominant outflow. This implies that spikes, visible on top of the JCMT CO line profiles and testifying the presence of individual gas clumps, will not be modelled.
  
 To retrieve \Mdot$_{\rm{CO}}$, we follow the same modelling strategy as outlined in Sect.~\ref{Sec:CO_analysis}. The input stellar and CSE parameters are listed in Table~\ref{Tab:outcome_other_RSGs}. For $\alpha$~Ori and $\mu$~Cep, the CO outer envelope size can be estimated from radio interferometric observations, while for VX~Sgr we use the photodissociation radius following the results of \citet{Groenewegen2021A&A...649A.172G}. For each of the sources, the individual modelling could be optimised by fine-tuning some of the input parameters described in Sect.~\ref{Sec:CO_analysis}. However, we refrain from doing so since we aim to assess the predictive power of Eq.~(\ref{Eq:Mdot_new}) in relation to our overall modelling strategy of retrieving \Mdot$_{\rm{CO}}$. Given the complexity of the line profiles and the absolute flux calibration accuracy of $\sim$30\%  \citep{Kemper2003A&A...407..609K}, the fits shown in Fig.~\ref{Fig:other_RSGs} prove that for all sources the main CO component is well reproduced. In contrast to the analysis of the RSGC1 RSGs, the main uncertainty on the retrieved \Mdot$_{\rm{CO}}$ values arises from uncertainties on the distance; translating into an \Mdot$_{\rm{CO}}$ uncertainty of a factor of $\sim$2\,--\,3.

 The gas mass-loss rates retrieved from this analysis, \Mdot$_{\rm{CO}}$, are listed in  Table~\ref{Tab:outcome_other_RSGs} and can  be compared to the  \Mdot-values predicted using Eq.~(\ref{Eq:Mdot_new}) and using Eq.~(\ref{Eq:Beasor}), the latter once following the results of \citet{Beasor2020MNRAS.492.5994B}. With the exception of the cool extreme supergiant VY~CMa, Eq.~(\ref{Eq:Mdot_new}) predicts the gas mass-loss rate for the other 3 red supergiants well --- the average difference only being $\sim$30\%. This result supports the use of  Eq.~(\ref{Eq:Mdot_new}) for the prediction of mass-loss rates for M-type red supergiants with effective temperature $\ga$3200\,K.
 Eq.~(\ref{Eq:Beasor}) predicts the gas mass-loss rates well for  $\mu$~Cep and VX~Sgr, but fails for $\alpha$~Ori and VY~CMa.

 \begin{table*}[htpb]
 	\caption{Stellar and CSE parameters for the red supergiants observed in various CO rotational lines by \citet{DeBeck2010A&A...523A..18D}. }
 	\label{Tab:outcome_other_RSGs}
 	\vspace*{-1.5ex}
 	\setlength{\tabcolsep}{.2mm}
 	\begin{tabular}{l|cccccc|cccccc}
 		\hline
 		\multicolumn{1}{c}{}	& \multicolumn{6}{|c|}{} & \multicolumn{6}{c}{} \\[-2ex]	
 		\multicolumn{1}{c}{}	& \multicolumn{6}{|c|}{Stellar parameters} & \multicolumn{6}{c}{CSE parameters} \\
 		\hline
 		\multicolumn{1}{c|}{}	& \multicolumn{6}{|c|}{} & \multicolumn{6}{c}{} \\[-2ex]	
 		%\xrowht[()]{8pt}
 		\multicolumn{1}{l|}{Star} & $\log(L_{\rm{bol}}/L_\odot)$ & $T_{\rm{eff}}$ & Spectral  & \Rstar & $M_{\rm{ini}}$ & $D$ & $v_{\rm{LSR}}$$^{(c)}$ &  $v_\infty$$^{(c)}$ & $R_{\rm{out}}$$^{(c)}$ & \Mdot$_{\rm{CO}}$$^{(c)}$ & \Mdot$^{(c)}$ & \Mdot$^{\rm{Beasor}}$ \\
 		& & & type & & & & & & & & Eq.~(\ref{Eq:Mdot_new}) & Eq.~(\ref{Eq:Beasor})\\
 		& & [K] &  & [R$_\odot$] &  [$M_\odot$] & [pc] & [km~s$^{-1}$]  & [km~s$^{-1}$] & [\Rstar] & [10$^{-6}$\,M$_\odot$/yr] & [10$^{-6}$\,M$_\odot$/yr] & [10$^{-6}$\,M$_\odot$/yr] \\
 		\hline
 		& & & & & & & & & & & \\[-2ex]
 		$\alpha$ Ori & 4.94$^{(a)}$ & 3600$^{(a)}$ &  M2Iab$^{(b)}$ & 764$^{(a)}$ & 19.5$^{(a)}$ & 168$^{(a)}$  & 3.4 & 13 & 800$^{(j)}$ & 0.4 & 0.6 & 0.07\\
 		$\mu$ Cep & 5.13$^{(d)}$  & 3551$^{(d)}$  & M2Ia$^{(b)}$ & 974$^{(d)}$ & 17.5$^{(d)}$  & 790$^{(i)}$ & 31 & 21 & 600$^{(k)}$ & 3.0 & 6 & 1.6\\
 		VX~Sgr & 5.01$^{(b)}$ & 3535$^{(b)}$  & M4eIa$^{(b)}$  & 855$^{(b)}$ &11$^{(e)}$ & 1570$^{(b)}$ & 6.5 & 22 & 3400$^{(l)}$ & 25 & 28 & 13\\ 		
 		\hdashline
 		& & & & & & & & & & &  & \\[-2ex]
 		VY~CMa& 5.48$^{(g)}$ & 2800$^{(g)}$ & M2/4II$^{(b)}$ & 1680$^{(f)}$ &25$^{(h)}$& 1200$^{(h)}$ & 22.6$^{(g)}$ & 35$^{(g)}$ & 950$^{(g)}$ & 80$^{(g)}$ & 5 &  1.4 \\
 		\hline
 	\end{tabular}
 	\tablefoot{Listed are the stellar luminosity $L_\star$, the effective temperature $T_{\rm{eff}}$, the spectral type, the stellar radius \Rstar, the initial mass $M_{\rm{ini}}$, the distance $D$, the local standard of rest velocity $v_{\rm{LSR}}$,  the terminal wind velocity $v_\infty$, the radius of the CO envelope $R_{\rm{out}}$,  the gas mass-loss rate \Mdot$_{\rm{CO}}$,  the mass-loss rate as predicted from Eq.~(\ref{Eq:Mdot_new}), and the mass-loss rate predicted using the luminosity-\Mdot-relation of \citet{Beasor2020MNRAS.492.5994B} (Eq.~(\ref{Eq:Beasor})).\\
 		\tablefoottext{a}{From \citet{Joyce2020ApJ...902...63J}. Initial mass estimate ranges between 18--21\,\Msun.}\\
 		\tablefoottext{b}{From \citet{DeBeck2010A&A...523A..18D}.}\\
 		\tablefoottext{c}{Current work.}\\
 		\tablefoottext{d}{From \citet{Montarges2019MNRAS.485.2417M}. Initial mass estimate ranges between 15-20\,\Msun.}\\
 		\tablefoottext{e}{From \citet{Tabernero2021A&A...646A..98T}. Initial mass estimate ranges between 10-12\,\Msun.} \\
 		\tablefoottext{f}{From \citet{Zhang2012A&A...544A..42Z}.}\\
 		\tablefoottext{g}{From \citet{Decin2006A&A...456..549D}; current mass-loss rate. For fitting the complex CO line profiles, \citet{Decin2006A&A...456..549D} invoked different shells with varying mass-loss rate.}\\
 		\tablefoottext{h}{From \citet{Wittkowski2012A&A...540L..12W}.  Initial mass estimate ranges between 15-35\,\Msun. }\\
 		\tablefoottext{i}{Estimates of $\mu$ Cep's distance vary between 390$\pm$140 and 1818$\pm$661\,pc \citep[][and references therein]{Montarges2019MNRAS.485.2417M}. We here take the average value of the distance estimate of \citet{Montarges2021Natur.594..365M} (based on physical considerations on the relative size of the MOLsphere, $D$\,=\,$641^{+148}_{-144}$\,pc) and of \citet{Davies2020MNRAS.493..468D} (based on the average parallax of neighbouring OB stars, under the assumption that the RSG is part of the same association; $D$\,=\,$940^{+140}_{-40}$\,pc). I.e., $D$\,=\,790\,pc. For a change of distance from 790\,pc to 390\,pc, the retrieved \Mdot$_{\rm{CO}}$ changes from 3$\times$10$^{-6}$ to 1$\times$10$^{-6}$\,\Msun/yr.}\\
 		\tablefoottext{j}{The wind of Betelgeuse has various components.	\citet{OGorman2012AJ....144...36O} proposes that the S2 flow of $\alpha$ Ori extends out to a radius of 17\arcsec\ (or 800\,\Rstar), although the measured intensity distribution of CO emission as a function of projected radius extends to $\sim$8.5\arcsec\ ($\sim$400\,\Rstar). Changing the radius of the CO envelope from 800\,\Rstar\ to 400\,\Rstar\  changes the retrieved \Mdot$_{\rm{CO}}$ from 0.4$\times$10$^{-6}$\,\Msun/yr to 0.5$\times$10$^{-6}$\,\Msun/yr.}\\
 		\tablefoottext{k}{\citet{Montarges2019MNRAS.485.2417M} have used the NOEMA interferometer to get a channel map of the  CO(2-1) emission of $\mu$~Cep  at a spatial resolution of 0\farcs92$\times$0\farcs72, with maximum recoverable scale (MRS) being 8\arcsec. Emission is detected up to $\sim$3\farcs5 from the central star (or $\sim$600\,\Rstar).} 		\\
 		\tablefoottext{l}{For a mass-loss rate around 2$\times$10$^{-5}$\,\Msun/yr, the predicted CO photodissociation radius is $\sim$3500\,\Rstar\ \citep{Groenewegen2021A&A...649A.172G} or a full extent of 17\arcsec. The only CO interferometric data currently available for VX~Sgr have been obtained in the framework of the ALMA {\sc atomium} large program. However, the MRS of those data is only $\sim$8--10\arcsec\ and hence cannot be used to estimate the CO envelope size. For that reason, we resort to the predictions of \citet{Groenewegen2021A&A...649A.172G}.}
 	}
 \end{table*}

% /Users/leen/leda/ALMA/RSGs/programs/plot_other_rsgs.pro
% turn ps figure, crop and save as pdf
\begin{figure*}[htp]
 	\begin{minipage}[t]{\textwidth}
 	\centering
 	\includegraphics[width=.75\textwidth]{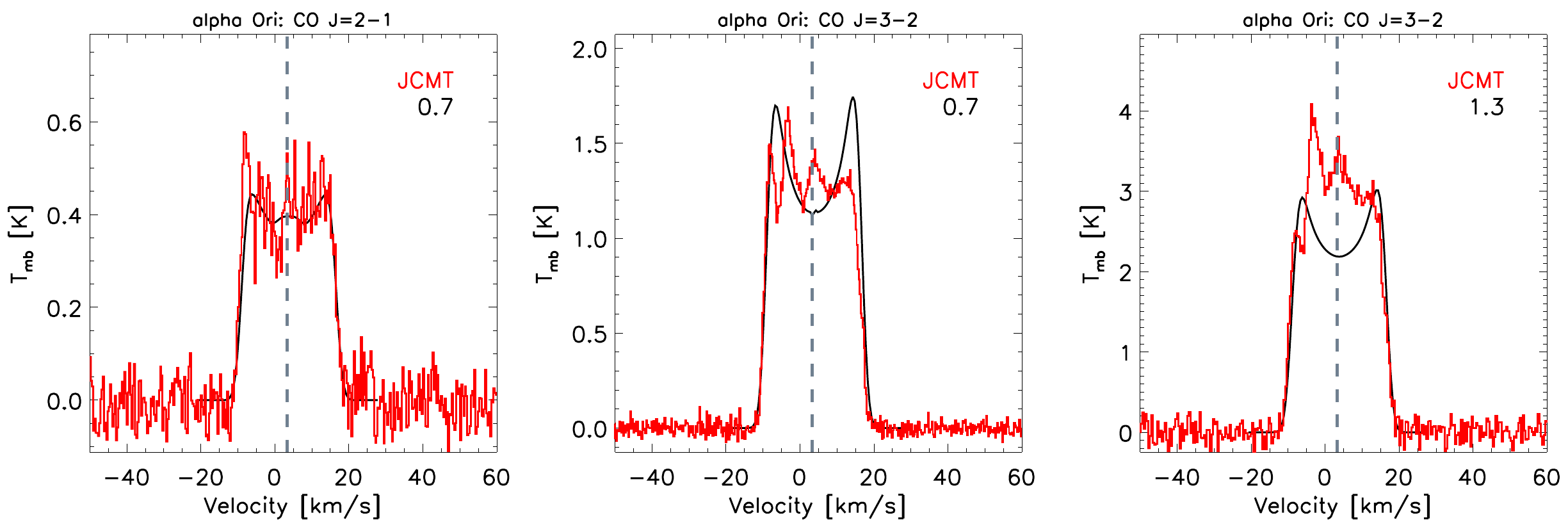}\\
    \end{minipage}\\
 	\begin{minipage}[t]{\textwidth}
 	\centering \hspace*{-3ex}
 		 \includegraphics[width=.77\textwidth]{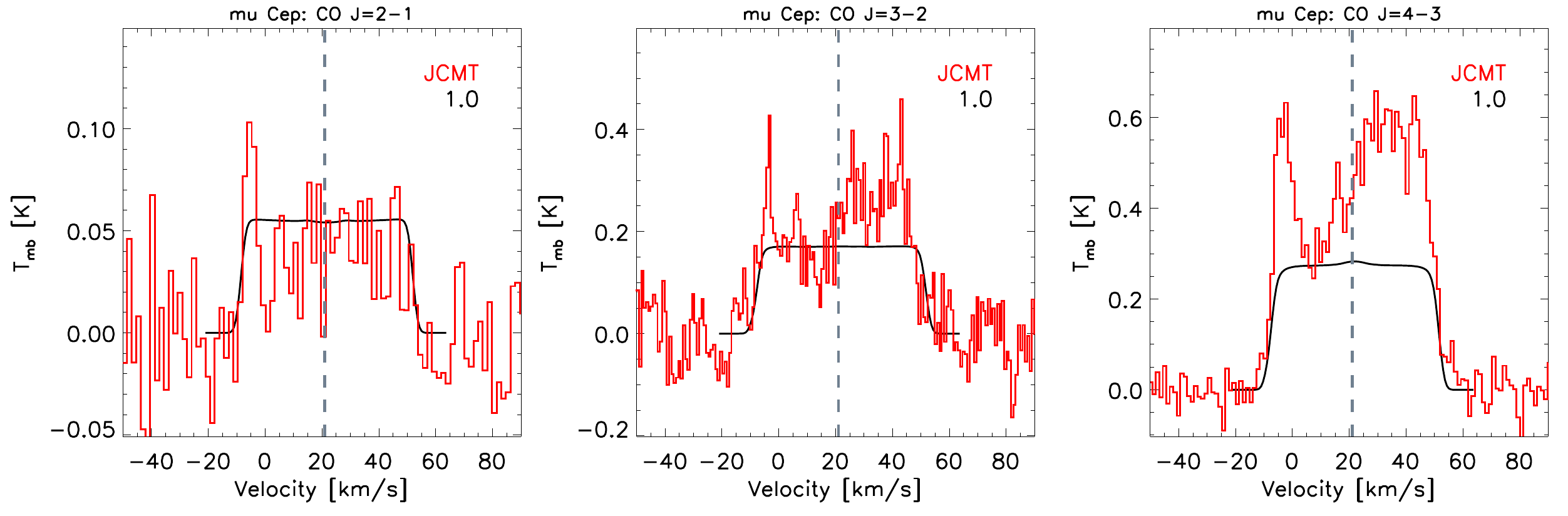}\\
  \end{minipage}\\
  \begin{minipage}[t]{\textwidth}
 	\centering
 	\includegraphics[width=.75\textwidth]{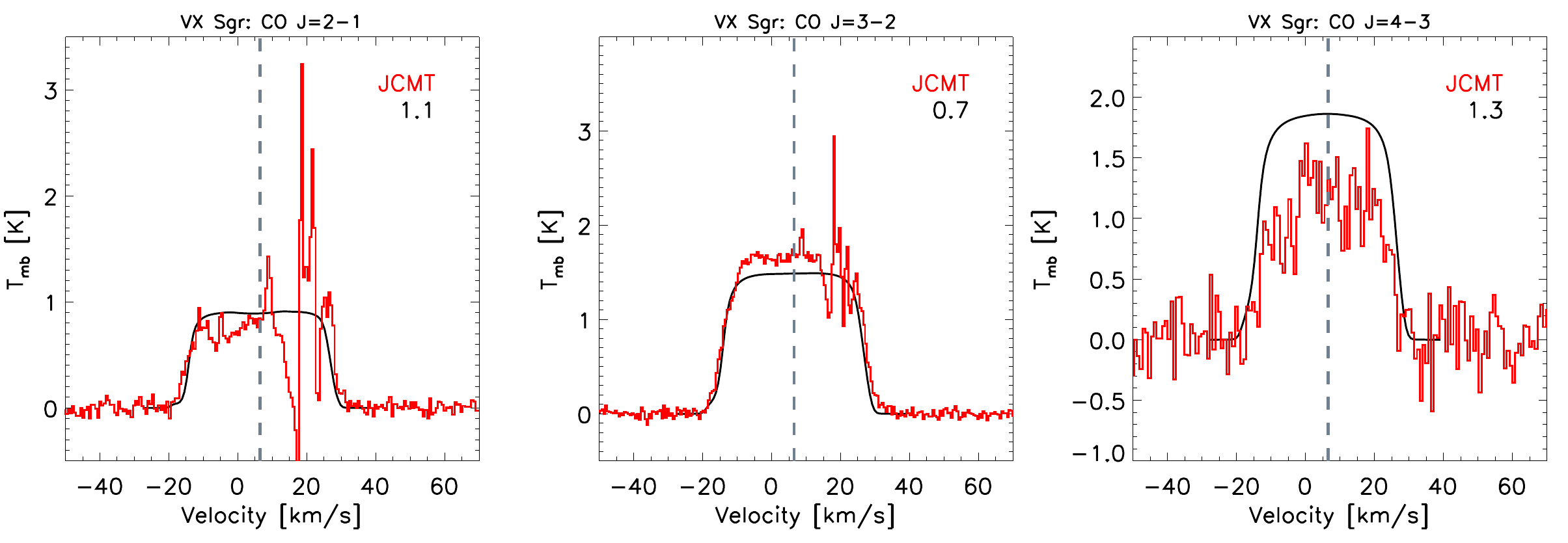}
   \end{minipage}
\caption{CO rotational line transitions observed towards $\alpha$~Ori (top panels), $\mu$ Cep (middle panels), and VX~Sgr (bottom panels) with the JCMT. The CO(2-1), CO(3-2), and CO(4-3) data are plotted as red histograms. Synthetic line profiles are overplotted as black solid lines. The vertical dashed grey lines indicate the LSR velocity. Accounting for the absolute flux uncertainty of $\sim$30\%, the JCMT data have been scaled by a factor, indicated in the upper right corner of each panel.}
\label{Fig:other_RSGs}
 \end{figure*}

The derived \Mdot$_{\rm{CO}}$ values can be compared to \Mdot$_{\rm{SED}}$ retrievals to determine the gas-to-dust ratio of the winds of those Galactic RSGs. Assuming a wind speed of 10\,km/s and gas-to-dust ratio of 100,  \citet{Verhoelst2009A&A...498..127V} derived that \Mdot$_{\rm{SED}}$ equals 6.3$\times$10$^{-8}$\,\Msun/yr (for a distance of 131\,pc) in the case of $\alpha$~Ori, and 2.9$\times$10$^{-7}$\,\Msun/yr (for a distance of 830\,pc) in the case of $\mu$~Cep. Similarly we use the values derived by \citet{Liu2017MNRAS.466.1963L} and \citet{Shenoy2016AJ....151...51S} for VX~Sgr and VY~CMa, respectively. Scaling for the difference in adopted distance and wind speed, we derive a gas-to-dust ratio of 297, 544, 226, and 18 for $\alpha$~Ori, $\mu$~Cep, VX~Sgr, and VY~CMa, respectively. The gas-to-dust ratio derived for VY~CMa might seem very low. However, it should be noted that \citet{Decin2006A&A...456..549D} derived that VY~CMa underwent a phase of high mass loss ($\sim$3.2$\times$10$^{-4}$\,\Msun/yr; see Table~\ref{Tab:outcome_other_RSGs}) some 1000\,yr ago. This phase took some 100\,yr and was preceded by a low mass-loss phase ($\sim$1$\times$10$^{-6}$\,\Msun/yr) taking some 800 yr. The current mass-loss rate (of $\sim$8$\times$10$^{-5}$\,\Msun/yr) was used to derive the gas-to-dust value, which obviously will change during other mass-loss phases. Moreover, the assumption of a spherical symmetry for the circumstellar envelope of VY CMa is obviously a simplification on the basis of which the dust mass-loss rate has been derived by \citet{Shenoy2016AJ....151...51S}; see discussion in that paper.

 \end{appendix}
 \end{document}